\documentclass[a4paper,fleqn,usenatbib]{mnras}

\usepackage{newtxtext,newtxmath}

\usepackage[T1]{fontenc}
\usepackage{ae,aecompl}

\usepackage{graphicx}	
\usepackage{amsmath}	
\usepackage{amssymb}	
\usepackage{dcolumn}

\title[Ionized outflows in local luminous AGN]{Ionized outflows in local luminous AGN: what are the real densities and outflow rates?}

\author[R. Davies et al.]{R. Davies,$^{1}$
D.~Baron,$^{2}$
T. Shimizu,$^{1}$
H.~Netzer,$^{2}$
L.~Burtscher,$^{3}$
P.~T.~de~Zeeuw,$^{1,3}$ \newauthor
R.~Genzel,$^{1}$
E.K.S.~Hicks,$^{4}$
M.~Koss,$^{5}$
M.-Y.~Lin,$^{6}$
D.~Lutz,$^{1}$
W.~Maciejewski,$^{7}$ \newauthor
F.~M\"uller-S\'anchez,$^{8}$
G.~Orban~de~Xivry,$^{9}$
C.~Ricci,$^{10,11}$
R.~Riffel,$^{12}$
R.A.~Riffel,$^{13}$ \newauthor
D.~Rosario,$^{14}$
M.~Schartmann,$^{1}$
A.~Schnorr-M\"uller,$^{12}$
J.~Shangguan,$^{1}$
A.~Sternberg,$^{2}$ \newauthor
E.~Sturm,$^{1}$
T.~Storchi-Bergmann,$^{12}$
L.~Tacconi,$^{1}$
and
S.~Veilleux$^{15}$
\\
$^1$ Max-Planck-Institut f\"ur extraterrestrische Physik, Postfach 1312, 85741, Garching, Germany\\
$^2$ School of Physics and Astronomy, Tel-Aviv University, Tel Aviv 69978, Israel\\
$^3$ Leiden Observatory, Leiden University, PO Box 9513, 2300 RA, Leiden, the Netherlands\\
$^4$ Department of Physics \& Astronomy, University of Alaska Anchorage, AK 99508-4664, USA\\
$^5$ Eureka Scientific, 2452 Delmer Street Suite 100, Oakland, CA 94602-3017, USA\\
$^6$ Institute of Astronomy and Astrophysics, Academia Sinica, Roosevelt Rd, Taipei 10617, Taiwan\\
$^7$ Astrophysics Research Institute, Liverpool John Moores University, IC2 Liverpool Science Park, 146 Brownlow Hill, L3 5RF, UK\\
$^8$ Physics Department, University of Memphis, Memphis, TN 38152, USA\\
$^{9}$ Space Sciences, Technologies, and Astrophysics Research Institute, Universit\'e de Li\`ege, 4000 Sart Tilman, Belgium\\
$^{10}$ N\'ucleo de Astronom\'ia de la Facultad de Ingenier\'ia, Universidad Diego Portales, Av. Ej\'ercito Libertador 441, Santiago, Chile\\
$^{11}$ Kavli Institute for Astronomy and Astrophysics, Peking University, Beijing 100871, China\\
$^{12}$ Departamento de Astronomia, Universidade Federal do Rio Grande do Sul, IF, CP 15051, 91501-970 Porto Alegre, RS, Brazil\\
$^{13}$ Departamento de F\'isica, Universidade Federal de Santa Maria, 97105-900 Santa Maria, RS, Brazil\\
$^{14}$ Department of Physics, Durham University, South Road, Durham, DH1 3LE, UK\\
$^{15}$ Department of Astronomy and Joint Space-Science Institute, University of Maryland, College Park, MD 20742-2421, USA
}

\date{Accepted XXX. Received YYY; in original form ZZZ}

\pubyear{2020}

\hypersetup{draft}
\begin{document}
\label{firstpage}
\pagerange{\pageref{firstpage}--\pageref{lastpage}}
\maketitle

\begin{abstract}
We report on the determination of electron densities, and their impact on the outflow masses and rates, measured in the central few hundred parsecs of 11 local luminous active galaxies.
We show that the peak of the integrated line emission in the AGN is significantly offset from the systemic velocity as traced by the stellar absorption features, indicating that the profiles are dominated by outflow. In contrast, matched inactive galaxies are characterised by a systemic peak and weaker outflow wing.
We present three independent estimates of the electron density in these AGN, discussing the merits of the different methods.
The electron density derived from the [SII] doublet is significantly lower than than that found with a method developed in the last decade using auroral and transauroral lines, as well as a recently introduced method based on the ionization parameter.
The reason is that, for gas photoionized by an AGN, much of the [SII] emission arises in an extended partially ionized zone where the implicit assumption that the electron density traces the hydrogen density is invalid.
We propose ways to deal with this situation and we derive the associated outflow rates for ionized gas, which are in the range 0.001--0.5\,M$_{\sun}$\,yr$^{-1}$ for our AGN sample.
We compare these outflow rates to the relation between $\dot{M}_{\rm out}$ and $L_{\rm AGN}$ in the literature, and argue that it may need to be modified and rescaled towards lower mass outflow rates.
\end{abstract}




\section{Introduction}
\label{sec:intro}

The evidence that outflows, whether driven by star formation or AGN, play a fundamental role in the evolution of galaxies is now long undisputed \citep{vei05} and numerous reiews have been written about different aspects of the subject.
Some of the most striking evidence comes from observations of galaxy clusters, in which mechanical feedback from radio jets heats the intra-cluster medium \citep{fab12}. This is associated with the more massive black holes in classical bulges and elliptical galaxies.
Large surveys have shown this appears to be distinct from radiative feedback which occurs at accretion rates above 1\% Eddington and is primarily associated with less massive black holes fed by secular process or minor mergers \citep{hec14}.
Through studies of ultrafast outflows, which have speeds in excess of $\sim0.1c$, a theoretical framework has been developed about the impact of the wind on the inter-stellar medium, and how its dependency on the bulge mass might lead to the observed relation between black hole mass and global properties of the stellar bulge \citep{kin15}.
A broader framework, encompassing cosmological models of galaxy evolution, highlights the role that outflows are expected to have in order for such models to reproduce the observed galaxy scaling relations \citep{som15}.
Through sensitive observations at millimetre wavelengths, it has been realised that cool outflows of neutral and molecular gas play perhaps an even more significant role than the ionized outflows \citep{vei20}.
However, the amount of gas in each phase of the outflow, whether the outflow escapes the host galaxy, or if it has a significant impact on the global star formation rate, are not yet firmly established although some efforts have been made in this direction \citep{mor17,fio17}.

As emphasized by \citet{har18}, for ionized outflows, a large part of the uncertainty is directly linked to the density of gas in the outflow.
The reason is straightforward to show because, at a fixed (i.e. the observed) line luminosity, the derived mass of ionized gas, and hence outflow rate, is inversely proportional to the adopted density.
The line luminosity is 
$L_{\rm line} = \gamma_{\rm line} \, n_{\rm e} \, n_{\rm p} \, V f$
where $\gamma_{\rm line}$ is the appropriate volume emissivity, $n_{\rm e} \sim n_{\rm p}$ is the electron or equivalently ion density, $V$ is the volume, and $f$ the filling factor.
An additional implicit assumption is that $n_{\rm e} \sim n_{\rm H}$, that is the clouds are fully ionized. 
In this case, the mass of ionized gas in that volume is 
$M_{\rm out} = \mu m_{\rm H} \, n_{\rm p} \, V f$
where $\mu m_{\rm H}$ is the effective atomic mass.
Together these show that the dependencies of the derived ionized gas mass are
$M_{\rm out} \propto L_{\rm line} \, / \, (\gamma_{\rm line} \, n_{\rm e})$
and equivalently the dependencies for the outflow rate are 
$\dot{M}_{\rm out} \propto L_{\rm line} \, v_{\rm out} \, / \, (\gamma_{\rm line} \, n_{\rm e} \, r_{\rm out})$.
Thus, a reliable assessment of the density in the outflow, that is appropriate to the spatial scales being measured, is an essential ingredient for deriving the outflow mass and rate.

In the literature, a wide range of different densities (either assumed or measured), covering several orders of magnitude, have been used when deriving quantities related to ionized outflows driven by active galactic nuclei (AGN). 
These include, at low and high redshift:
100\,cm$^{-3}$ \citep{liu13,rif13,har14,kak16,rup17};
200\,cm$^{-3}$ \citep{fio17};
500\,cm$^{-3}$ \citep{sto10,car15,rif15};
1000-1500\,cm$^{-3}$ \citep{sch16b,per17,for19,shi19};
5000\,cm$^{-3}$ \citep{mue11};
and in some instances densities of 10$^4$--10$^5$\,cm$^{-3}$ have been reported \citep{hol11,rosm18,san18,bar19}.
Among this plethora of values, lower densities are often adopted or measured for larger scales of 1--10\,kpc, while the higher densities apply to smaller 0.1--1\,kpc scales.
This tendency is also reflected in spatially resolved studies \citep[e.g.][]{bar18,fre18,kak18,shi19,nas19,hin19} which tend to show that $n_{\rm e}$ decreases with radius.

In this paper, we make use of the high quality spectroscopic data available for the LLAMA (Local Luminous AGN with Matched Analogues) survey \citep{dav15} to derive electron densities in the outflowing gas.
These are then used to estimate the outflow rates, which are compared to well known relations between AGN luminosity and outflow rate.
The paper begins with a description of the sample and observations in Sec.~\ref{sec:obs}, together with estimates of the systemic velocity and a discussion of how the stellar continuum is subtracted.
In Sec.~\ref{sec:ism_outflow} we argue, based on the line ratios and profiles, that the entire line profile in these AGN is dominated by outflow, and any systemic component is sub-dominant.
Because of this, when deriving densities, we integrate over the complete emission lines.
The outflow densities are derived using three independent methods, which are summarized in Sec.~\ref{sec:density}.
They are the standard method of the [SII]\,$\lambda$6716/6731 doublet ratio; a method developed by \citet{hol11} which makes use of auroral and transauroral lines in the ratios 
[SII]\,$\lambda$(6716+6731) / [SII]\,$\lambda$(4069+4076) and 
[OII]\,$\lambda$(3726+3729) / [OII]\,$\lambda$(7320+7331);
and a method recently introduced by \citet{bar19} that is based on the definition of the ionization parameter.
We discuss the merits of the various methods and present the densities derived from our sample of AGN for each of them.
Using the density measure that we argue is most appropriate, in Section~\ref{sec:rate} we assess whether the derived outflow rates extend the lower luminosity end of the $L_{\rm AGN} - \dot{M}_{\rm out}$ relation proposed by \citet{fio17}.
We finish with our conclusions in Section~\ref{sec:conc}.


\begin{table*}
\caption{Summary of Observations}
\label{tab:obs}
\begin{tabular}{llrrll}
\hline
Object & type $^a$ & $v_{\rm sys}$ (km\,s$^{-1}$) $^b$ & \# obs & dates & seeing (\arcsec)$^c$\\
\hline
ESO\,137-G034 & Sy\,2      & 2763 & 3 & 2015.05.18, 2015.05.20, 2015.06.23 & 1.03, 1.54, 0.80 \\
MCG-05-23-016 & Sy\,1.9    & 2513 & 2 & 2014.01.21 & 1.06, 0.95 \\ 
NGC\,2110     & Sy\,2 (1h) & 2329 & 2 & 2013.11.24 & 0.58, 0.50 \\ 
NGC\,2992     & Sy\,1.8    & 2328 & 2 & 2014.02.25, 2014.02.26 & 0.73, 0.76 \\ 
NGC\,3081     & Sy\,2 (1h) & 2385 & 2 & 2014.02.19 & 0.58, 0.80 \\
NGC\,5506     & Sy\,2 (1i) & 1963 & 1 & 2016.03.03 & 0.60 \\
NGC\,5728     & Sy\,2      & 2775 & 2 & 2015.05.12, 2015.07.15 & 0.81, 0.78 \\
NGC\,7582     & Sy\,2 (1i) & 1598 & 2 & 2016.07.26, 2016.08.08 & 0.38, 0.72 \\ 
ESO\,021-G004 & Sy\,2      & 2834 & 1 & 2016.08.01 & 0.41 \\
NGC\,1365     & Sy\,1.8    & 1630 & 2 & 2013.12.10 & 1.34, 1.18 \\
NGC\,7172     & Sy\,2 (1i) & 2568 & 2 & 2015.08.11 & 1.13, 1.44 \\
ESO\,093-G003 & inactive   & 1828 & 2 & 2014.01.21, 2014.03.20 & 1.44, 1.24 \\ 
ESO\,208-G021 & inactive   & 1075 & 2 & 2013.12.11, 2014.01.21 & 0.61, 1.05 \\ 
NGC\,0718     & inactive   & 1729 & 1 & 2015.12.04 & 0.60 \\ 
NGC\,1079     & inactive   & 1455 & 1 & 2013.11.22 & 0.80 \\ 
NGC\,1947     & inactive   & 1189 & 2 & 2013.12.22, 2014.02.07 & 0.67, 1.21 \\ 
NGC\,3175     & inactive   & 1099 & 2 & 2014.03.08, 2013.03.09 & 1.49, 0.63 \\ 
NGC\,3351     & inactive   &  781 & 2 & 2014.02.20 & 1.08, 1.06 \\ 
NGC\,3717     & inactive   & 1744 & 2 & 2014.03.21 & 1.37, 1.22 \\ 
NGC\,3749     & inactive   & 2746 & 2 & 2014.03.21, 2014.03.31 & 1.58, 1.46 \\ 
NGC\,4224     & inactive   & 2623 & 1 & 2015.05.12 & 0.79 \\ 
NGC\,4254     & inactive   & 2418 & 1 & 2016.06.01 & 0.42 \\ 
NGC\,5037     & inactive   & 1916 & 2 & 2015.05.12, 2016.02.03 & 1.18, 0.76 \\ 
NGC\,5921     & inactive   & 1484 & 2 & 2015.06.15, 2015.06.23 & 0.82, 0.86 \\ 
NGC\,7727     & inactive   & 1837 & 1 & 2015.08.08 & 0.70 \\ 
IC\,4653      & inactive   & 1535 & 2 & 2015.05.18, 2016.06.03 & 0.93, 0.55 \\ 
\hline
\end{tabular}

$^a$ Classifications are taken from \citet{dav15} and references therein; 1i indicates the broad lines have been observed in the near-infrared, and 1h that hidden broad lines have been detected via polarisation measurements.\\
$^b$ We report heliocentric systemic velocity. The uncertainty, including systematic effects, is 13\,km\,s$^{-1}$; for internal referencing, the uncertainty is typically 3\,km\,s$^{-1}$. NGC\,1365 and NGC\,5506 are special cases, described in Sec.~\ref{sec:systemic}.\\
$^c$ Seeing is that reported for the beginning of the observations by the seeing monitor at the VLT. In the few cases this is not available, an alternative measurement close in time is used.
\end{table*}

\section{Sample and Observations}
\label{sec:obs}

\subsection{Sample}

We make use of the LLAMA sample of active and inactive galaxies.
\citet{dav15} provide the rationale for the sample, and a detailed description of its selection.
The key aspect is that these are taken from the all-sky flux limited 14-195\,keV 58-month Swift BAT survey \citep{bau13} in such a way as to create a volume limited sample of active galaxies that is as unbiased as possible, for detailed study using optical spectroscopy and adaptive optics integral field near-infrared spectroscopy. 
The sole selection criteria were 
$z < 0.01$ (corresponding to a distance of $\sim$40\,Mpc), 
$\log{L_{\rm 14-195keV}}$ [erg\,s$^{-1}$] $>$ 42.5 (using redshift distance), and
$\delta < 15^{\circ}$ so that they are observable from the VLT. 
This yielded 20 AGN. 
A set of inactive galaxies were selected to match them in terms of 
host galaxy type, 
mass (using H-band luminosity as a proxy), 
inclination, 
presence of a bar, and 
distance.
More details, as well as a comparison of the active and inactive galaxies, are given in \citet{dav15}.
Although small, this volume limited sample is sufficient for detailed studies of emission line ratios, the molecular and ionized gas kinematics and distributions, as well as the stellar kinematics and populations, in the nuclear and circumnuclear regions.
And the ability to compare the results to a matched sample of inactive galaxies has been essential in many of the studies so far, including the analysis presented here.
These studies include:
the physical properties of, and extinction to, the broad line region \citep{sch16a};
the respective roles of host galaxy and environment in fuelling AGN \citep{dav17};
the molecular gas content and depletion time on kiloparsec scales \citep{rosd18};
the nuclear stellar population and kinematics \citep{lin18};
the black hole masses and location in the $M_{\rm BH}-\sigma*$ plane \citep{cag20};
and the nuclear star formation histories \citep{bur20}.

For this paper, we have limited the subsample of active galaxies to include only Sy\,1.8--2 types, which gives 11 objects.
This sub-selection ensures that any broad line emission is weak enough that it does not limit our ability to fit the complex profiles of the narrow line profles or to separate them from the stellar absorption features.
For the analysis using the auroral and transauroral lines, we had to limit the AGN further, omitting those for which the full set of necessary [SII] and [OII] lines could not be measured reliably (ESO\,021-G004, NGC\,1365, and NGC\,7172).
This does not introduce a line ratio bias because these are simply the three objects with the faintest line fluxes.
The subsample of inactive galaxies with Xshooter observations contains 18 objects.
We exclude three -- NGC\,1315, NGC\,2775, and NGC\,5845 -- which have no measurable line emission.
Thus, there are 11 active and 15 inactive galaxies in the sample analysed in this paper.

\subsection{Observations and Data Reduction}

Details of the observations, data reduction, and spectral extraction, for all these objects are given in \citet{bur20}.
We have simply re-used those spectra.
We list in Table~\ref{tab:obs} the number of observations that were performed for each object, with their dates.
Very briefly, we used the integral field unit (IFU) in Xshooter and integrated for approximately 1\,hr for each observation of each source.
The instrument re-arranges the 3 slitlets across the IFU to form a single slit, which is dispersed as a single 2D spectrum.
The ESO pipeline version 2.6.8 \citep{mod10} was used to process the data in its default mode. The main steps included bias subtraction, removal of the inter-order background, flatfielding, subtraction of the sky exposure, and rectification. 
Wavelength calibration was performed using arclines, for which the pipeline provides wavelengths in air. For this reason, the analysis in this paper also uses wavelengths of spectral lines and features defined in air rather than vacuum (being inconsistent in this respect leads to an offset of 80\,km\,s$^{-1}$ when measuring systemic velocity).
Because we used Xshooter in its IFU mode, the 3 slitlets across the IFU were then put back together to create a datacube for each arm. 
A 1D spectrum was extracted in a 1.8\arcsec$\times$1.8\arcsec\ aperture from the UVB and VIS arms which cover 300--560\,nm and 560--1024\,nm respectively.
This aperture corresponds to 3$\times$3 spatial pixels in the IFU field of view (i.e. covering its full width), and was chosen over a 1$\times$1 pixel aperture to minimize the impact of imperfect positioning of the target in the IFU, variations in seeing, and to increase the signal-to-noise.
The location of the extraction aperture on the object was determined in the NIR arm (although those data are not used in the analysis presented here), which allows the best centering because it is least affected by extinction and, for the AGN, includes bright non-stellar continuum.
For the UVB and VIS arms, the extraction aperture location was adjusted with respect to the NIR according to the mean differential atmospheric refraction during the observation.
Observations of the telluric and flux calibrator stars confirmed the validity of this approach.
Flux calibration, and correction of telluric absorption, were performed on the extracted 1D spectra.
A comparison of the count rates from one particular calibration star that was observed multiple times over a period of 21 months indicates that the uncertainty in flux calibration is \la 2\% in the wavelength range relevant to this work.
Finally, the calibrated 1D spectra from the UVB and VIS arms were merged by scaling the overlapping region to provide a consistent flux calibration over 300\,nm to 1.0\,$\mu$m.
The spectral resolution is $R=8600$ in the UVB arm and $R=13500$ in the VIS arm. We have not corrected for this difference, partly because the unresolved linewidths in both arms are rather smaller than those discussed in this paper, and also because the difference does not affect the measurement of line fluxes and ratios, nor of outflow velocities, that are pertinent to our analysis.
Specifically, within the uncertainties of the small numbers analysed here, the distributions of the stellar velocity dispersion for both the active and inactive galaxies are the same, with values for individual objects in the range $\sigma_* \sim 100$--200\,km\,s$^{-1}$.
In all cases the measured dispersion is at least a factor 3 greater than the instrumental broadening (more typically a factor 6-12), and so any correction for that will have a negligible effect.

For each object, the resulting spectra from each observation were analysed separately in order to provide independent estimates of each measured property.
By doing this, we are able to confirm that the uncertainties derived for these properties are consistent with the differences between the values measured from each spectrum.
When more than one observation is available for an object, the resulting values reported in this paper are the mean of the values derived from the individual observations weighted by their respective uncertainties.

\subsection{Stellar Continuum Fitting and Subtraction}
\label{sec:contsub}

Many of the emission lines we measure are superimposed on stellar absorption features, which need to be removed.
Since our sole aim is to remove the stellar features, adopt a simple approach for removing them, using a theoretical library of individual stellar spectra from \citet{coe14} covering a range of temperature and surface gravity.
The advantage is that the high resolution of the library enables us to retain the full resolution of the spectra.

We fit intermediate length $\sim$500\,\AA\ segments independently.
This was done in order to avoid difficulties asociated with a single very long wavelength baseline (e.g. due to irregularities in the continuum shape that arise during data processing, or assumptions such as applying only a single value of extinction to all the stellar populations), while still enabling each segment to include multiple spectral features and allowing strong emission lines to be masked or avoided.
We therefore performed this step separately in advance of fitting the emission lines. 
Nevertheless, when fitting individual lines or doublets, we do still include a linear function that allows for adjustment of the remaining continuum baseline locally over the $<100$\,\AA\ regions pertinent to those lines.

We limited the range of template parameters to 
temperatures $4000 \le T_{\rm eff}\,(K) \le 7000$ and 
surface gravities $2.5 \le \log{g\,(cm\,s^{-2})} \le 4$, with solar abundances.
This range was selected for two reasons. It provides enough flexibility to account for the variety of continua seen in the data. But it also avoids creating a fit with overly deep H$\alpha$ absorption, which is important in that spectral segment because the hot star contribution is less well constrained than in the segment with H$\beta$. The consistency of the fits to different spectral segments in the same object was checked by comparing the relative contributions of different stellar types for each segment.
The fitting was done using the Penalised Pixel Fitting routine pPXF \citep{cap04,cap17}.

\subsection{Emission and Absorption Line Fitting}
\label{sec:linefitting}

We have characterised the emission and absorption lines using one or more Gaussian functions. We emphasize that multiple Gaussian components are used only to match the observed line profiles, and we do not consider this in terms of a decomposition into different kinematic components. Indeed, in Sec.~\ref{sec:ism_outflow} we argue that for the observations presented here, one cannot attribute a physical meaning to the individual Gaussian components for these line profiles.

As a first step, we subtract the fitted stellar continuum over an intermediate wavelength range, as described in Sec.~\ref{sec:contsub}.
The emission lines were then fitted using Gaussian components together with a local linear continuum.
For the strong lines (e.g. Figs.~\ref{fig:sii_linefits} and~\ref{fig:oiii_linefits_s}) this was done simultaneously in a single fitting process.
For the weaker lines discussed in Sec.~\ref{sec:den_ta} (see Figs.~\ref{fig:ta_linefits1} and~\ref{fig:ta_linefits2}), there are other emission lines close around the line of interest, and so the continuum is best constrained using separate line-free regions.
Therefore, for all the lines discussed in that section, the continuum was fitted separately using regions that were offset on both sides of the line of interest.
The impact of the uncertainty of this linear fit was included in the uncertainty of the emission line fit.
In each case the number of components used to fit the emission lines was kept to a minimum. The number was only increased if visual inspection of the residuals from the [SII] and [OIII] line fits showed coherent structure above the noise level that could be substantially reduced by including another component in the fit.
As such, the number of components required is simply a reflection of the complexity of the profile shape.
We found that only one or two components were needed for each line in the inactive galaxies, while the active galaxies required more typically three components (and for NGC\,5728 four components).
In order to ensure that the local continuum level and slope was well matched, we included a linear function in the fit. 
This operates on the short wavelength range associated with the individual emission lines and so is independent of the stellar continuum fit described above, the purpose of which was to compensate for the complex shape associated with absorption features.
We used the routine {\it mpfitfun} \citep{mar09}; and we propagated the uncertainties of the parameters provided by this routine to the relevant properties using Monte Carlo techniques.
For doublet lines, we have applied constraints to the fit based on two main premises: (i) that the separation of the two lines is set by atomic physics, and (ii) to make sure that the profiles of the two lines in each doublet are the same.
Hence when deriving a doublet line ratio, we are measuring a single ratio for the full line profile.
In addition, when fitting the four doublets described in Sec.~\ref{sec:den_ta}, we expect that the kinematic properties of the lines will be similar although not necessarily exactly like. Based on the properties of the highest signal-to-noise doublet, [SII]\,$\lambda$6716,6731, we have therefore constrained the kinematic structure of the weaker doublets.
Specifically for the these three doublets, we have fixed the relative centering and FWHM of the various Gaussian components, while allowing their relative strength to vary between each doublet to provide some flexibility in terms of differences between the doublets.

Table~\ref{tab:obs} shows that three of our targets (MCG-05-23-016, NGC\,2992, and NGC\,1365) are type 1.8--1.9, and therefore have measurable broad line emission at H$\alpha$ and one case also H$\beta$.
Detailed fits to these and other HI lines have been performed by \citet{sch16a} in order to derive the extinction to the broad line region (BLR) as well as constrain the excitation.
Here, we use a single Moffat function to represent the BLR, and fitted it simultaneously with the narrow lines. The example of NGC\,1365, discussed later in Sec.~\ref{sec:systemic}, is shown in Fig.~\ref{fig:n1365}; the other two objects are more straightforward to fit and show no broad H$\beta$.
The BLR properties we find are consistent with those reported by \citet{sch16a}.

\subsection{Uncertainties}
\label{sec:errors}

This section outlines how we have derived uncertainties throughout the paper, and what is included in the values quoted.
In general, one can consider three types of uncertainty: interpretational, systematic, and random. 

Interpretational uncertainties are related to assumptions or simplifications made when analysing the data. They are discussed at the places where they arise in the paper.
In terms of the systemic velocity of the galaxies, they are related to whether the spectral lines used do indeed trace the stellar population, or if there is a contribution from the ISM which might bias the measured velocity. Sec.~\ref{sec:systemic} shows that this is negligible.
In terms of outflow rate, interpretational uncertainties dominate. In Sec.~\ref{sec:rate} we highlight that the different classes of simple outflow models differ intrinsically by a factor 3. In addition, the adopted extinction law has an impact on de-reddened line fluxes. Another common assumption is that all the gas emission within the beam or aperture comes from clouds that are identical in terms of their size, mass, internal structure, and emergent line emission.
How this simplification propagates into uncertainties on the derived parameters is unknown. 
None of these effects are explicitly included in the quantitative uncertainties given here (nor in the literature more generally), but their existence should still be borne in mind.

Systematic uncertainties, in the context here, are associated with calibration and processing of the data.
They play an important role for the systemic velocity and, as we show below, dominate the uncertainties in that case.
However, for most of the analysis in this paper, we deal with velocities that are referenced internally within our own data.
As such, they affect all velocity measurements in the same way and so can be ignored.

Random (or statistical) uncertainties depend on the data quality, which is largely determined by the photon and read noise.
They also depend on how well the data processing procedures minimize these effects in the final data product.
These uncertainties are based on formal error propagation and so it should be borne in mind that they should always be qualified as ``under the assumptions made''.
This is particularly important here because the spectral resolution and signal-to-noise of the emission lines we measure mean that these uncertainties are small.

When deriving the systemic velocity, we have used three methods to assess the systematic and random uncertainties.
The first uses the uncertainties estimated internally by {\it mpfitfun} for each of the three absorption lines. We have combined them to estimate a single uncertainty for each spectrum; and then further combined the values for all the spectra of each object. This yields typical uncertainties of 3\,km\,s$^{-1}$.
As an alternative, for each spectrum, we have calculated the standard error of the measured velocity offset of each line from their median velocity. Because only three values are available, this provides an estimate that is larger than the true uncertainty. Combining the values for all the spectra of each object yields uncertainties around 5\,km\,s$^{-1}$.
These are the relevant numbers to use when comparing the velocities of the emission lines to the systemic velocity.
The third method we have used includes systematic effects, because it explicitly takes into account variations due to different observation times, dates, conditions, and calibrations.
For the objects with two or more spectra, we have considered the difference in the estimated systemic velocity between them.
The standard deviation of these differences is 13\,km\,s$^{-1}$.
This is the value that should be used when comparing the systemic velocity to other (external) measurements.

When estimating uncertainties on properties of the emission lines, we have also made use of the uncertainties derived internally by {\it mpfitfun}.
These correspond to parameters of the component Gaussians and the linear continuum.
Since, as described in Secs.~\ref{sec:contsub} and~\ref{sec:linefitting}, the continuum is fit in two steps, it is important to note that the residual structure from the initial fit is taken into account when evaluating the noise in the spectrum, and that the residual offset and slope are removed by the second local fit.
The parameter uncertainties are used to create 1000 realisations of the line profile.
In this Monte Carlo approach we implicitly make the conservative assumption that these uncertainties are uncorrelated.
The suite of line profiles then allows an estimate of the uncertainty on the higher level properties used in our analysis, which focus on line fluxes, ratios, and velocities.
We have performed a cross-check on this process using a simple and robust independent estimate of the zero moment and its error.
We define a spectral region where the fitted line flux is above 5\% of its peak, and measure the line flux within this region.
We also measured the noise of the residual which, scaled by the square root of the number of pixels, yields the uncertainty of the line flux.
While this method is simple and robust, it cannot take into consideration the constraints included in the fitting procedure, for example that the profiles of the two lines in the doublet are tied so that they are the same
Thus, while one can expect the error estimates to be generally comparable, one should also expect that there are some differences.
This is what we find for all the lines on which we have performed the comparison, namely the [SII], [OII], and [OIII] doublets, and the H$\beta$ line.
We conclude that while the uncertainties due to random noise in the data are very small, this is a reflection of the data quality in the spectra.

\subsection{Systemic Velocities}
\label{sec:systemic}

A measure of each object's systemic (heliocentric) velocity, $v_{\rm sys}$, that is independent of the emission lines, is an important aspect of the analysis presented here.
For all except two of the objects, we have used the Ca\,II triplet lines since these are strong in a wide variety of stellar types, relatively narrow, and generally free from strong line emission, making them an ideal tracer of internal stellar kinematics \citep{dre84}.
And in contrast to the Ca\,II~H and K lines, they are not contaminated by interstellar absorption. 
This is illustrated for measurements of star cluster F in M\,82 by \citet{gal99}, who found that the Ca\,II~K line had a velocity that differed by 150\,km\,s$^{-1}$ from the triplet lines.
The former matched Na\,I~D while the latter velocities were similar to Mg\,I~b.
We fitted a Gaussian profile separately to each of the Ca\,II triplet lines at 8498\,\AA, 8542\,\AA, and 8662\,\AA, including a local estimate of the continuum as part of the fit as described in Sec.~\ref{sec:linefitting}. These can be seen in Fig.~\ref{fig:caii_linefits}.

In the case of NGC\,1365 and NGC\,5506, the Ca\,II features were filled with strong Paschen H\,I emission.
Instead, for these two objects we have explored other features to estimate $v_{\rm sys}$. These are specifically the Mg\,I~b triplet around 5175\,\AA, the Na\,I~D doublet around 5890\,\AA, and the K\,I doublet around 7680\,\AA. The Mg\,I lines are strong in the spectra of K-M type stars, and generally absent from interstellar absorption spectra. The Na\,I, and especially K\,I, lines are present in the spectra of cooler stars but need to be treated cautiously because they can also arise from, and in some cases are dominated by, interstellar absorption \citep{hob74,sch04,raj14}. This means that any estimate of systemic velocity using these lines may be biased by a contribution from a neutral outflow \citep{rup05a,rup05b}.

For NGC\,1365, inspection of these regions (see Fig.~\ref{fig:n1365}) shows that the latter are dominated by the ISM while the Mg\,I lines are too weak to fit reliably. We have instead used the systemic velocity adopted by \cite{ven18}, because this yields a symmetric stellar velocity field over the central arcminute of the galaxy (G.~Venturi, priv. comm.), which is a robust way to derive $v_{\rm sys}$. As a visual confirmation of this value, we note that the Mg\,I lines are consistent with the wavelengths it implies. In contrast the Na\,I and K\,I lines are seen at velocities corresponding to $-110$\,km\,s$^{-1}$ and $+50$\,km\,s$^{-1}$. That absorption can be both redshifted and outflowing is possible if it is tracing the receding bicone against an extended stellar continuum. It is notable from Fig.~\ref{fig:n1365} that the velocity of the fit to the broad emission line matches that of the blueshifted absorption component, while the narrow emission lines are more associated with the redshifted absorption component. Such an association between between ionised and neutral components in an outflow has been analysed in detail for a post starburst E+A galaxy using spatially resolved data by \citet{bar20}.
In closing, we note that many references in the NASA/IPAC Extragalactic Database (NED) give redshifts closer to 1660\,km\,s$^{-1}$ which matches the velocity we measure for the narrow emission lines. As our analysis in Sec.~\ref{sec:ism_outflow} shows, this offset arises because the peak of the emission lines do not necessarily trace the systemic velocity. 

For NGC\,5506 the optical spectrum (see Fig.~\ref{fig:ngc5506}) shows an almost featureless continuum.
This suggests that one should be cautious of the systemic velocity of 1750\,km\,s$^{-1}$ quoted by \citet{bos15} because it was based on fitting the stellar continuum.
Instead, we have searched for a well-defined narrow stellar feature that can be robustly measured.
The only clear features are the Na\,I and K\,I absorption, which appear to be composed of a series of narrow lines. Based on our spectra of the other galaxies, estimating the systemic velocity from the Na\,I line in our sample yields a modest bias of $<10$\,km\,s$^{-1}$ although the scatter for individual objects is hgher. In the absence of other tracers, we estimate $v_{\rm sys} = 1962$\,km\,s$^{-1}$ using that doublet, and assess whether it is consistent with expectations. 
Crucially, the emission line profiles -- especially [OIII] -- show a distinctive dip at this velocity. In Sec.~\ref{sec:ism_outflow} we show that many of the AGN line profiles have a similar form, with the dip at systemic velocity. As we argue there, in models this represents the transition between the approaching and receding sides of bicone outflows. As such, we take this as confirmation that the systemic velocity is close to our estimate.
In closing, it is worth noting that there are a large number of redshift measurements in NED, typically giving 1800-1850\,km\,s$^{-1}$ based on optical emission lines. However, the emission line peaks are blueshifted with respect to our estimate of $v_{\rm sys}$; and fitting our data with a single Gaussian would yield a velocity for the line peak of 1810\,km\,s$^{-1}$, consistent with the published values such as 
1829\,km\,s$^{-1}$ \citep{vau91},
1853\,km\,s$^{-1}$ \citep{kee96}, and 
1828\,km\,s$^{-1}$ \citep{kos17}.

The systemic velocities we measure are given as heliocentric velocities in Table~\ref{tab:obs}.
Most of these are consistent with the values reported in the literature by NED.
However, in addition to NGC\,1365 and NGC\,5506 where we have referred to the literature to estimate $v_{\rm sys}$, there were three galaxies where the difference exceeded 100\,km\,s$^{-1}$.
We have checked these in order to confirm our measurements as follows:

\paragraph*{ESO\,021-G004}
The only redshift in NED is 2960\,km\,s$^{-1}$ from an HI measurement.
The HI source was matched to ESO\,021-G004 by \citet{doy05}, but its coordinates put it about 65\arcsec\ from the galaxy centre, beyond the optical size of the galaxy. Therefore, while the HI source may indeed be associated with this galaxy, we recommend caution with the cross-identification.

\paragraph*{NGC\,5037}
The value adopted by NED is based on an HI measurement from \citet{pis11}. This is likely a mis-identification, since the coordinates of the galaxy do not match those listed for it in their Table~2.
Instead, the velocity of 1904\,km\,s$^{-1}$ given in \citet{vau91} and the 1890\,km\,s$^{-1}$ reported by \citet{men08} are consistent with the velocity we find.

\paragraph*{IC\,4653}
A redshift of 1890\,km\,s$^{-1}$ is given in \citet{vau91}, but a more recent measurement of 1551\,km\,s$^{-1}$ based on the Mg\,II feature was reported by \citet{weg03}. Although both of these have sizeable uncertainties (87\,km\,s$^{-1}$ and 55\,km\,s$^{-1}$ respectively), it is difficult to account for such a large difference between them. We note that for this inactive galaxy, our data are characterised by a narrow line profile for both stellar absorption and line emission, which are centered at very similar velocities and show no indication for an outflow. The systemic velocity implied by these features is consistent with that reported by \citet{weg03}.

\begin{figure*}
\includegraphics[width=18cm]{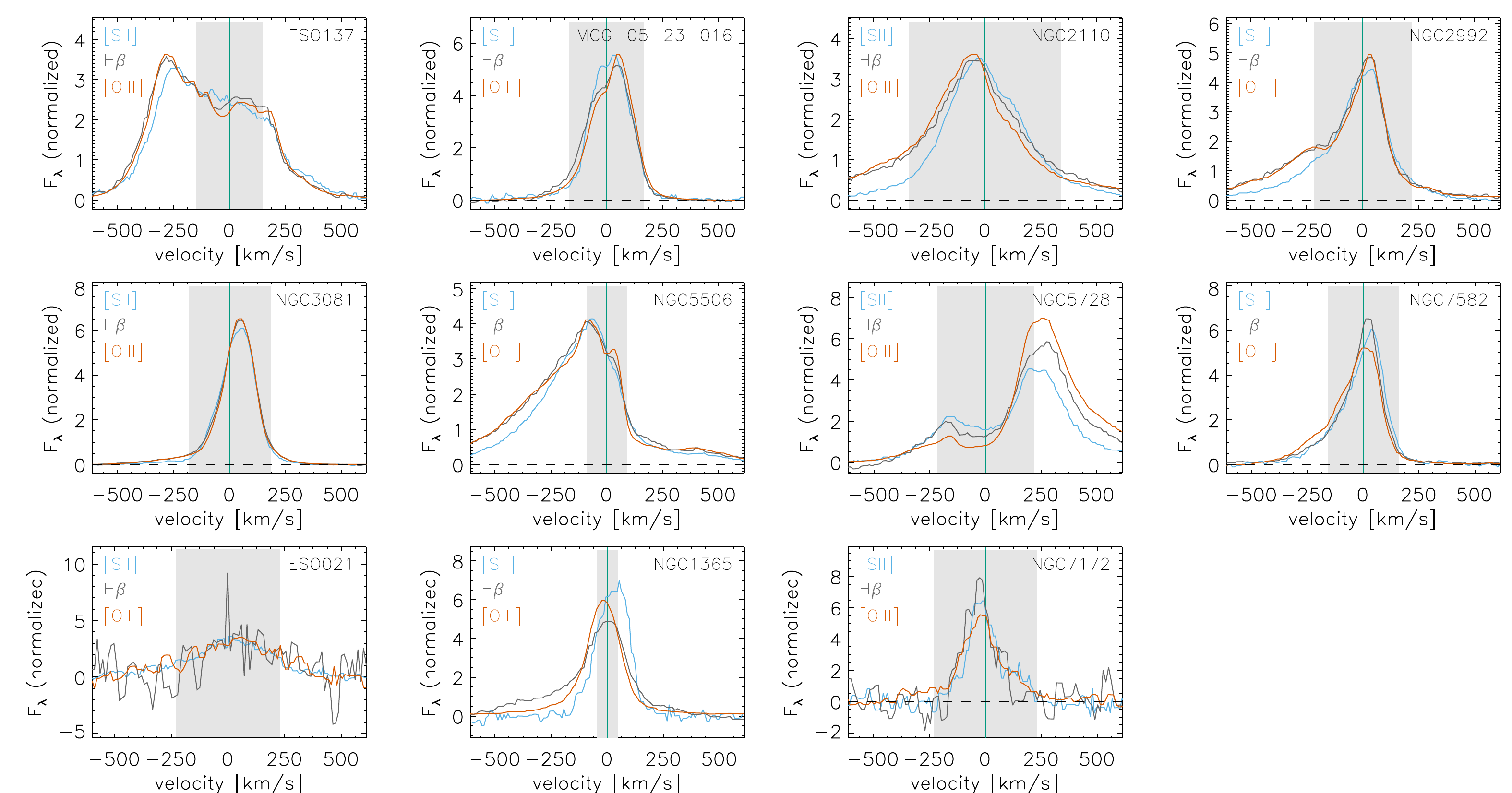}
\caption{Comparison of the central part of the line profiles of the AGN as a function of velocity, normalised so that they have the same flux within $\pm$250\,km\,s$^{-1}$.
The H$\beta$, [OIII], and [SII] profiles are shown in grey, red, and blue respectively. 
For visualisation purposes, the [SII] profile is a combination of the short side of the 6716\,\AA\ line and the long side of the 6731\,\AA\ line, scaled to match where they overlap.
These plots are of the data only, and not the fits.
For reference, the FWHM of the stellar absorption profile is indicated by the shaded grey region.
It is clear from this plot that the profiles show a variety of shapes that are generally inconsistent with the `systemic + outflow' decomposition often adopted. We argue that the whole profile is dominated by, and hence traces, outflow.
For reference, similar plots of the line profiles for the inactive galaxies are shown in Fig.~\ref{fig:profile_inactive} of the Appendix.}
\label{fig:profile}
\end{figure*}

\section{Interstellar medium vs outflow}
\label{sec:ism_outflow}

In this section we discuss whether the emission line profiles shown in Fig.~\ref{fig:profile} are dominated by the interstellar medium (ISM) or the outflow; and whether, for the AGN, there is in fact any measurable ISM component.
When referring to the ISM, we mean the ambient gas in the circumnuclear disk of the host galaxy.
When referring to the outflow, we mean any gas that is not just photoionized by the AGN, but also kinematically disturbed by it.
Spatially resolved studies \citep{fis17,fis18} have shown that there may be threshold radii, with gas closest to the AGN being driven out, beyond this a region where gas is still kinematically disturbed by the AGN, and finally a region where gas is photoionized by the AGN but remains undisturbed kinematically.
\citet{fis18} reported that for their sample of AGN, kinematically disturbed gas is seen out to $\sim$1.1\,kpc.
Our line profiles are integrated over a 1.8\arcsec\ box which corresponds to a much smaller radius of $\sim$150\,pc.
Even though, based on L$_{\rm [OIII]}$, the AGN in our sample are an order of magnitude less luminous, we would still expect that at these radii the kinematics should be dominated by the AGN rather than the host galaxy.
In the following, we examine whether the excitation and kinematical properties of the AGN emission lines, in comparison to the matched inactive galaxies, support this asssumption.

\begin{figure*}
\includegraphics[width=8.5cm]{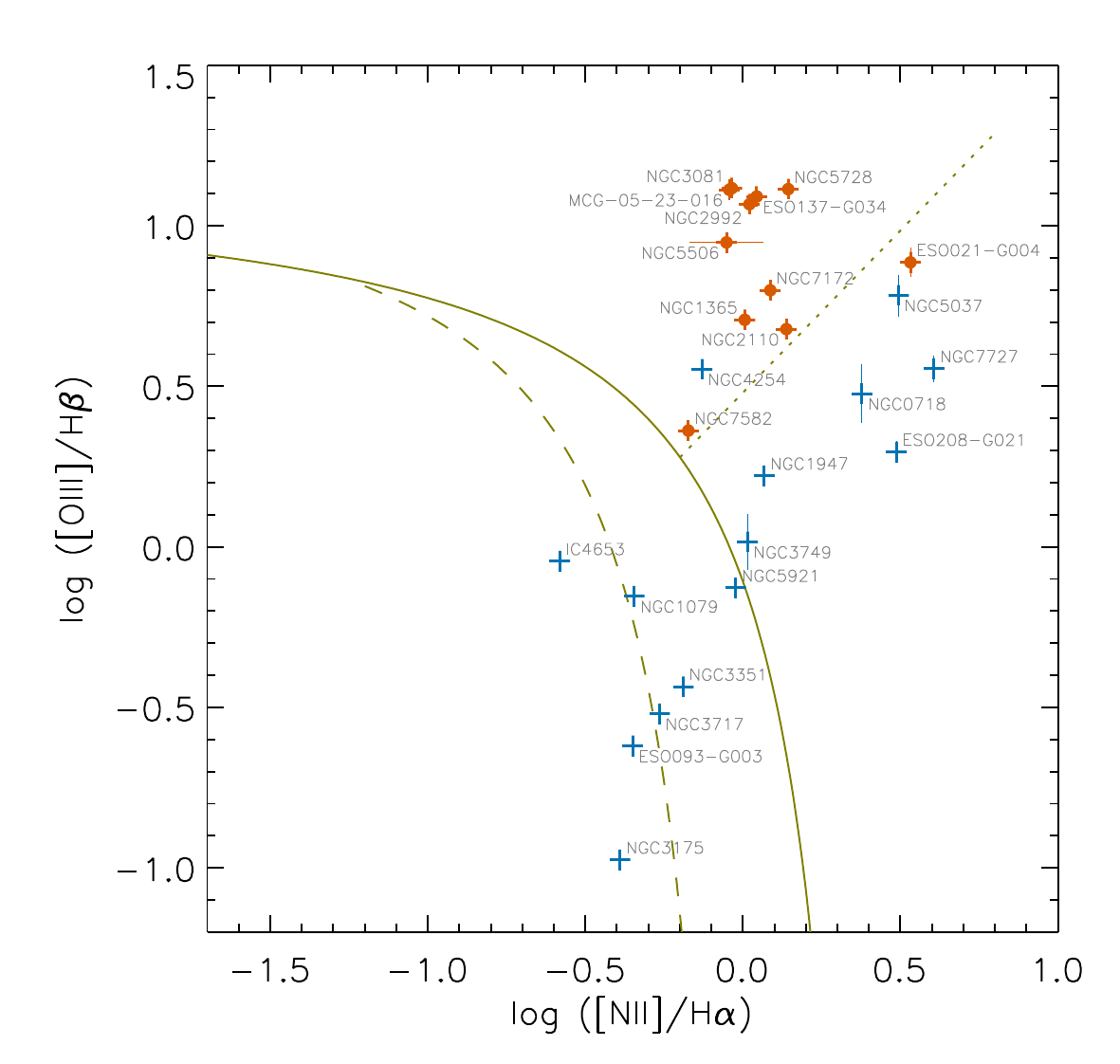}
\includegraphics[width=8.5cm]{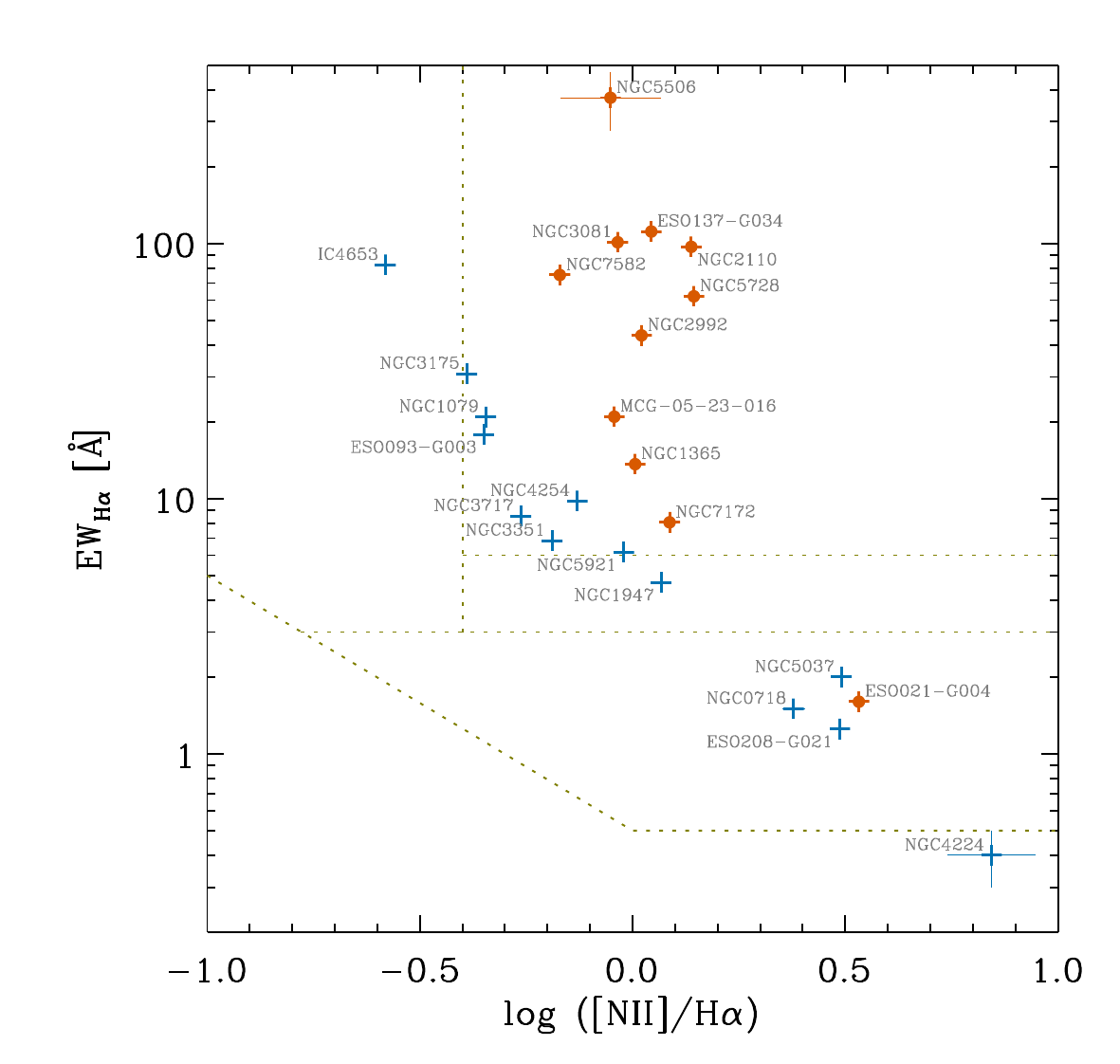}
\caption{Comparison of measured emission line quantites for AGN (red with filled circles) and inactive galaxies (blue).
Left: standard [NII]/H$\alpha$ versus [OIII]/H$\beta$ diagnostic ratios, with the \citet{kew01} extreme starburst line (solid) and \citet{kau03} classification line (dashed), as well as the \citet{cid10} Seyfert/LINER coarse separation (dotted).
Among the inactive galaxies, NGC\,4224 is omitted because H$\beta$ was not detected.
Right: The WHAN plot of the equivalent width of H$\alpha$ versus the [NII]/H$\alpha$ ratio. The dotted lines are shown for reference to Figs.~1 and~6 of \citet{cid11}. The key point here is that the AGN and inactive galaxies form distinct sequences.
These panels show that the emission lines in the AGN sample are dominated by the AGN photoionization.}
\label{fig:bpt}
\end{figure*}

\begin{figure*}
\includegraphics[width=8.5cm]{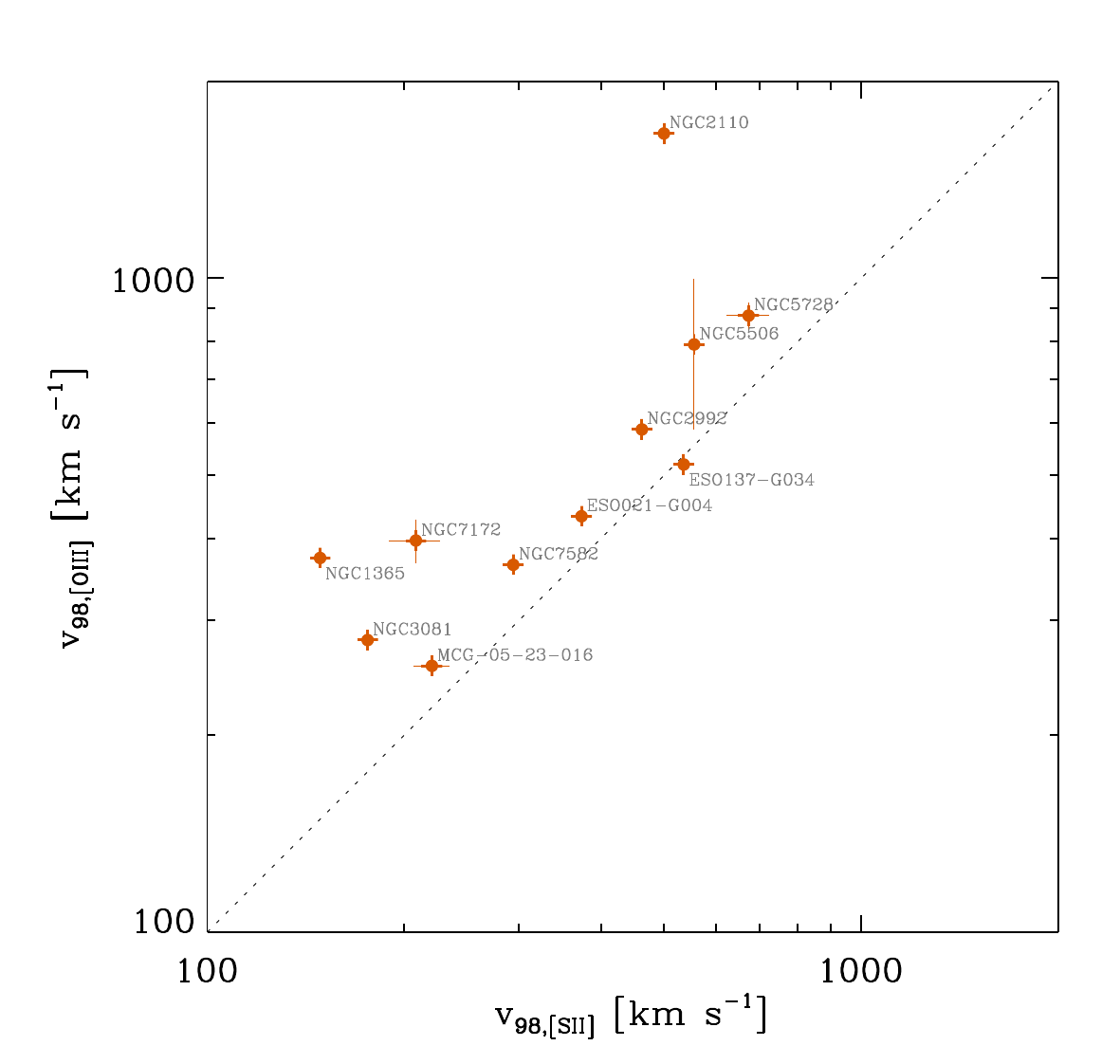}
\includegraphics[width=8.5cm]{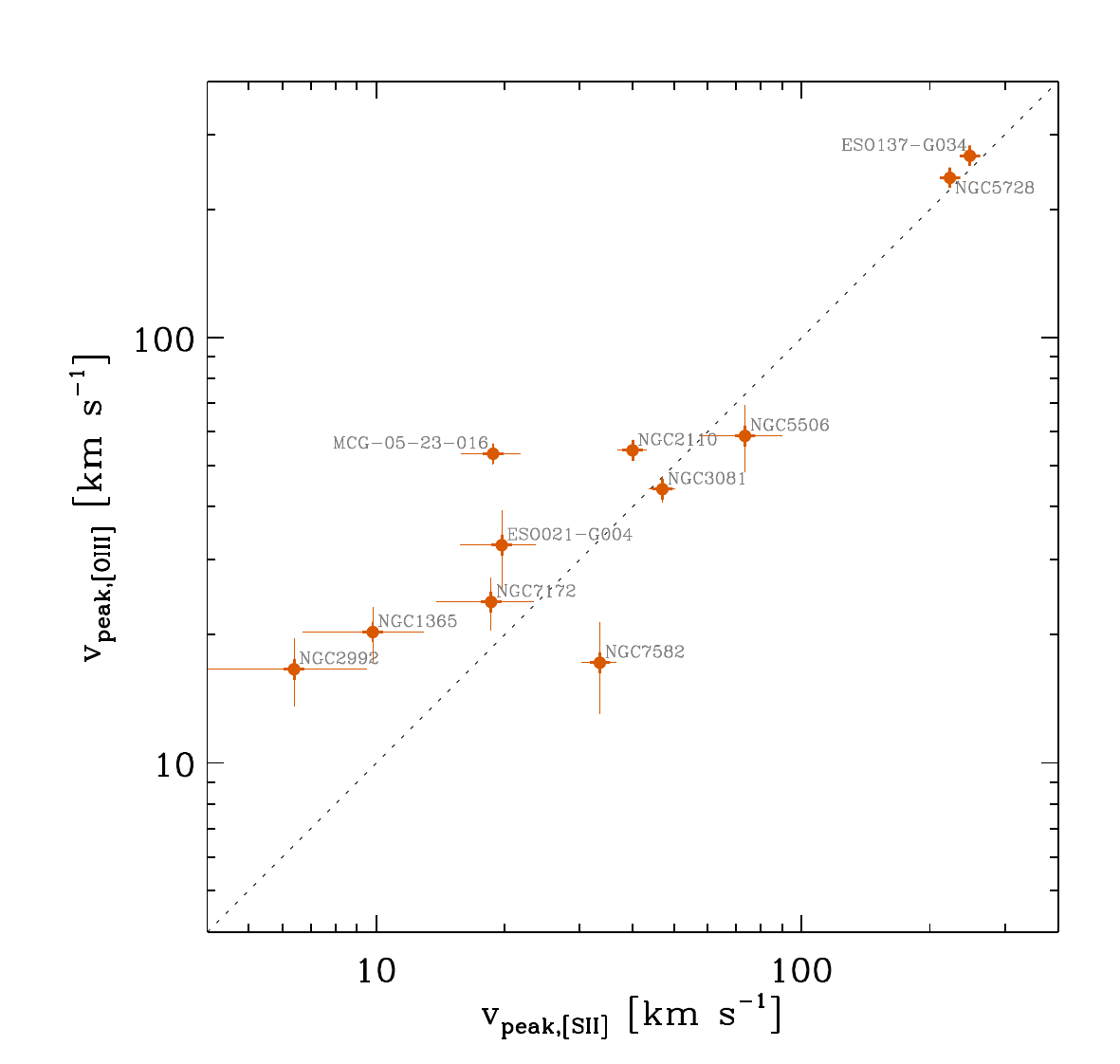}
\caption{Comparison of the [SII] and [OIII] emission lines for the AGN, in terms of peak offset $v_{\rm peak}$ from systemic (left) and outflow velocity $v_{\rm 98}$ (right). The dotted lines indicating a 1:1 correspondance show that in each case the two lines are very similar. These show that there is little difference between the kinematics traced by these two emission lines. The left panel reflects the well-known result that [OIII] has broader wings than [SII] \citep{vei91}; in this case the velocity difference is about $\sim$25\%. Only NGC\,2110 is an outlier, and inspection of Fig.~\ref{fig:profile} shows that its [OIII] line does have very broad wings. In the right panel, the peak velocity offsets are well correlated to 10-20\,km\,s$^{-1}$ which is the expected level of systematics.}
\label{fig:sii_oiii}
\end{figure*}

\begin{figure}
\includegraphics[width=8.5cm]{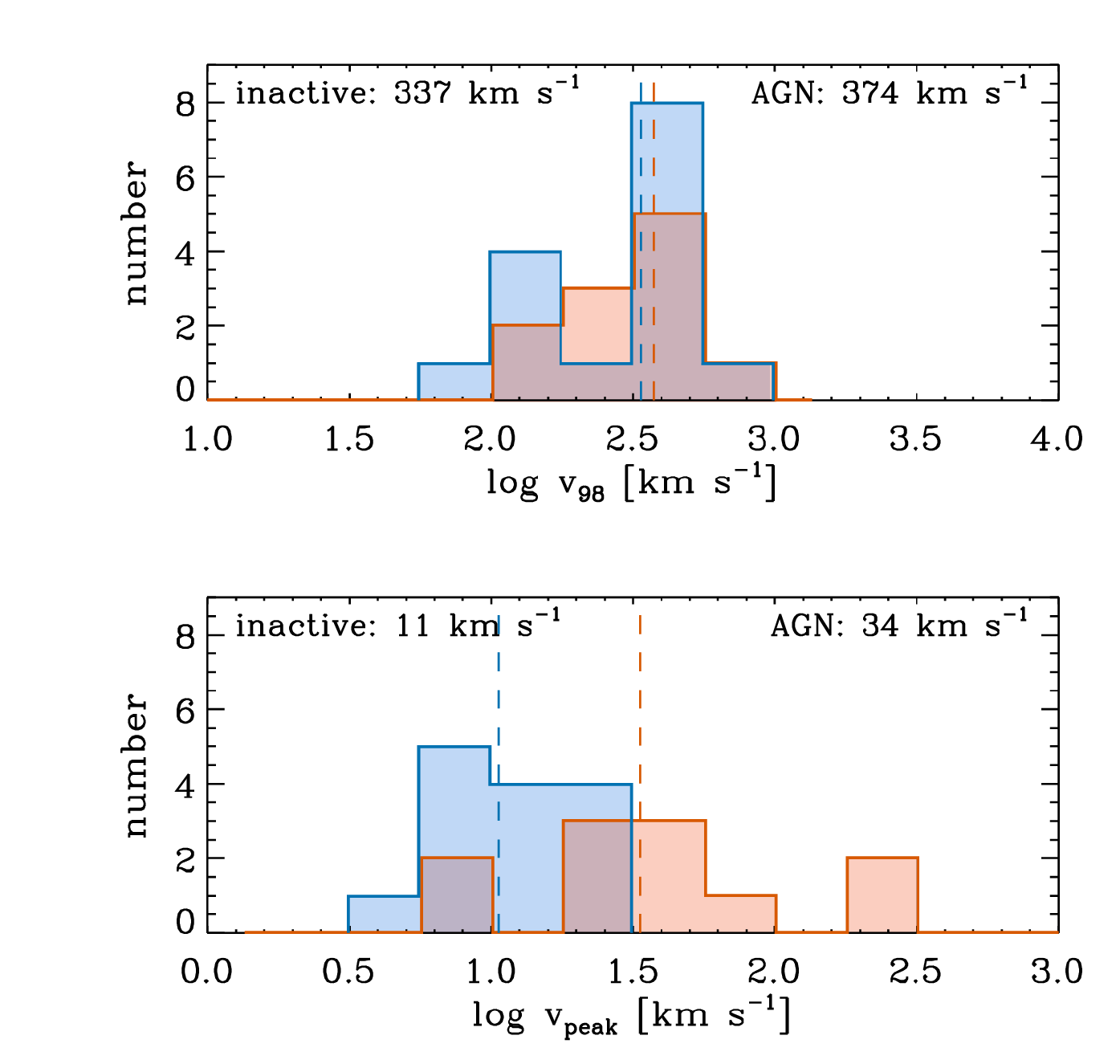}
\caption{Comparison of kinematic emission line properties for AGN (red) and inactive galaxies (blue), with median values indicated by the vertical dashed lines.
Upper: maximum velocity $v_{\rm 98,[SII]}$ of the [SII] line from systemic.
Lower: velocity offset $v_{\rm peak,[SII]}$ of the peak of the [SII] emission line from systemic (as measured by stellar absorption features).
Both quantities are given as absolute values.
These show that $v_{\rm 98}$ is similar for the active and inactive galaxies, while $v_{\rm peak}$ distinguishes between them rather well. 
This is because the high value of $v_{\rm 98}$ for the inactive galaxies is due to an outflow wing on the line profile that is distinct from the systemic component; while for the AGN the outflow is the dominant part and so affects $v_{\rm peak}$ as well.}
\label{fig:v90}
\end{figure}

\begin{figure*}
\includegraphics[width=12cm]{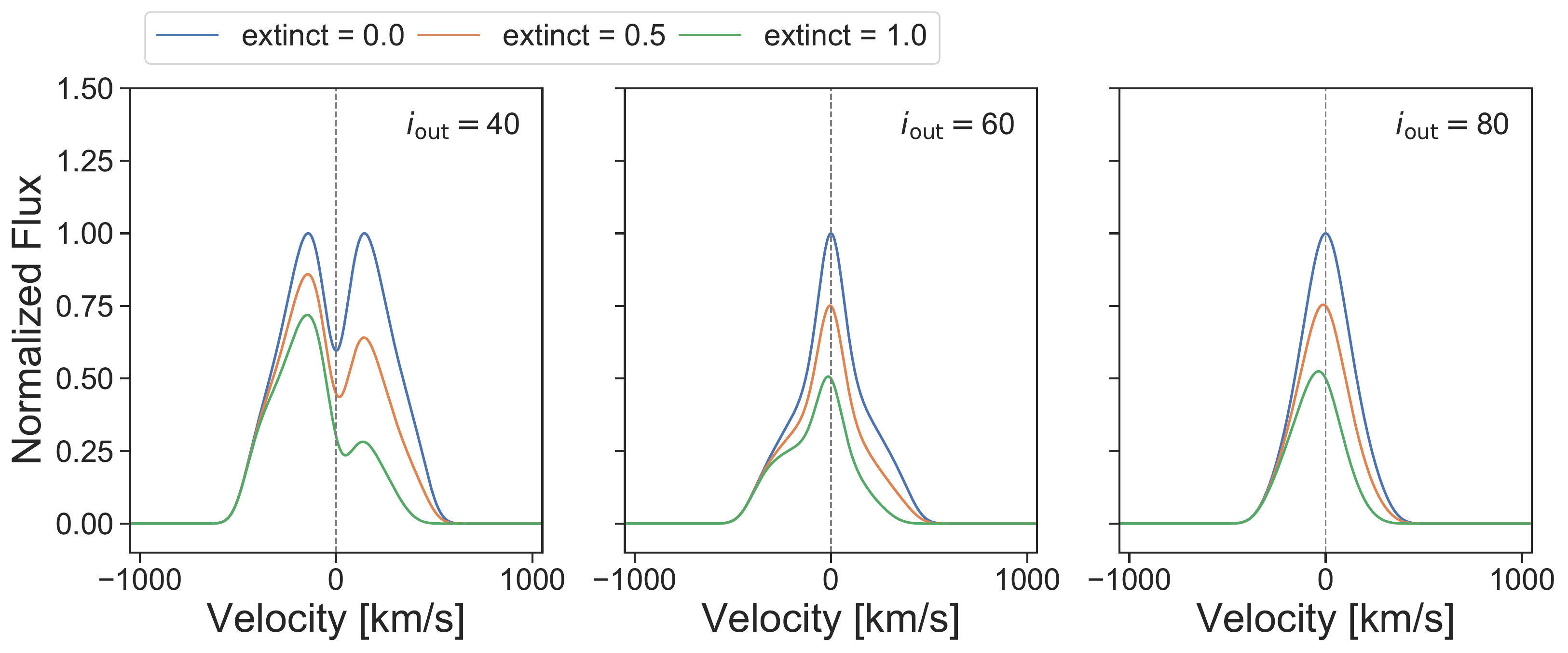}
\caption{Examples of the variety of line profiles that can be produced by a biconical outflow, depending on orientation (panels are for $i_{\rm out} = 40^\circ$, $60^\circ$, $80^\circ$ to the line of sight) and dust obscuration (colours denote a fraction $extinct = 0$, 0.5, 1 of the emission from behind the host galaxy disk is blocked).
These particular examples are for a bicone with inner and outer half-opening angles of $20^\circ$ and $40^\circ$, with a velocity profile that accelerates to 500\,km\,s$^{-1}$ at a turnover radius and then decelerates.
A host galaxy disk oriented at $60^\circ$ provides obscuration. This can affect not just the rear cone, but also produce more complex effects in the line proifile if part of each cone is obscured.
A full library at lower velocity resolution is presented by \citet{bae16}.
The examples here include double-peaked profiles, as well as rather narrower profiles, both with and without prominent wings.
They demonstrate that even the narrower profiles among our AGN are consistent with an outflow origin.}
\label{fig:profile_models}
\end{figure*}

\subsection{Excitation}

The left panel of Fig.~\ref{fig:bpt} shows the standard [NII]/H$\alpha$ versus [OIII]/H$\beta$ ratios for our sample.
For this plot, we used the flux integrated over the full line profile, and so it simply shows that the whole of the line emission is dominated by AGN photoionization rather than other processes.
Although some AGN lie close to the border with the LINER region, it is known that there is considerable overlap between Seyferts and LINERs at that boundary \citep{kew06}.
The line equivalent width should be considered a third dimension of the standard line ratio plot because it shows that, at high [NII]/H$\alpha$ ratios, AGN photoionization is generally associated with much brighter line emission than would be expected from (post-AGB) stellar photoionization \citep{sta08}.
\citet{cid10,cid11} proposed its use in an alternative and complementary `WHAN' diagnostic plot of the H$\alpha$ equivalent width versus the [NII]/H$\alpha$ ratio.
This is shown for our sample in the right panel of Fig.~\ref{fig:bpt}, where the AGN and inactive galaxies form distinct sequences.
The location of the AGN in these two plots confirms that their emission lines are indeed AGN dominated.
The only exception is ESO\,021-G004, which has a remarkably low H$\alpha$ equivalent width.
This is surprising because the sample selection ensures that all the AGN are of similar moderate luminosity, and Table~\ref{tab:outflow} shows that ESO\,021-G004 is unremarkable in this respect.
Instead, our estimate of the extinction to the narrow lines based on the H$\alpha$/H$\beta$ ratio given in Table~\ref{tab:bpt} indicates that this object (together with NGC\,7172) has $A_{\rm V} > 3$\,mag and hence its intrinsic line luminosity is a factor 15 higher than that observed.

The inactive galaxies occupy a rather different locus on these plots, following a relatively narrow track that extends from pure star-forming galaxies to LINERs photoionized by post-AGB stars.
This sequence clarifies that NGC\,4254, while it is formally among the Seyferts in the left panel of Fig.~\ref{fig:bpt} (bearing in mind that there is overlap with LINERs across that boundary), does indeed fit better among the inactive galaxies.
These objects were selected to be matched to the AGN host galaxies \citep{dav15,rosd18} and so are similar on large scales.
An analysis of the stellar population in the central $\sim$300\,pc shows that they are also similar on small scales \citep{bur20}.
These studies imply that if the AGN were to become inactive, we would expect them to look like the inactive sample, with similar photoionization properties and line strengths.
This difference is apparent also in the kinematics, as discussed next.

\subsection{Velocities}

To quantify the line kinematics, we define two properties which are measured relative to the systemic velocity defined by stellar absorption features:
\begin{description}
\item[$v_{\rm peak}$] is the velocity offset from systemic of the peak of the emission line. It has the same role as $v_{\rm int}$ (line centroid) used by \citet{bae16} in their analysis of line profiles in biconical outflows. We use it to determine whether or not the core of the line is tracing the ambient ISM or the outflow. 
\item[$v_{\rm 98}$] is the velocity above (or below) which one finds 98\% of the line flux. We compare the absolute values of these and use whichever is the larger (we do not distinguish between the red and blue wings, and so $v_{\rm 98}$ as used here does not have a sign). This is motivated by the need to reliably estimate the maximum outflow velocity. It is similar in concept to other commonly used metrics \citep{rup11,liu13,vei13,per19}, and is an appropriate measure for the complex line profiles encountered here.
\end{description}
We have derived these for [SII], rather than the standard outflow tracer [OIII], because it is well detected in both active and inactive galaxies, allowing a comparison of these subsamples.
In order to assess the level of bias in using [SII], Fig.~\ref{fig:sii_oiii} compares $v_{\rm peak}$ and $v_{\rm 98}$ for the two lines in the AGN. 
With the exception of NGC\,2110, there is good agreement between them, the most notable difference being that $v_{\rm 98}$ for the [OIII] line is about 25\% larger (see also a comparison of the line profiles in Fig.~\ref{fig:profile}).

The distributions of $v_{\rm 98}$ and $v_{\rm peak}$ are shown in Fig.~\ref{fig:v90}.
The active and inactive galaxies have rather similar distributions of $v_{\rm 98}$.
The reason is that for the inactive galaxies, it traces the edge of an outflow wing in the line profile that is distinct from the dominant systemic component. 
In contrast, for the active galaxies, the outflow is the dominant part of the line profile, and so $v_{\rm 98}$ traces the edge of the bulk of the line emission.
This is clarified by the distribution of $v_{\rm peak}$.
For inactive galaxies, the median absolute offset of the line peak from systemic is only 11\,km\,s$^{-1}$.
Such small values are consistent with irregularities in a distribution that basically traces host galaxy rotation.
It strongly suggests that the line peak for inactive galaxies is tracing the ambient ISM, as expected.
In contrast, the offsets for almost all the AGN are larger than the median of the inactive galaxies; and for more than half of the AGN they are greater than the maximum 28\,km\,s$^{-1}$ for the inactive galaxies.
The median of 34\,km\,s$^{-1}$ for the AGN is a significant offset for the line peak and, if associated with host galaxy rotation, would imply a highly asymmetric (one-sided) line distribution.
Instead, we conclude that in the central few hundred parsecs of AGN, even the peak of the emission lines is tracing outflow.

\subsection{Outflowing versus ambient gas}

The line profiles of some AGN, typified by NGC\,5728 and ESO\,137-G034, are characterised by a wide double-peaked profile either side of systemic, which can be understood in terms of the approaching and receding sides of an outflow.
In these objects any systemic contribution to the line is negligible.
A library of line profiles from biconical outflows has been modelled by \citet{bae16}, covering a variety of orientations, opening angles, and differential extinction.
Although at lower resolution than our spectra, they show clearly that, in addition to complex profiles such as those discussed above, one can expect some outflows to have rather narrow unremarkable profiles because of their more edge-on orientation.
This is particularly important for Seyfert~2s which are by definition closer to edge-on.
We have reproduced some examples at a higher resolution in Fig.~\ref{fig:profile_models}.
These cover inclinations from 40--80$^\circ$ and have the emission from behind a host galaxy disk blocked by varying amounts.
They qualitatively match the profiles we observe, for example the red profile in the left panel ($i_{\rm out} = 40^\circ$, with half of the emission behind the disk blocked) is similar to ESO\,137-G034, while the green profile in the right panel ($i_{\rm out} = 80^\circ$, with all of the emission behind the disk blocked) is more like NGC\,7582.
Generally, these model profiles include not only double-peaked profiles, but also narrower profiles, both with and without prominent wings.
This helps understand some of the less remarkable profiles such as those for MCG-05-23-016 or NGC\,3081, both of which clearly show some characteristics of outflow.
In particular, MCG-05-23-016 shows a distinct break at systemic typical of a double-peak from a  more inclined bicone, while for NGC\,3081 the profile is smoother but its peak is far offset from systemic suggesting that the approaching side may be obscured behind the galaxy disk.

As a final check about whether there is a measurable systemic component, we have examined the [SII] doublet ratio, and also the [OIII]/H$\beta$ ratio (which directly affects the derived density for the {\it logU} method) as a function of velocity.
There are clearly trends with velocity (similar to those reported by \citealt{vei91}), especially between the red-shifted and blue-shifted emission, reflecting differences in the approaching and receeding sides of the outflow.
However, even at the spectral resolution and signal-to-noise of these data, there is no evidence for changes in either ratio associated with the systemic velocity.
Together with the assessment of the profile shape above, this suggests that integrating over the full line profiles will not lead to any bias in the resulting density due to a (sub-dominant) systemic component in the line profile.

Our conclusion in this section is that, while a `systemic + outflow' decomposition can be appropriate in many cases, it cannot be meaningfully applied to the nuclear observations of active galaxies presented here.
Instead, the multiple Gaussian components can only be considered together, in terms of a convenient way to characterise a complex line profile.
Indeed, the observed emission lines are fully dominated by an AGN photoionized outflow.
The reasons include: 
(i) that our spectra are extracted from small apertures that trace gas within $\sim$150\,pc from moderately luminous AGN, 
(ii) our ability to make an independent measure of the systemic velocities from the stellar continuum, 
(iii) the complexity of the line profiles in Fig.~\ref{fig:profile} which either have a dip at the systemic velocity or continue smoothly across it, and
(iv) that we know it is easy to find reasonable parameter sets for biconical outflow models that reproduce both the dramatic and the more unremarkable profiles.
Our interpretation is that in every case, the whole profile is probing outflowing gas.
As such, for the analysis of the outflow density in Sec.~\ref{sec:density} we use the full integrated line flux.


\begin{figure*}
\includegraphics[width=17.5cm]{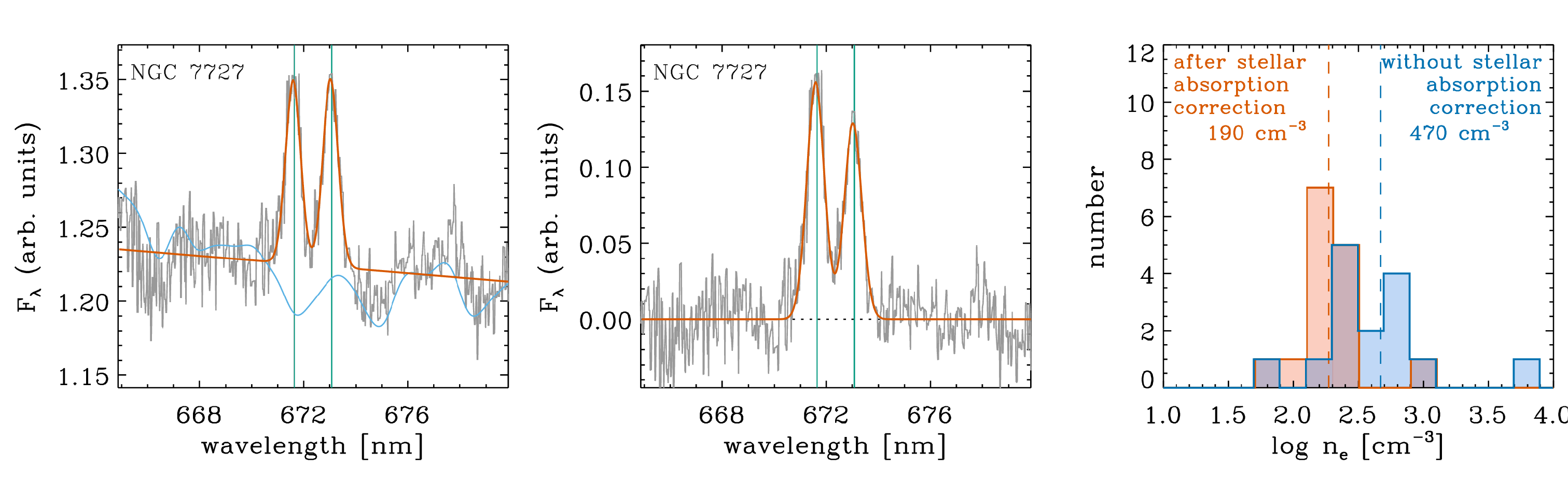}
\caption{Bias in density measurement for a weak [SII] doublet. 
Left: the observed spectrum (grey line) for NGC\,7727 overplotted with the stellar continuum (blue line) and a fit to the emission line doublet approximating the stellar continuum with only a linear function (red line).
The vertical green lines indicate the line centres for the doublet.
Centre: as for the left panel but after subtracting the fitted stellar continuum.
Right: the derived densities for all the inactive galaxies without (blue) and after (red) correction for the stellar continuum.}
\label{fig:n7727}
\end{figure*}

\begin{table*}
\caption{Measurements of [SII] doublet, and derived densities, for active and inactive galaxies.}
\label{tab:siidata}
\begin{tabular}{l D{;}{\pm}{-1} D{;}{\pm}{-1} D{;}{\pm}{-1} D{;}{\pm}{-1} D{;}{\pm}{-1}}
\hline
Object & \multicolumn{1}{c}{$v_{\rm 98,[SII]}$} 
       & \multicolumn{1}{c}{$v_{\rm peak,[SII]}$} 
       & \multicolumn{1}{c}{EW$_{\rm [SII]}$ $^a$}
       & \multicolumn{1}{c}{[SII]\,doublet}
       & \multicolumn{1}{c}{$\log{n_{\rm e}}$} \\
       & \multicolumn{1}{c}{km\,s$^{-1}$} 
       & \multicolumn{1}{c}{km\,s$^{-1}$} 
       & \multicolumn{1}{c}{\AA}
       & \multicolumn{1}{c}{ratio}
       & \multicolumn{1}{c}{cm$^{-3}$} \\
\hline
ESO\,137-G034 & 536;5   & 248;3  & 39.6;0.3 & 0.96;0.01 & 2.76;0.01 \\
MCG-05-23-016 & 221;14  &  19;3  &  4.9;0.1 & 0.98;0.01 & 2.74;0.01 \\
NGC\,2110     & 499;7   &  40;3  & 51.4;0.7 & 1.09;0.02 & 2.54;0.03 \\
NGC\,2992     & 461;14  &   6;3  & 16.3;0.3 & 1.02;0.02 & 2.66;0.02 \\
NGC\,3081     & 176;3   &  47;3  & 23.4;0.6 & 0.99;0.03 & 2.72;0.05 \\ 
NGC\,5506     & 555;11  &  74;16 &111.8;3.3 & 1.17;0.04 & 2.37;0.08 \\
NGC\,5728     & 673;50  & 223;4  & 18.9;0.5 & 1.11;0.07 & 2.49;0.14 \\
NGC\,7582     & 294;7   &  34;3  &  9.5;0.2 & 0.96;0.03 & 2.77;0.04 \\
ESO\,021-G004 & 374;6   &  20;4  &  2.0;0.2 & 1.22;0.03 & 2.26;0.07 \\
NGC\,1365     & 149;3   &  10;3  &  1.6;0.1 & 1.24;0.02 & 2.22;0.04 \\
NGC\,7172     & 209;19  &  19;5  &  2.0;0.3 & 1.25;0.19 & 2.20;0.73 \\
\\
ESO\,093-G003 & 140;5   &  15;3  & 3.4;0.1 & 1.18;0.04 & 2.36;0.08 \\ 
ESO\,208-G021 & 548;18  &  11;4  & 1.7;0.1 & 1.17;0.06 & 2.38;0.14 \\
NGC\,0718     & 471;128 &  10;3  & 0.8;0.1 & 1.16;0.06 & 2.41;0.13 \\
NGC\,1079     & 152;3   &   6;3  & 3.5;0.1 & 1.27;0.02 & 2.13;0.04 \\
NGC\,1947     & 537;44  &  10;3  & 4.1;0.3 & 1.25;0.07 & 2.20;0.19 \\
NGC\,3175     & 106;4   &   9;3  & 4.1;0.1 & 1.23;0.05 & 2.23;0.12 \\
NGC\,3351     & 248;23  &   0;3  & 1.2;0.1 & 1.22;0.04 & 2.27;0.09 \\
NGC\,3717     & 474;28  &  28;3  & 1.9;0.1 & 1.23;0.03 & 2.25;0.06 \\
NGC\,3749     & 407;31  &  11;4  & 0.9;0.1 & 1.26;0.08 & 2.17;0.25 \\
NGC\,4224     & 595;39  &  16;9  & 0.9;0.1 & 0.83;0.04 & 2.98;0.07 \\
NGC\,4254     & 144;12  &  28;4  & 1.0;0.2 & 1.12;0.17 & 2.49;0.35 \\
NGC\,5037     & 337;6   &  28;4  & 1.8;0.1 & 1.30;0.03 & 2.02;0.11 \\
NGC\,5921     & 392;13  &   7;3  & 1.9;0.1 & 1.16;0.03 & 2.39;0.06 \\
NGC\,7727     & 332;7   &  25;4  & 1.1;0.3 & 1.21;0.04 & 2.29;0.10 \\
IC\,4653      &  69;3   &   3;3  &12.9;0.1 & 1.37;0.01 & 1.72;0.06 \\
\hline
\end{tabular}

Note: Quoted uncertainties are from formal error propagation based on the random noise in the spectrum.\\
$^a$ EW$_{\rm [SII]}$ is for the 6716\,\AA\ line in the doublet.
\end{table*}

\section{Density and mass measurements}
\label{sec:density}

In this Section we explore three independent ways to measure the density of the ionized gas.
These include the most commonly used method based on the [SII] doublet ratio, an alternative proposed by \citet{hol11} that uses a combination of [SII] and [OII] line ratios, and a method recently introduced by \citet{bar19} that is based on the ionization parameter.
We indicate their main limitations and merits, and calculate the density ranges for our sample using each method.
We compare these ranges, and use photoionization models to understand the differences between them and the impact on the implied ionized gas mass.

\subsection{[SII] doublet ratio method}
\label{sec:den_sii}

The most commonly used electron density tracer (here referred to as the {\it doublet} method) uses the [SII]\,$\lambda$6716,6731\,\AA\ doublet, because it only requires a measurement of the ratio of two strong emission lines in a convenient and clean part of the optical spectrum, and the physics of the excitation and de-excitation means that density -- covering a range commonly found in H\,II regions -- dominates the emitted line ratio \citep{ost89}.
An additional advantage is that, because the lines are necessarily close in wavelength, the derived density is unaffected by extinction.

There are, however, situations where this ratio can give misleading results.
Because the two lines are separated by only 14.4\,\AA, corresponding to 650\,km\,s$^{-1}$, deblending the doublet can become unreliable at moderate spectral resolution. This is particularly important for complex line profiles. 
As an example, in a detailed study of the ionized and molecular gas in the circumnuclear region of NGC\,5728, \citet{shi19} compared several methods of measuring the electron density using both high and moderate resolution data.
They showed that for data with R$<4000$, blending of the [SII] line profiles in this object leads to an increasing discrepancy in the derived density at smaller radii -- with an order of magnitude under-estimation at radial scales below 500\,pc.
For our sample, Fig.~\ref{fig:sii_linefits} shows that the high spectral resolution and signal-to-noise of our data mean that the line profiles of both active and inactive galaxies can be robustly determined.

It is well known that [SII] cannot probe high densities because collisional de-excitation dominates above $10^4$\,cm$^{-3}$ where the ratio saturates at $\sim$0.45, its asymptotic value.
We note that this effect should not bias our measurements because none of the electron densities derived from the [SII] doublet in our sample exceed 1000\,cm$^{-3}$.

A less well known bias can arise from the impact of the stellar continuum when the equivalent width of the [SII] lines is low.
This is illustrated in Fig.~\ref{fig:n7727} for NGC\,7727.
The apparent line ratio of 0.98 (left panel) is rather less than the actual line ratio of 1.21 (centre panel) due to the stellar absorption feature under the 6716\,\AA\ line.
It seems likely that this is dominated by Fe\,I \citep{barbuy18}. 
Since the feature is weak, it only affects lines with equivalent width $\la$10\,\AA. 
But the impact on the resulting derived densities can be significant.
For the inactive galaxies in our sample, for which the median equivalent width is 1.5\,\AA, the right panel of Fig.~\ref{fig:n7727} shows that failing to correct for this effect leads to a factor 2.5 over-estimation of the typical density.
In contrast, for the stronger lines in the AGN the bias is negligible.

An additional caveat is that this method yields the electron density in the region where the line is emitted.
In AGN, the emission from low ionization transitions such as [SII], [NII], and [OI] is enhanced (compared to HII regions around stars) by an extended partially ionized zone \citep{ost89}.
In such regions of a cloud, where the gas is mostly neutral, the electron density will not necessarily be comparable to the hydrogen gas density. In some circumstances this can become a critical issue and we return to it in Sec.~\ref{sec:den_comp}.

We have measured the [SII] doublet lines in the inactive galaxies as described in Sec.~\ref{sec:linefitting}, constraining the profiles of the two lines in the doublet to be the same.
The profile fits for the active and inactive galaxies are shown in Fig.~\ref{fig:sii_linefits} and the measured properties of the [SII] doublet lines are summarised in Table~\ref{tab:siidata}.
The median density for the inactive galaxies is 190\,cm$^{-3}$, while that for the active galaxies is 350\,cm$^{-3}$.
A quantitatively similar result using the [SII] doublet was reported by \cite{min19} who, for spatially resolved data of 9 nearby AGN, reported that the mean density of 250\,cm$^{-3}$ in the outflows was higher than the mean of 130\,cm$^{-3}$ for the circumnuclear disks.

In terms of density and outflow velocity, NGC\,4224 has characteristics more like an AGN outflow than the inactive galaxy that it is. 
It is excluded from the left panel of Fig.~\ref{fig:bpt} because H$\beta$ was not detected.
However, its ratio $\log{[NII]/H\alpha} \sim 0.85$ would put it well into the LINER region.
We have found no evidence that this object might be an AGN, and indeed it has a very low H$\alpha$ equivalent width of $<$0.5\,\AA.
Although it is a spiral galaxy classified as Sa \citep{dav15}, it is one of the few inactive galaxies in our sample for which the central optical spectrum in the IFU field of view shows no indication of any stellar population younger than $\sim3$\,Gyr \citep{bur20}.
And despite being a member of the Virgo Cluster \citep{bin85}, the outer parts of the galaxy show no signs of being disturbed \citep{but15}.
The line characteristics indicate that there is an outflow, and we speculate that it could be driven either by post-AGB stars or a fossil AGN, perhaps as a result of rapid variability \citep{mat03,for14,flu19}.

\begin{figure}
\includegraphics[width=8.5cm]{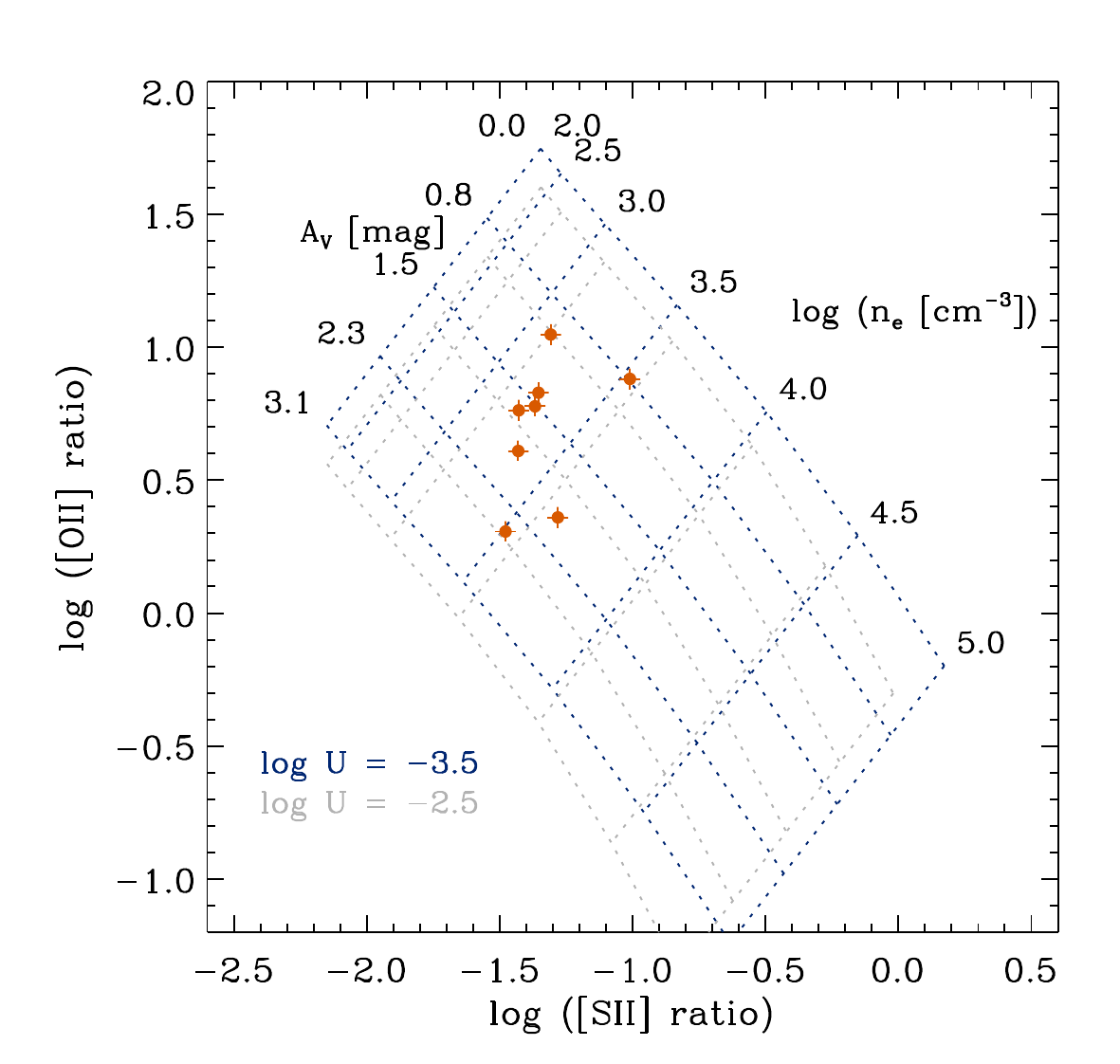}
\caption{Location of the measured [OII] and [SII] ratios for the AGN in our sample, in comparison to photoionization models that trace a grid of extinction $A_{\rm V}$ versus electron density $n_{\rm e}$. Models are shown for solar metallicity and a standard AGN spectral energy distribution; but for two different ionization parameters covering the range found for the AGN here.}
\label{fig:den_ta}
\end{figure}

\begin{figure}
\includegraphics[width=8.5cm]{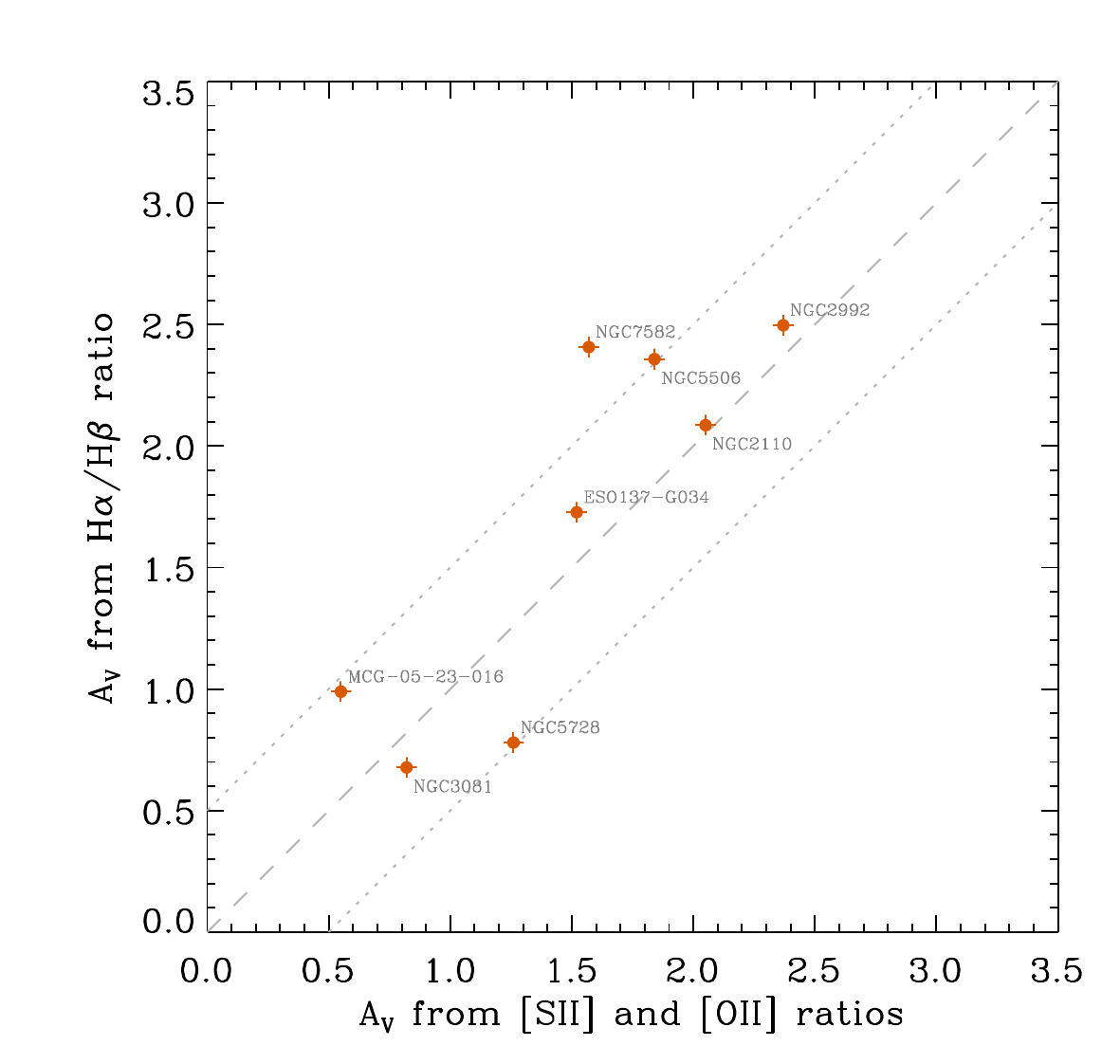}
\caption{Comparison of extinction $A_{\rm V}$ derived from the [SII] and [OII] line ratios to that derived from the $H\alpha/H\beta$ ratio.
The dashed line indicates a 1:1 ratio and the dotted lines a $\pm$0.5\,mag range.}
\label{fig:Av}
\end{figure}

\begin{table*}
\caption{Measurements of the [SII] and [OII] line ratios used by the auroral/transauroral method, and derived densities and extinctions for active galaxies}
\label{tab:den_ta}
\begin{tabular}{l D{;}{\pm}{-1}  D{;}{\pm}{-1} D{;}{\pm}{-1} D{;}{\pm}{-1}}
\hline
Object & \multicolumn{1}{c}{log\,[SII]\,ratio}
       & \multicolumn{1}{c}{log\,[OII]\,ratio}
       & \multicolumn{1}{c}{log\,$n_{\rm e}$ (cm$^{-3}$)}
       & \multicolumn{1}{c}{$A_{\rm V}$ (mag)} \\
       & \multicolumn{1}{c}{(4069+4076)/(6716+6731)}
       & \multicolumn{1}{c}{(3726+3729)/(7320+7331)} &  & \\
\hline
\\
ESO\,137-G034 & -1.43;0.01 & 0.76;0.02 & 3.13;0.02 & 1.52;0.04 \\
MCG-05-23-016 & -1.01;0.02 & 0.88;0.02 & 3.46;0.02 & 0.55;0.04 \\
NGC\,2110     & -1.28;0.01 & 0.36;0.01 & 3.61;0.02 & 2.05;0.03 \\
NGC\,2992     & -1.48;0.02 & 0.31;0.03 & 3.43;0.03 & 2.37;0.06 \\
NGC\,3081     & -1.31;0.02 & 1.05;0.01 & 3.01;0.02 & 0.82;0.04 \\
NGC\,5506     & -1.43;0.03 & 0.61;0.02 & 3.27;0.03 & 1.84;0.05 \\
NGC\,5728     & -1.35;0.03 & 0.83;0.02 & 3.16;0.04 & 1.26;0.06 \\
NGC\,7582     & -1.36;0.04 & 0.78;0.05 & 3.28;0.05 & 1.57;0.11 \\
\hline
\end{tabular}

Note: Quoted uncertainties are from formal error propagation based on the random noise in the spectrum.
\end{table*}

\subsection{Auroral and Transauroral line method}
\label{sec:den_ta}

To avoid the limitations of the [SII] doublet ratio, \citet{hol11} proposed an alternative method that has now been applied to a variety of different types of objects, and been shown to be sensitive to higher densities \citep{hol11,rosm18,san18,shi19}.
We will refer to this as the {\it TA} method because it uses the transauroral lines [SII]\,$\lambda$4069,4076 and the auroral lines [OII]\,$\lambda$7320,7331 (each of the lines within this 11\,\AA\ separation doublet is itself a doublet, but their separations of $\sim$1\,\AA\ are sufficiently small that at the spectral resolution here they can be considered single lines) which have higher critical densities.
These are used together with the stronger lines to give the ratios
[SII]\,$\lambda$(4069+4076) / [SII]\,$\lambda$(6716+6731) and
[OII]\,$\lambda$(3726+3729) / [OII]\,$\lambda$(7320+7331).

The way these lines are used to estimate density differs fundamentally from the {\it doublet} method.
Rather than providing a direct measure of $n_{\rm e}$ in the gas where the line emission originates, the ratios of summed doublet fluxes are compared to those produced in photoionization models.
By design, such models take account internally of how $n_{\rm e}$ (on which the emitted lines depend) varies through the cloud, and hence how the resulting cumulative line ratios are related to $n_{\rm H}$.
In this way the method traces $n_{\rm H}$ rather than $n_{\rm e}$ in a cloud;
but typically one equates $n_{\rm H}$ and $n_{\rm e}$ as if the gas were fully ionized.
The [OII] lines arise predominantly in the fully ionized gas, but they are far apart in wavelength and so the impact of extinction needs to be addressed.
This is done using the [SII] lines, which originate from different regions of the cloud and span a shorter wavelength range, so that their ratio has a different dependency on density and extinction.
The pair of ratios then provides a reference basis for photoionization model grids in which density and extinction are approximately orthogonal.

The primary advantages of the method are that it is sensitive to higher densities, which is now known to be important when measuring outflow rates; and it uses the summed flux in each doublet and thus is less sensitive to the details of the line profile.
However, it too has some limitations.
Most critically from the observational perspective, the auroral and transauroral lines are very weak.
Figs.~\ref{fig:ta_linefits1}-\ref{fig:ta_linefits2} show that their flux is typically only about 10\% of the strong [SII] and [OII] lines, and so signal-to-noise can be a major restriction.
For the same reason, the continuum level fitted around the lines can have a significant impact on the measured line flux and so subtracting the stellar continuum is mandatory.

The conversion of the measured ratios to a density is also not as straightforward as for the [SII] doublet ratio because it depends on photoionization models. 
\citet{hol11} assessed the impact of changes in the spectral index and ionization parameter, and argued that they are relatively unimportant.
Our own calculations (described below) indicate that these should not be ignored: Fig.~\ref{fig:den_ta} shows that the ionization parameter can change the derived density by a factor 2--3; and changing the metallicity has a comparable impact.
In addition, the wide wavelength range required to cover all the lines means that the effects of extinction must be included when fitting the models to the data.
And the choice of extinction model will also have an impact on density derived from the line ratios.

For the AGN in which all the necessary lines can be measured, we have fitted the [SII] and [OII] doublets in a similar way as before, including the various constraints described in Sec.~\ref{sec:linefitting}.
The resulting profile fits for each doublet are shown in Figs.~\ref{fig:ta_linefits1}-\ref{fig:ta_linefits2}.
As before we propagated the uncertainties in fitted parameters to the summed fluxes and hence ratios using Monte Carlo techniques.

To estimate densities from the line ratios, we have performed calculations using CLOUDY v17 \citep{fer17}.
We calculated a grid of photoionization models covering a large range in ionization parameter and
hydrogen density, with details and assumptions similar to those presented in Appendix A of \citet{bar19}. 
We adopted a standard AGN spectral energy distribution (SED) with a mean energy of an ionizing photon of 2.56~Ryd (SED~2 in Table~A1 of \citealt{bar19}; see also \citealt{net13}), although we note that the shape of
the ionizing SED has a negligible effect on our conclusions in this section. The assumed metallicity, on the
other hand, has a significant effect on the derived densities, which vary by a factor of 2--3 for a metallicity range of 0.5--2 times solar. We present in Sec.~\ref{sec:mass} evidence that the metallicity is close to solar, and thus we assume solar metallicity. We considered a model grid with seven values of $\log{n_{\rm H}}$ in the range 2.0 to 5.0, and examined eight values for ionization parameter $U$ ranging from -3.8 to -2. The separate grids in ionization parameters were created to match those calculated for our AGN (see Sec.~\ref{sec:den_U} and Table~\ref{tab:bpt}), which were derived from the [OIII]/H$\beta$ and [NII]/H$\alpha$ ratios as described in Sec.~\ref{sec:den_U} following \citet{bar19}.
Finally, we adopted the extinction law of \citet{car89}, taking $A_{\rm V} = 3.1\,E(B-V)$ and assuming the dust is in a foreground screen.
The photoionization models we consider are dusty, and thus dust is also mixed with the ionized gas
(see \citealt{bar19} for additional details). However, its effect is taken into account internally within the models (we use \emph{emergent} line luminosities), and thus we do not need to account for it separately. 
In addition, the column density of the internal dust is small compared to the column derived for the dusty screen in our sources. 
Two of the resulting model grids are shown in Fig.~\ref{fig:den_ta}, for $\log{U} = -2.5$ and~$-3.5$, representing the range of $U$ we find for the individual AGN.
The location of our AGN with respect to these models is shown in Fig.~\ref{fig:den_ta}, and the implied $n_{\rm e}$ and $A_{\rm V}$ are given in Table~\ref{tab:den_ta}.
The median density is $n_{\rm e} = 1900$\,cm$^{-3}$ and the 1$\sigma$ range covers 1200--3000\,cm$^{-3}$, significantly higher than that derived with the {\it doublet} method.

As a consistency check of our photoionization modelling, we compare the resulting extinction to that derived from the H$\alpha$ and H$\beta$ lines, assuming an intrinsic ratio H$\alpha$/H$\beta = 3.1$ appropriate for the narrow line region and using the \citet{car89} extinction curves as above (and taking into account the BLR as described in Sec.~\ref{sec:linefitting}).
The resulting extinction, in the range 1--3\,mag, are compared in Fig.~\ref{fig:Av}.
Here, the dashed line indicates a 1:1 ratio, while the dotted lines are offset by 0.5\,mag each.
The values from the two methods are comparable to within about 0.5\,mag, providing support for our derivation of $n_{\rm e}$ and $A_{\rm V}$ from the [OII] and [SII] ratios.

\begin{table*}
\caption{Measurements of the H$\alpha$ and H$\beta$ fluxes, the [NII]/H$\alpha$ and [OIII]/H$\beta$ line ratios, and the derived ionization parameters and densities for active galaxies}
\label{tab:bpt}
\begin{tabular}{l D{;}{\pm}{-1} D{;}{\pm}{-1} c D{;}{\pm}{-1} D{;}{\pm}{-1} D{;}{\pm}{-1} D{;}{\pm}{-1}}
\hline
Object & \multicolumn{1}{c}{F$_{\rm H\alpha}$}
       & \multicolumn{1}{c}{F$_{\rm H\beta}$}
       & $A_{\rm V}$ (H$\alpha$/H$\beta$)
       & \multicolumn{1}{c}{log\,[NII]/H$\alpha$}
       & \multicolumn{1}{c}{log\,[OIII]/H$\beta$}
       & \multicolumn{1}{c}{log\,U $^a$}
       & \multicolumn{1}{c}{log\,$n_{\rm e}$ $^b$} \\
       & \multicolumn{1}{c}{$10^{-15}$ erg\,s$^{-1}$\,cm$^{-2}$}
       & \multicolumn{1}{c}{$10^{-15}$ erg\,s$^{-1}$\,cm$^{-2}$}
       & mag &  &
       & \multicolumn{1}{c}{cm$^{-3}$} 
       & \multicolumn{1}{c}{cm$^{-3}$} \\
\hline
\\
ESO\,137-G034 &  49.4;0.6 &  9.25;0.13 & 1.7 & 0.044;0.005 & 1.091;0.006 & -2.64;0.10 & 2.86;0.14 \\
MCG-05-23-016 &   6.8;0.1 &  1.60;0.02 & 1.0 &-0.043;0.005 & 1.112;0.006 & -2.58;0.10 & 3.68;0.14 \\
NGC\,2110     &  41.1;0.5 &  7.85;0.10 & 1.7 & 0.138;0.005 & 0.678;0.005 & -3.31;0.10 & 4.63;0.14 \\
NGC\,2992     &  10.2;0.2 &  1.45;0.07 & 2.5 & 0.021;0.005 & 1.067;0.002 & -2.68;0.11 & 2.78;0.16 \\
NGC\,3081     &  68.2;0.2 &17.8\ \ ;0.4& 0.7 &-0.035;0.001 & 1.117;0.010 & -2.57;0.10 & 3.49;0.14 \\
NGC\,5506     & 173\ \ \ ;45&26.3\ \ ;0.5& 2.4 &-0.052;0.118& 0.948;0.008 & -2.87;0.12 & 4.03;0.14 \\
NGC\,5728     &  41.8;1.8 &10.4\ \ ;0.3& 0.8 & 0.144;0.018 & 1.114;0.011 & -2.62;0.10 & 3.48;0.14 \\
NGC\,7582     &  60.9;0.3 &  9.39;0.08 & 2.3 &-0.173;0.002 & 0.362;0.003 & -3.54;0.10 & 4.83;0.16 \\
ESO\,021-G004 &   0.7;0.03&  0.09;0.01 & 3.1 & 0.532;0.019 & 0.886;0.046 & -3.02;0.12 & 3.10;0.28 \\
NGC\,1365     &   2.7;0.04&  0.51;0.01 & 1.7 & 0.006;0.006 & 0.707;0.010 & -3.24;0.10 & 3.70;0.18 \\
NGC\,7172     &   3.1;0.06&  0.36;0.02 & 3.2 & 0.087;0.008 & 0.799;0.019 & -3.14;0.10 & 4.01;0.14 \\
\hline
\end{tabular}

Note: Quoted uncertainties of line fluxes and ratios are from formal error propagation and based on the random noise in the spectrum.\\
$^a$ Uncertainty in $\log{U}$ is dominated by the 0.1\,dex scatter around the relation with line luminosity in \citet{bar19}.\\
$^b$ Uncertainty of log\,$n_{\rm e}$ includes that of the 14--195\,keV luminosity as well as 0.09\,dex due to the relation between the 14--195\,keV and bolometric luminosities given in \citet{win12}.
\end{table*}

\subsection{Ionization Parameter method}
\label{sec:den_U}

The {\it logU} method was proposed by \citet{bar19} and is based on the definition of the ionization parameter, the number of ionising photons per atom,
$U = Q_{\rm Lyc} \, / \, 4\pi \, r^2 \, c \, n_{\rm H}$
where $Q_{\rm Lyc}$ is the ionising photon rate from a source, $r$ is the distance from that source, and $n_{\rm e} \sim n_{\rm H}$ is the local number density of atoms or equivalently electrons.
The speed of light, $c$, is included to make the parameter dimensionless.
These authors showed that one can re-arrange this to give the electron density $n_{\rm e}$ in terms of the AGN luminosity, the distance from the AGN, and the ionization parameter such that
$n_{\rm H} \propto L_{\rm AGN} \, r^{-2} \, U^{-1}$.
At the same time, they showed that $U$ can be derived rather reliably (i.e. with a scatter of 0.1\,dex) from the strong line ratios N[II]/H$\alpha$ and [OIII]/H$\beta$.
That these lines are readily measurable for many AGN makes the method widely applicable.
A similar approach was adopted by \citet{rev18a} and \citet{rev18b} when fitting multiple components with different densities to a suite of line ratios in spatially resolved data for Mkn\,573 and Mkn\,34, with the difference that they left $U$ as a free parameter and derived it as part of the fit.

For our sample, the absorption corrected 14--195\,keV luminosity has been calculated by \citet{ric17} based on 0.3--150\,keV broadband X-ray data.
Taking those values, adjusted to our adopted distances, we have used the relation in \citet{win12} to recover the AGN bolometric luminosity.
Uncertainties due to the AGN luminosity and this relation, which combine to be about 0.14\,dex, will propagate directly into the derived density; but errors in the distance to the source do not affect the derived density because the impact when converting flux to luminosity for $L_{\rm AGN}$ is cancelled by the $1/r^2$ term when converting aperture size from arcseconds to parsecs.
We have already shown the line ratios in Fig.~\ref{fig:bpt}, and we report their values in Table~\ref{tab:bpt}.\footnote{These differ from those reported in \citet{bur20} by typically $<$10\%, which is attributable to the different resolution and use of single versus multiple Gaussian profiles adopted.}
Also given in Table~\ref{tab:bpt} is $\log{U}$, derived using the equations in \citet{bar19}.
The final parameter is the distance $r$ from the AGN.
It is clear that this method is most suited to spatially resolved data, where one can take a measurement at a known projected distance from the AGN.
In our case, we have only an aperture centered on the AGN, and the luminosity distribution within this can be complex.
For the purposes of this analysis, we aim to estimate a reasonable upper limit to $r$ which will therefore lead to a estimate of $n_{\rm e}$ that is towards the lower end of the likely range.
As the projected radius we therefore take the distance of 0.9\arcsec\ from the centre to the edge of the aperture, which typically corresponds to $\sim$150\,pc (this yields $r$ a factor 1.3 higher than it would be under the assumption of a uniform luminosity distribution).
To account for the projection effects, we note that the AGN are Seyfert~2 (see Table~\ref{tab:obs}), and hence oriented more towards edge-on than face-on.
For those objects in our sample for which the orientation of the ionized outflow has been estimated, it lies in the range 10$^\circ$--49$^\circ$ for the Sy\,2s from edge-on, and a slightly higher 55$^\circ$ for the Sy\,1.8 \citep{hje96,fri10,mue11,fis13,shi19}.
We therefore adopt an inclination of 45$^\circ$ from edge-on (this yields $r$ a factor 1.4 higher than it would be for a fully edge-on outflow).
We emphasize that in both cases, uncertainties are likely to be towards smaller values of $r$ and hence higher values for $n_{\rm e}$ than those we derive (specifically, a reasonable range for $n_{\rm e}$ would include values a factor 3--4 higher, but not lower).
The resulting densities are reported in Table~\ref{tab:bpt}.
They have a median value of 4800\,cm$^{-3}$ and a 1$\sigma$ range of 1050--22000\,cm$^{-3}$.
Given the large scatter in values, this is consistent with that found using the {\it TA} method, and again much higher than that derived with the {\it doublet} method.
It is likely that the scatter is a combination of the uncertainty of the AGN luminosity together with an observational effect related to our use of aperture measurements within which the characteristic distance from the AGN to the line emitting gas is not known.
The aperture effect is not a limitation of the method itself, but the impact of the way we apply it here.
It creates an uncertainty that is much more acute than for spatially resolved data, and simply means that here we should make use of the derived densities in a statistical sense rather than focussing on individual values.

\begin{figure*}
\includegraphics[width=16cm]{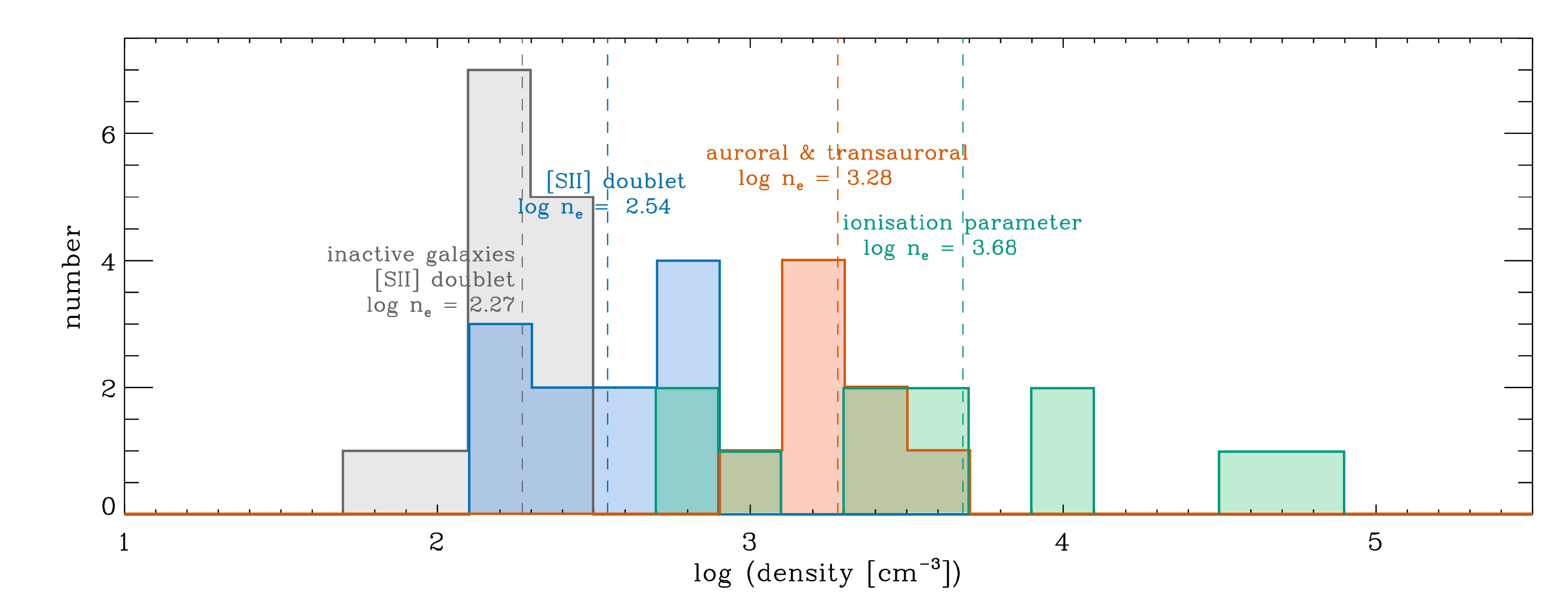}
\caption{Comparison of the densities measured for the AGN using the different methods: [SII] doublet ratio (blue), using the auroral and transauroral [SII] and [OII] line ratios (red), and based on the ionization parameter (green).
In addition, the inactive galaxies, for which the [SII] doublet method was used, are shown in grey.
For the doublet and TA methods, sample spread dominates over individual measurement errors.
Although the {\it TA} and {\it logU} methods measure $n_{\rm H}$, in fully ionised gas this is the same as $n_{\rm e}$. We have therefore labelled median values for all the methods as electron densities.}
\label{fig:ne_comp}
\end{figure*}

\begin{figure*}
\includegraphics[width=18cm]{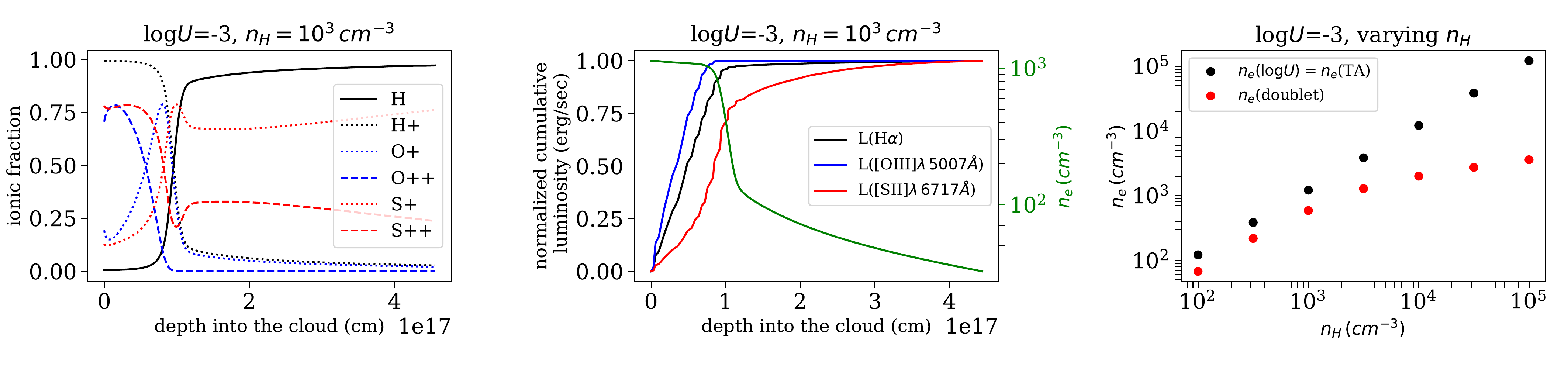}
\caption{Photoionization model showing how different emission lines trace different properties of a constant density cloud (see \citealt{bar19} for details of the models).
Left: relative population of various ions as a function of distance into the cloud. One can define the ionization front to be where most of the hydrogren is neutral.
Centre: Resulting line emission, and also electron density, as a function of depth into the cloud. This shows that much of the [SII] line originates from mostly neutral gas where the electron density is lower.
Right: Comparison of electron density as measured from the model using the methods described in the text, as a function of hydrogen density in the cloud. At $n_{\rm e} > 10^3$\,cm$^{-3}$ the discrepancy between the density derived from the [SII] doublet ratio and that found using the other methods (which trace the actual density) is significant.}
\label{fig:photoionization}
\end{figure*}

\subsection{Comparison of densities}
\label{sec:den_comp}

The distributions in Fig.~\ref{fig:ne_comp} show that the electron densities derived for the three methods cover 1$\sigma$ ranges of: 210--580\,cm$^{-3}$ for the {\it doublet} method, 1200--3000\,cm$^{-3}$ for the {\it TA} method, and 1050--22000\,cm$^{-3}$ for the {\it logU} method.
The densities from both the {\it TA} and {\it logU} methods are consistent, athough the scatter of the latter is somewhat larger as discussed above. 
Both are significantly higher than the densities found with the {\it doublet} method.
The reason for this difference cannot be due to saturation of the [SII] ratio at high densities, because the ratio is not close to its asymptotic limit; nor is it due to blending of the line profiles because our data are both high resolution and high signal-to-noise.
Instead, it is related to the physics of the photoionization.
It has been known for a long time that the hard radiation field from an AGN produces photons that can penetrate deep into clouds, leading to a partially ionized zone \citep{ost89}.
This is much more extensive than in gas photoionized by OB stars, and is responsible for the excess emission from low excitation transitions seen in AGN spectra (which lead \citealt{vo87} to propose the use of species such as [SII], in terms of its flux ratio with respect to H$\alpha$, as a diagnostic for AGN).
\citet{bar19} pointed out that because the electron density decreases further into the cloud, different excitation states of species arise from different regions within the cloud, specifically that while the H$\alpha$ and [OIII] are emitted throughout most of the ionized cloud, much of the [SII] is emitted close behind the ionization front where the electron density drops dramatically.
Because the ionization fraction falls rapidly below 10\%, most of the gas in this region is neutral.

To illustrate this point, Fig.~\ref{fig:photoionization} shows several results from the photoionization models. 
In the left panel, we show the ionized fraction of hydrogen, oxygen, and sulphur, for a
model with $\log{U} = -3$ and $\log{n_{\rm H}} = 3$, as a function of depth into the cloud. 
One can see that both H$^+$ and O$^{++}$, which are responsible for H$\alpha$ and [OIII] emission respectively, peak within the ionized cloud. 
On the other hand, S$^+$, which is responsible for the [SII] emission, peaks after the ionization front where more than half of the hydrogen is already neutral. 
In the middle panel, we compare the cumulative line emission of H$\alpha$, [OIII], and [SII], and show the electron density as a function of depth into the cloud. 
The H$\alpha$ and [OIII] line luminosities saturate near the ionization front, and one can see that their emission traces high electron density of $n_{\rm e} \sim 10^3 = n_{\rm H}$. 
On the other hand, the [SII] emission extends far into the neutral part of the cloud, reaching 80\% of the cumulative line luminosity where the electron density has already dropped by a factor of four. 
Similar trends are also observed in models with different ionization parameters.

In the right panel of Fig.~\ref{fig:photoionization} we compare the electron densities derived using the {\it TA}, {\it logU}, and {\it doublet} methods in our models with $\log{U} = -3$. 
To estimate the electron densities, we calculate the line luminosities predicted by the models and use the respective ratios discussed in Sec.~\ref{sec:den_sii}--\ref{sec:den_U} to estimate
the electron densities, in order to match our observational approach. 
The electron densities derived using the {\it TA} and {\it logU} methods are similar in all the models we examined, regardless of the ionization parameter. 
This is expected since both methods trace the hydrogen density in the cloud, which is assumed to be constant in our models. 
We find a significant difference between the electron densities derived using either of those methods, and those derived using the {\it doublet} method.
This difference increases as the hydrogen density in the cloud increases, and becomes significant even below the critical density $n_{\rm H} \sim 10^4$\,cm$^{-3}$ of the [SII] lines. 
This is because, independent of the hydrogen density, the [SII] lines trace low electron density regions within the ionized cloud. 
We therefore suggest that the {\it doublet}-based electron densities provide a biased view of the ionized
cloud, and that they should not be used above $n_{\rm e} \sim 10^3$\,cm$^{-3}$, since then the hydrogen density (and thus the electron density in the ionized part of the cloud) can be a factor of 3--100 larger. 
This invalidates key assumptions in the use of the [SII] doublet to estimate the density of the ionized cloud, and thus its mass.

\begin{table}
\caption{Effective Emissivities}
\label{tab:gamma}
\begin{tabular}{D{.}{.}{-1} D{.}{.}{-1} D{.}{.}{-1} D{.}{.}{-1}}
\hline
\multicolumn{1}{c}{$\log{U}$} & \multicolumn{3}{c}{$\gamma_{\rm eff}$ $^a$ (10$^{-25}$\,cm$^3$erg\,s$^{-1}$)} \\
      & \multicolumn{1}{c}{H$\alpha$} & \multicolumn{1}{c}{[SII] 6716\,\AA $^b$ } & \multicolumn{1}{c}{[OIII] 5007\,\AA} \\ 
\hline
-2.00 & 1.75 & 0.46 & 7.03 \\
-2.26 & 2.02 & 0.58 & 7.41 \\
-2.51 & 2.25 & 0.79 & 7.19 \\
-2.77 & 2.45 & 1.13 & 6.35 \\
-3.03 & 2.57 & 1.60 & 4.81 \\
-3.29 & 2.53 & 2.12 & 2.87 \\
-3.54 & 2.55 & 2.86 & 1.33 \\
-3.80 & 2.26 & 3.17 & 0.41 \\
\hline
\end{tabular}

$^a$ See Sec.~\ref{sec:mass} for an explanation of how these were calculated.\\
$^b$ The values for [SII] emissivity are only valid for $n_{\rm e} < 10^3$\,cm$^{-3}$.
\end{table}

\begin{figure*}
\includegraphics[width=13cm]{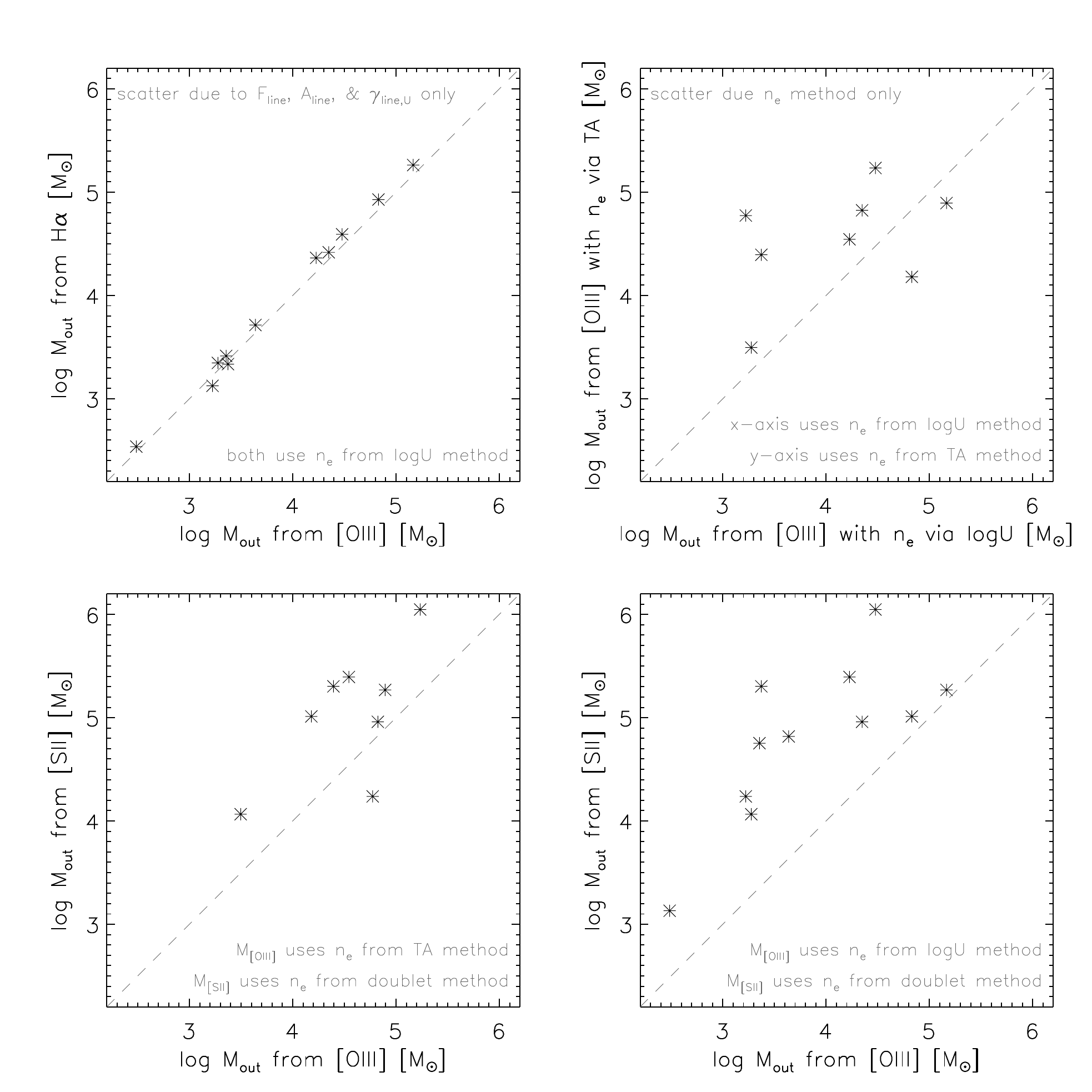}
\caption{Comparison of the derived masses using different tracers and density estimates.
Top left: when using the same $n_{\rm e}$ method, the masses from the H$\alpha$ line and [OIII] line are very similar.
Top right: when using the same line, the masses using $n_{\rm e}$ from the {\it logU} and {\it TA} methods scatter around a 1:1 line.
Bottom: an offset is found for the masses derived from the [SII] line with $n_{\rm e}$ from the {\it doublet} method, and the [OIII] line with $n_{\rm e}$ from the {\it TA} (left) or {\it logU} (right) method.
This offset is consistent with the rapid increase in discrepancy for the {\it doublet} method at $n_{\rm e} > 10^3$\,cm$^{-3}$, where our calculation of $\gamma_{\rm [SII]}$ is no longer valid.
See the text for a discussion of these panels.}
\label{fig:mass_comp}
\end{figure*}

\subsection{Ionized gas masses}
\label{sec:mass}

As noted in Sec.~\ref{sec:intro}, given a particular transition with volume emissivity $\gamma_{\rm line}$ and
emission line luminosity $L_{\rm line}$, the ionized gas mass can be estimated using 
$M_{\rm ion} = \mu m_{\rm H} \, L_{\rm line} \, / \, (\gamma_{\rm line} \, n_{\rm e})$, 
where $n_{\rm e}$ is the electron density in the \emph{line-emitting} region. 
The volume emissivity, $\gamma_{\rm line}$, depends on known atomic physics and on the physical properties of the cloud (for additional details see \citealt{bar19}). 
We have demonstrated in Sec.~\ref{sec:den_comp} that different emission lines are emitted in different parts of the ionized cloud, with those regions showing significantly different electron densities and temperatures. Therefore, the common practice of estimating the ionized gas mass using $L_{\rm H\alpha}$ or $L_{\rm [OIII]}$ and an electron density from the [SII] doublet is invalid and will result in an over-estimate of the ionized gas mass. 
As first argued by \citet{bar19}, it is necessary to use density tracers that match the emission line in question in order to obtain an unbiased estimate of the gas mass in the cloud. 
For example, when using $L_{\rm H\alpha}$ or $L_{\rm [OIII]}$, it is necessary to use {\it TA} or {\it logU} based electron densities; 
when estimating the ionized gas mass using the {\it doublet} based density, it is necessary to use the [SII] luminosity (although with an additional caveat described below about the valid density range). 
Mixing these up would result in an incorrect estimate of the ionized gas mass.

\citet{bar19} calculated the effective line emissivities for the H$\alpha$ and [OIII] emission lines, by taking the mean emissivity in the cloud weighted by
the electron density. 
Since, in those models, the H$\alpha$ and [OIII] lines are emitted in similar regions within the ionized cloud, the effective emissivities are expected to result in similar masses when using either the H$\alpha$ or [OIII] emission lines. 
However, if two transitions are emitted in different regions within the cloud, we no longer expect their masses to be equal to each other, even when using the appropriate effective emissivities and electron densities. 
This is because the mass of the gas that emits the different transitions may be different. 
For the example in  Fig.~\ref{fig:photoionization}, the mass of the [SII]-emitting gas is roughly a factor two larger than the mass in the H$\alpha$-emitting region. 
More generally we find that, for $n_{\rm e} \la 10^3$\,cm$^{-3}$, the gas mass ratio of the [SII]-emitting and H$\alpha$-emitting regions is 1--3, depending on the ionization parameter.

In this work, our goal is to calculate effective line emissivities for the H$\alpha$, [OIII], and [SII] emission lines, that will result in \emph{similar} ionized gas masses. To achieve this, it is necessary to scale each of the emissivities so that the mass traced by the different transitions is similar. 
We begin by defining an effective electron density $\langle n_{\rm e} \rangle_{\rm line}$ as the luminosity weighted mean electron density in the cloud. Its value will depend on the emission line used, and is a reasonable approximation to the values that would be derived observationally using the methods discussed in Sec.~\ref{sec:den_sii}--\ref{sec:den_U}.
We define the ionization front, separating the fully ionized and mostly neutral regions of the cloud, to be
where 80\% of the hydrogen is neutral\footnote{The exact definition of the ionization front (i.e., where the 50\%, 80\%, or 90\% of the hydrogen is neutral) has a negligible effect on the resulting ionized gas mass}. 
Defining the ionized gas mass $M_{\rm ion}$ to be the mass of the gas before the ionization front, we calculate the effective product 
$\langle \gamma_{\rm line} n_{\rm e} \rangle = \mu m_H \, L_{\rm line} \, / \, M_{\rm ion}$ 
using the integrated line luminosity $L_{\rm line}$ for each of the three emission lines H$\alpha$, [OIII], and [SII]. 
We can then define the new effective emissivities as 
$\langle \gamma_{\rm line} \rangle = \langle \gamma_{\rm line} n_{\rm e} \rangle \, / \, \langle n_{\rm e} \rangle_{\rm line}$.
Values for $\langle \gamma_{\rm line} \rangle$ for H$\alpha$, [OIII], and [SII] are given in Table~\ref{tab:gamma} as a function of $\log{U}$.
It is notable in this Table that even $\gamma_{\rm H\alpha}$ is rather different from its canonical value of $3.6\times10^{-25}$\,\,cm$^3$erg\,s$^{-1}$ \citep{ost89}.
By construction, $\langle \gamma_{\rm line} \rangle$ will yield a similar mass for the ionized gas independent of which tracer is used, as long as one uses the matching estimate of electron density.
Our constant density models verify the efficacy of this approach to 10--20\%.

Fig.~\ref{fig:mass_comp} compares the ionized masses derived using this technique for our AGN sample.
The far left panel shows we find a negligible ($<$0.1\,dex) difference in mass when using  L$_{\rm H\alpha}$ or L$_{\rm [OIII]}$.
Since these were both calculated using the same density tracer, it is a verification of our measurements and our approach to calculating $\gamma_{\rm \rm eff}$.
It also acts as {\it a posteriori} support for adopting solar metallicity.
The centre left panel compares the {\it TA} and {\it logU} methods for the same line.
Since our models indicate that these should yield the same density, the modest difference is likely due to a systematic effect in AGN luminosity or adopted distance from the AGN to the line emission, or due to complexities that our constant density models do not address.
The centre right panel shows a factor three difference between using the [OIII] line with the {\it TA} method, and the [SII] line with the {\it doublet} method.
The former is more robust, and as indicated in the right panel of Fig.~\ref{fig:photoionization}, the difference between this and the [SII] derived mass is a further indication that $n_{\rm e} > 10^3$\,cm$^{-3}$ where our calculated $\gamma_{\rm [SII]}$ can no longer be properly applied.
Similarly, the far right panel compares the [OIII] line and {\it logU} method with the [SII] line and {\it doublet} method.

Our summary for these three density estimators is then:
The {\it doublet} method is severely biased. It will yield incorrect masses unless used with the [SII] line luminosity and $\gamma_{\rm [SII]}$ given in Table~\ref{tab:gamma}; and even then it can only be used when $n_{\rm e} < 10^3$\,cm$^{-3}$.
The {\it TA} method is robust, but unfortunately in many cases impractical because it relies on measurements of very weak lines. Even in our sample of local luminous AGN we were unable to use it in about 1/3 of the cases.
The {\it logU} method is both robust and straightforward to apply. 
It has the additional advantage of emphasizing, by its definition, that the density is expected to decrease at increasing distance from the AGN.
It is, however, more suited to spatially resolved data and one needs to be careful when estimating the characteristic distance from the AGN to the outflow in aperture measurements.
It is this method, together with the [OIII] line luminosity and respective emissivity that we use in Sec.~\ref{sec:rate} when estimating the outflow rates for the AGN.

\begin{table*}
\caption{Measured and derived properties related to outflows in the active galaxies. $^a$}
\label{tab:outflow}
\begin{tabular}{lccccc D{.}{.}{4} cc}
\hline
Object & log\,L$_{\rm AGN}$ & R$_{\rm ap}$$^b$ & log\,L$_{\rm [OIII]}$ & v$_{\rm 98,[OIII]}$ & log\,$M_{\rm out}$ $^c$ & \multicolumn{1}{c}{$\dot{M}$} & $\dot{E}_{\rm kin}$ & log\,$\dot{E}_{\rm kin}/L_{\rm AGN}$ \\
       & erg\,s$^{-1}$ & pc & erg\,s$^{-1}$ & km\,s$^{-1}$ & M$_{\sun}$ & \multicolumn{1}{c}{M$_{\sun}$\,yr$^{-1}$} & erg\,s$^{-1}$ &  \\ 
\hline
 ESO\,137-G034 &  43.4 & 152 & 40.8 &  519 & 5.2 & 0.52 & 40.7 & -2.7 \\
 MCG-05-23-016 &  44.3 & 152 & 39.7 &  255 & 3.3 & 0.003& 37.8 & -6.4 \\
     NGC\,2110 &  44.5 & 148 & 40.4 & 1665 & 3.4 & 0.03 & 40.4 & -4.1 \\
     NGC\,2992 &  43.3 & 157 & 40.4 &  587 & 4.8 & 0.26 & 40.5 & -2.8 \\
     NGC\,3081 &  44.1 & 148 & 40.6 &  279 & 4.3 & 0.04 & 39.0 & -5.0 \\
     NGC\,5506 &  44.1 & 117 & 41.2 &  792 & 4.5 & 0.21 & 40.6 & -3.5 \\
     NGC\,5728 &  44.1 & 170 & 40.5 &  876 & 4.2 & 0.09 & 40.3 & -3.8 \\
     NGC\,7582 &  44.0 &  95 & 40.1 &  364 & 3.2 & 0.007& 38.4 & -5.6 \\
 ESO\,021-G004 &  43.3 & 170 & 39.4 &  432 & 3.6 & 0.01 & 38.8 & -4.5 \\
     NGC\,1365 &  43.0 &  78 & 38.6 &  373 & 2.5 & 0.002& 37.8 & -5.2 \\
     NGC\,7172 &  44.1 & 161 & 39.9 &  397 & 3.3 & 0.005& 38.4 & -5.7 \\
\hline
\end{tabular}

$^a$ As discussed in Sec.~\ref{sec:den_comp} we have adopted $n_{\rm e}$ derived using the {\it logU} method. 
As such, the uncertainty of $M_{\rm out}$ is dominated by the uncertainty of $n_{\rm e}$ which is typically 0.14\,dex (through $\log{U}$ and $L_{\rm AGN}$). This propagates directly into the outflow rate $\dot{M}$ and kinetic power $\dot{E}_{\rm kin}$ (for both of which the choice of outflow model can have a factor three impact). Because $L_{\rm AGN}$ impacts the ratio $\dot{E}_{\rm kin}/L_{\rm AGN}$ a second time, the uncertainty on that increases to 0.21\,dex.\\
$^b$ The radius of the outflow $r_{\rm out}$ is based on the aperture radius $R_{\rm ap}$ with an adjustment for projection as described in Sec.~\ref{sec:den_U}.\\
$^c$ For consistency with our use of L$_{\rm [OIII]}$, we have used $v_{\rm 98,[OIII]}$ rather than $v_{\rm 98,[SII]}$ although typically they differ by only $\sim$25\%.\\
\end{table*}

\section{Outflow Mass and Rate}
\label{sec:rate}

The expression for the outflow rate $\dot{M}_{\rm out}$ depends on the outflow scenario and geometry considered (see, for example, \citealt{lut19}).
One option is to assume a spherically or biconically symmetric flow with constant velocity and uniform density up to a given radius 
(e.g. \citealt{can12,bru15,bis17,fio17,roj20}).
In this scenario, the outflow rate decreases over time, and what one calculates is the initial outflow rate which is traced by the material at the maximum radius considered.
The outflow rate is then a factor 3 higher than that derived using the common alternative models of a time-averaged thin shell geometry, or a density profile scaling as $r^{-2}$ so that the outflow rate is constant over time.
(e.g. \citealt{rup05,gon17,vei17,flu19,lut19}. 
We have adopted the latter class of models, for which 
$\dot{M}_{\rm out} = M_{\rm out} \, v_{\rm out} \, / \, r_{\rm out}$
Similarly the kinetic power of the outflow is calculated as 
$\dot{E}_{\rm kin} = 1/2 \, \dot{M}_{\rm out} \, v^2_{\rm out}$.

We have used the [OIII] line luminosity when calculating the outflowing mass, and adopted the volume emissivity $\gamma_{\rm [OIII]}$ given in Table~\ref{tab:gamma}.
To be consistent with that, we have also used the maximum outflow velocity derived from the [OIII] line, $v_{\rm 98,[OIII]}$.
In terms of $n_{\rm e}$, we have taken that derived using the {\it logU} method.
The resulting outflowing masses and rates are given in Table~\ref{tab:outflow}.
We note that if we were to use the H$\alpha$ line luminosity to estimate mass, 
with $\gamma_{\rm H\alpha}$ as given in Table~\ref{tab:gamma}, we would get very similar outflowing masses as is shown in the left panel of Fig.~\ref{fig:mass_comp}; 
and if we were to use $v_{\rm 98,[SII]}$ as the outflow velocity, the left panel of Fig.~\ref{fig:sii_oiii} shows that the the outflow rates would be about 25\% lower.
As a cross-check, we compare the outflow rates for two objects to recently published values based on spatially resolved data.
Despite the different methods used, our outflow rate for NGC\,5728 compares well to that of 0.08\,$M_{\sun}$\,yr$^{-1}$ for both sides of the outflow within 250\,pc found by \citet{shi19}.
For NGC\,1365, the outflow rate on the south-east side within 380\,pc found by \citet{ven18} was 0.01\,$M_{\sun}$\,yr$^{-1}$.
This is about a factor 10 higher than ours because, as discussed above, the density from the [SII] doublet used by those authors, leads to an overestimate of the outflowing mass by that amount.

\begin{figure*}
\includegraphics[width=18cm]{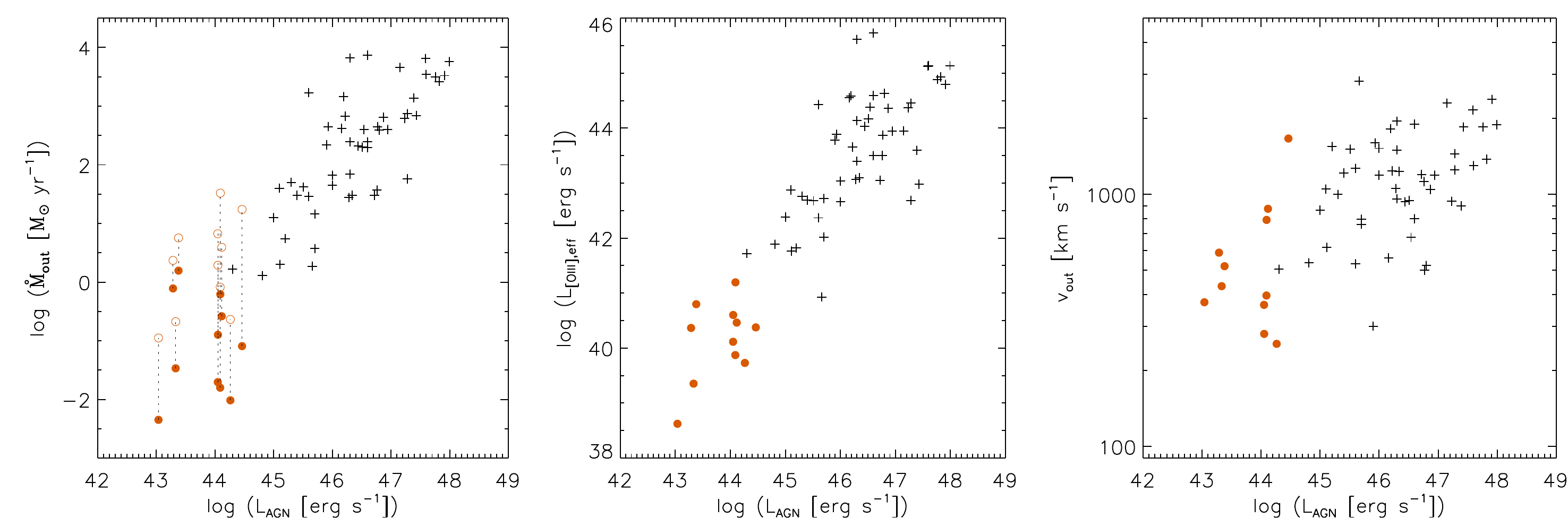}
\caption{Relations with $L_{\rm AGN}$ using data from \citet{fio17} (black plus points) and from Table~\ref{tab:outflow} for our AGN sample (red filled circles).
Left: relation with $\dot{M}_{\rm out}$. In this panel, the outflow rates for our AGN have been increased by a factor three to match the equation used by \citet{fio17}. The open circles indicate how those points would shift if in addition we used $n_{\rm e} = 200$\,cm$^{-3}$ as adopted by those authors for all their objects.
One needs to be cautious with this relation.
Centre: relation with $L_{\rm [OIII],eff}$, the luminosity that would be required to produce the given outflow rate. This covers 5 orders of magnitude, and is the primary driver of the outflow rate relation.
Right: relation with $v_{\rm out}$ (as given by \citealt{fio17}, and $v_{\rm 98,[OIII]}$ for our AGN). This covers barely a single order of magnitude and so does not contribute much to the outflow rate relation.}
\label{fig:fiore}
\end{figure*}

\begin{figure}
\includegraphics[width=8.5cm]{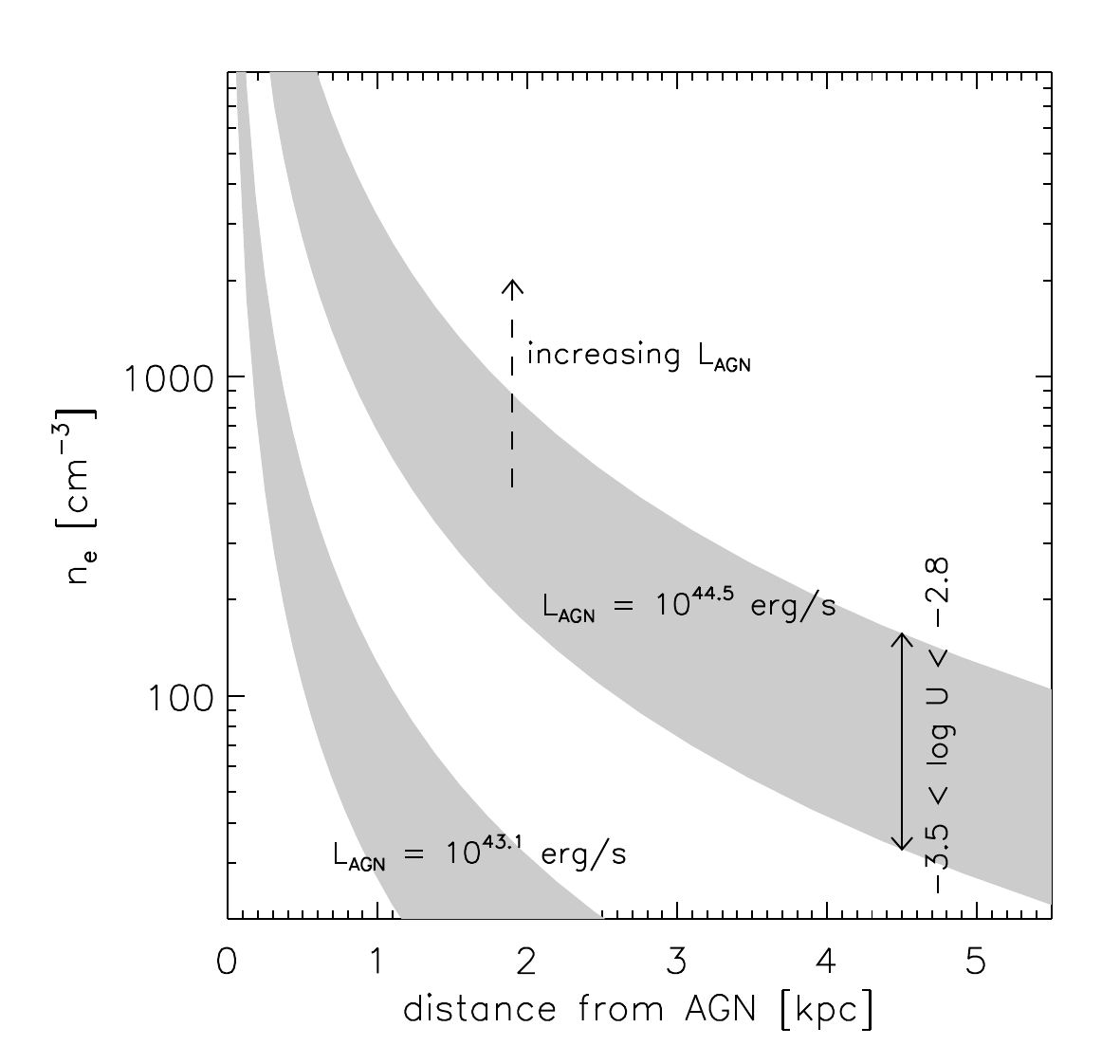}
\caption{Illustration of how one might estimate a reasonable range for the electron density in an outflow at a given distance from an AGN of known luminosity.
This figure shows two example ranges for $L_{\rm AGN}$ corresponding to the minimum and maximum in our sample. The grey regions indicate the uncertainty due to an unknown ionization parameter.
This figure should be used as a guide only, and emphasizes how much variation can be expected in outflow densities between objects and at different radii.}
\label{fig:sdss_density}
\end{figure}

\begin{figure}
\includegraphics[width=8.5cm]{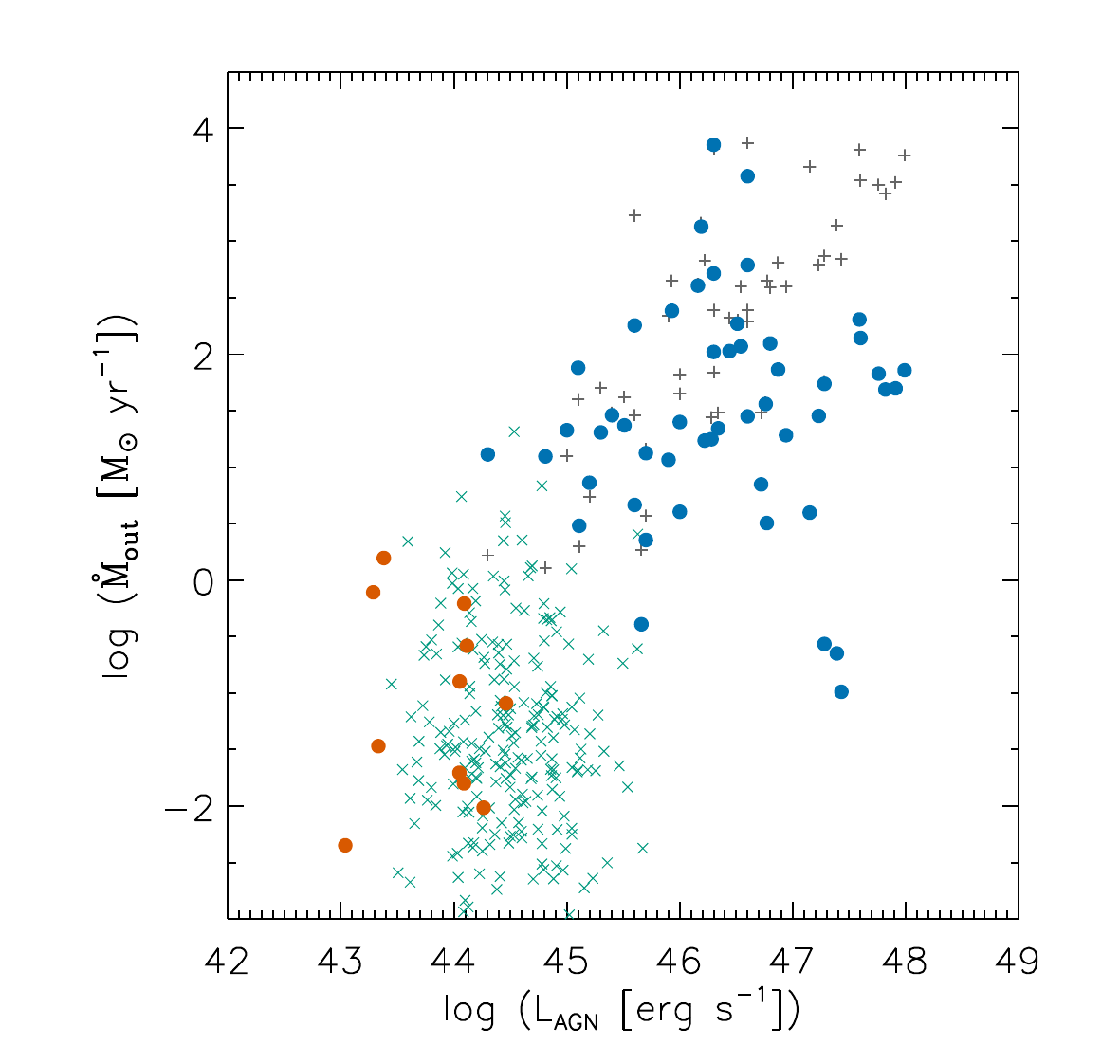}
\caption{A modification to the well known relation from \citet{fio17} in which we have adjusted the outflow rates according to our best estimate of the density as described in Sec.~\ref{sec:fiore_disc}. The original data are shown as grey plus points, while the adjusted points are drawn as filled blue circles. We have included our own points as filled red circles (again a factor three more than in Table~\ref{tab:outflow}, as described in Sec.~\ref{sec:fiore_disc}), as well as the data from \citet{bar19} as green crosses.
A correlation remains but there is more scatter than before, and with the original data tracing its upper limit.
The three outliers at high $L_{\rm AGN}$ and low $\dot{M}$ are at $z \sim 2.4$. One was already a low point on the original relation; the other two have very high estimated densities because $r_{\rm out}$ is small.}
\label{fig:fiore_alt}
\end{figure}

\subsection{Outflow rate relation at low luminosity}
\label{sec:fiore_disc}

Although the AGN in our sample are considered luminous with respect to local AGN, they are of only moderate luminosity when compared to samples taken from larger distances that include luminous quasars.
As such, they provide an opportunity to extend to lower luminosity the relation between AGN luminosity and outflow rate in the literature.
One of the most well known is that of \citet{fio17}, which is shown in the left panel of Fig.~\ref{fig:fiore}.
However, there are two important adjustments to our data required in order to effect such a comparison.
The first is that \citet{fio17} adopted a definition of outflow rate that is a factor three higher than the one we use above (as discussed at the start of this Section).
The second is that to deal with the multitude of values for $n_{\rm e}$ adopted in the literature by different authors in the data they compiled, \citet{fio17} re-scaled the mass of outflowing gas using a single electron density of $n_{\rm e} = 200$\,cm$^{-3}$.
This is a rather problematic issue, because that density is less than any of those we have measured in our AGN sample.
To indicate the major impact this has, we have plotted our derived outflow rates from Table~\ref{tab:outflow} as filled circles, and values adjusted to $n_{\rm e} = 200$\,cm$^{-3}$ as open circles.
It is immediately clear that the relatively low scatter in the relation over 5 orders of magnitude is at least partly due to the adoption of a single value for the density of all the outflows.
Using the densities we have derived, the low luminosity AGN are highly scattered and lie below where one would expect from the correlation.

Addressing this issue has prompted us to look also at the parameters $L_{\rm line}$ and $v_{\rm out}$ that contribute to the relation given that, for constant $n_{\rm e}$, 
$\dot{M}_{\rm out} \propto L_{\rm line} \, v_{\rm out} \, / \, r_{\rm out}$ as shown in Sec.~\ref{sec:intro} (and noting that there is no individual correlation between $L_{\rm AGN}$ and $r_{\rm out}$).
Using the data in \citet{fio17}, we have calculated (by `reverse engineering') the effective [OIII] luminosity $L_{\rm [OIII],eff}$ that would be required to produce the given outflow rate, and have verified this approach using the [OIII] luminosities available (F.~Fiore priv. comm.).
The individual correlations for $L_{\rm [OIII],eff}$ and $v_{\rm out}$ are shown in the centre and right panels of Fig.~\ref{fig:fiore}.
They show that it is the line luminosity $L_{\rm [OIII]}$ rather than the outflow speed $v_{\rm out}$ that must be the primary driver of the correlation over so many orders of magnitude.
It is a long established relation for X-ray selected AGN \citep{mul94,alo97,hec14}, which has a similar large scatter.
Using a sample selected on the basis of broad H$\alpha$, \citet{ste12} find a good (approximately linear) correlation with a scatter of 0.44\,dex;
and based on compiled data for a very hard X-ray selected sample, \citet{ber15} suggest that the scatter is 0.62\,dex.
The scatter for the \citet{fio17} data shown here is 0.67\,dex.
Although there will be additional uncertainty in our `reverse engineered' estimate of L$_{\rm [OIII]}$, the correlation covers the range required, and naturally accounts for the approximate proportionality between $\dot{M}_{\rm out}$ and L$_{\rm AGN}$ -- 
although some authors have noted a non-linear relation between narrow line and AGN luminosity \citep{net06,ste12}.

Given the issues above, we explore an alternative representation of the \citet{fio17} relation:
instead of fixing a constant density for all objects, we adopt an estimate of the density based on the ionization parameter method.
We use the given AGN luminosites $L_{\rm AGN}$ and outflow sizes $r_{\rm out}$, making the same assumptions as in Sec.~\ref{sec:den_U} about distribution of line luminosity within an aperture and projection.
We adopt a value of $\log{U} = -2.7$, calculated to be the median for the outflows in \citet{fio17} for which both [OIII] and H$\beta$ luminosities are available (F.~Fiore priv. comm.), and which should therefore be characteristic for AGN photoionized outflowing gas.
We can then make an order-of-magnitude estimate of $n_{\rm e}$ as illustrated in Fig.~\ref{fig:sdss_density}.
The key point of this figure is that it emphasizes the enormous variation that can be expected for $n_{\rm e}$ in different outflows.
It highlights how, in the central few hundred parsecs one can find densities in the range $10^3$--$10^4$\,cm$^{-3}$;
and that while in some AGN this may fall to $\sim$100\,cm$^{-3}$ at kiloparsec scales, for luminous quasars it may still be in the $\sim$1000\,cm$^{-3}$ range.

Bearing in mind the expected variation in outflow density, in Fig.~\ref{fig:fiore_alt} we have modified the relation of \citet{fio17} to take into account our best estimate of $n_{\rm e}$ for the individual objects, and overplotted both our data as well as that from \citet{bar19}.
The relation between AGN luminosity and outflow rate, which covers five orders of magnitude in $L_{\rm AGN}$, remains but with two important differences:
(i) the scatter is much larger than before, and (ii) the outflow rates are typically lower by about a factor three since the densities are generally that much higher.
The original relation appears to be more of an upper limit on the outflow rate, but not all AGN driven outflows reach that limit.
With the data available, the values we have estimated for $n_{\rm e}$ are still uncertain and so we would only claim that this modified relation should be considered as equally valid as the original relation of \citet{fio17}.
It does, however, underline that there remain unanswered questions about how the ionized gas outflow rate is related to AGN luminosity.

\section{Conclusion}
\label{sec:conc}

We present an analysis of the kinematics and electron density derived from emission lines integrated over the central 1.8\arcsec\ ($\sim$300\,pc) of local luminous active galaxies and matched inactive galaxies. 
The data are from Xshooter observations, taken as part of the LLAMA survey which comprises a complete volume limited sample of 14-195\,keV selected AGN together with inactive galaxies matched by their global properties. We find:
\begin{itemize}
\item
The peak of the emission lines in the AGN are offset from systemic (traced by stellar absorption features), indicating that the line profiles of the AGN are dominated by outflow. Any systemic component is highly sub-dominant. In contrast, the line profiles of the inactive galaxies are characterised by a core close to the systemic velocity with a wing tracing outflow.
\item
For the galaxies with weak emission lines (equivalent width of [SII]\,6716\AA\ $\la10$\AA), the [SII] doublet ratio method overestimates the electron density unless a correction is made for stellar absorption features. This is particularly important for the inactive galaxies where the impact on $n_{\rm e}$ is a factor two. After correcting for the stellar continuum, their median density is found to be $n_{\rm e} = 190$\,cm$^{-3}$.
\item
For the active galaxies, the densities found using three independent methods (the [SII] doublet ratio, a method using auroral and transauroral line ratios, and a method based on the ionization parameter) differ. The median densities found with these methods are 350\,cm$^{-3}$, 1900\,cm$^{-3}$, and 4800\,cm$^{-3}$ respectively.
The latter two have comparable ranges and are both significantly larger than the density found with the [SII] doublet ratio.
An explanation for this difference lies in the ionization structure of the clouds. Specifically, in the conditions considered here, most of the [SII] line emission originates in a partially ionized zone where $n_{\rm e}$ drops rapidly to below 10\% of $n_{\rm H}$ and most of the gas is neutral.
This invalidates the assumptions under which the [SII] doublet is used to estimate density.
\item
The definition of `ionized mass' ought to be treated carefully. Specifically, the extensive region in which most of the gas is neutral increases the total mass of the cloud by a factor 2--4 with respect to the fully ionized gas.
\item
The implied outflow rates are 0.001--0.5\,M$_{\sun}$\,yr$^{-1}$, and span an order of magnitude more range than the AGN luminosities of $(1-30)\times10^{43}$\,erg\,s$^{-1}$. Over this limited range of luminosity, the scatter in outflow rate is the dominant effect, overwhelming the relation between the two quantities.
\item
The AGN in this sample extend the lower luminosity end of the relation between outflow rate and AGN luminosity in the literature. However, caution needs to be applied when making such comparisons because of uncertainties in the outflow density. Adjusting the data points in the relation using our best estimate of $n_{\rm e}$ shows that there may be more scatter than previously thought, and that the outflow rates may be at least a factor of a few lower than previously estimated.
\end{itemize}


\section*{Acknowledgements}
The authors thank the referee for careful reading of the manuscript and providing feedback that helped to improve it.
We thank all the staff of the Paranal Observatory who were involved in carrying out the service mode observations used in this paper.
And we extend particular thanks to Fabrizio Fiore for providing additional data for the ionized outflows.
We also thank Marina Trevisan for a discussion about features in the stellar continuum.
And we thank the referee for suggestions that have helped improve the manuscript.
R.R. thanks CNPq, CAPES, and FAPERGS.
This research has made use of the NASA/IPAC Extragalactic Database (NED), which is funded by the National Aeronautics and Space Administration and operated by the California Institute of Technology. 


\section*{Data Availability}
The data underlying this article are available in the ESO Science Archive Facility, at http://archive.eso.org/cms.html.
The datasets were derived from observations that are in the public domain, under Program ID 095.B-0059(A) and 092.B-0083(A).



\appendix

\section{Individual fits for active and inactive galaxies}

\begin{figure*}
\includegraphics[width=17.5cm]{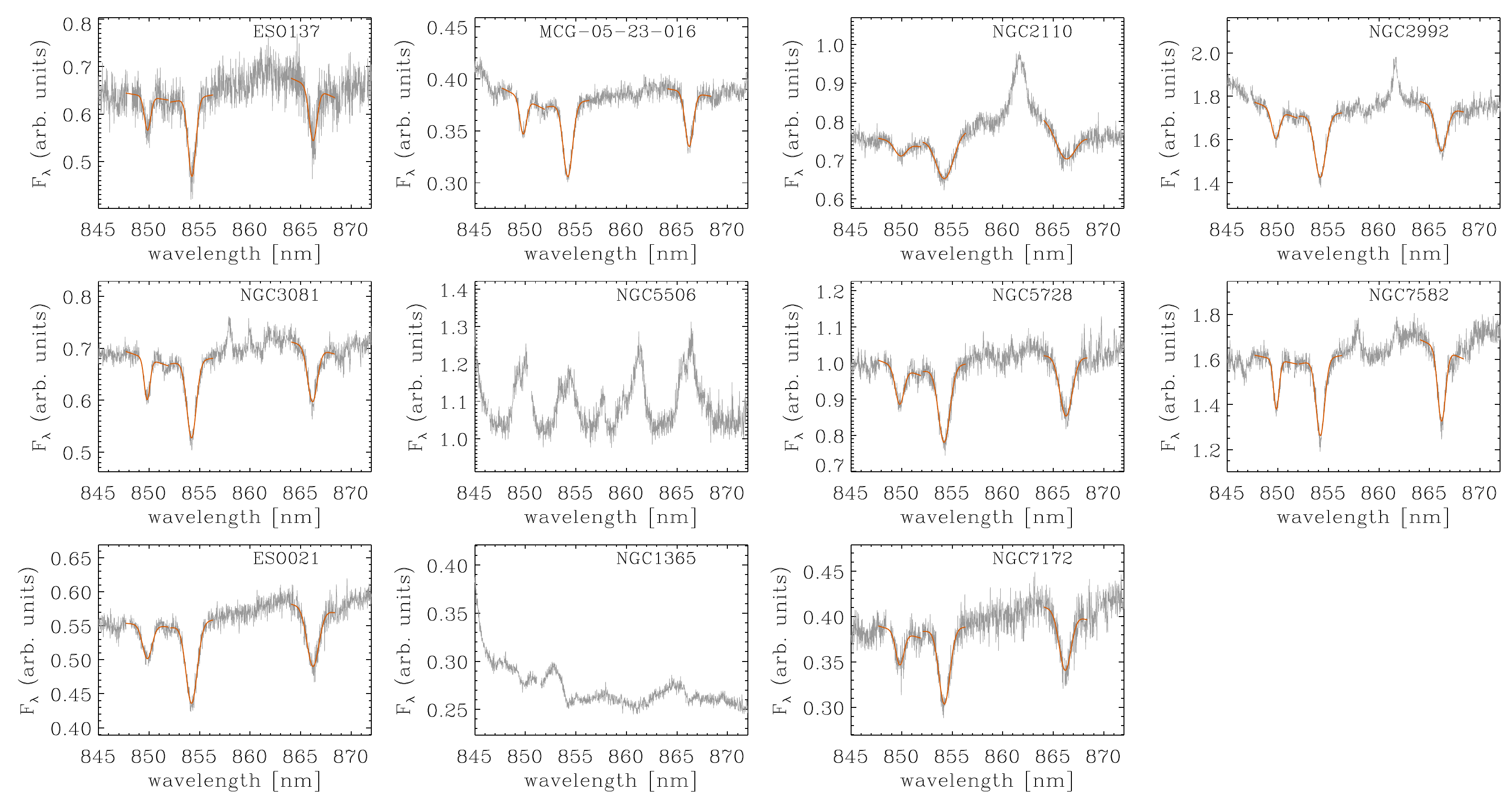}

\vspace{5mm}

\includegraphics[width=17.5cm]{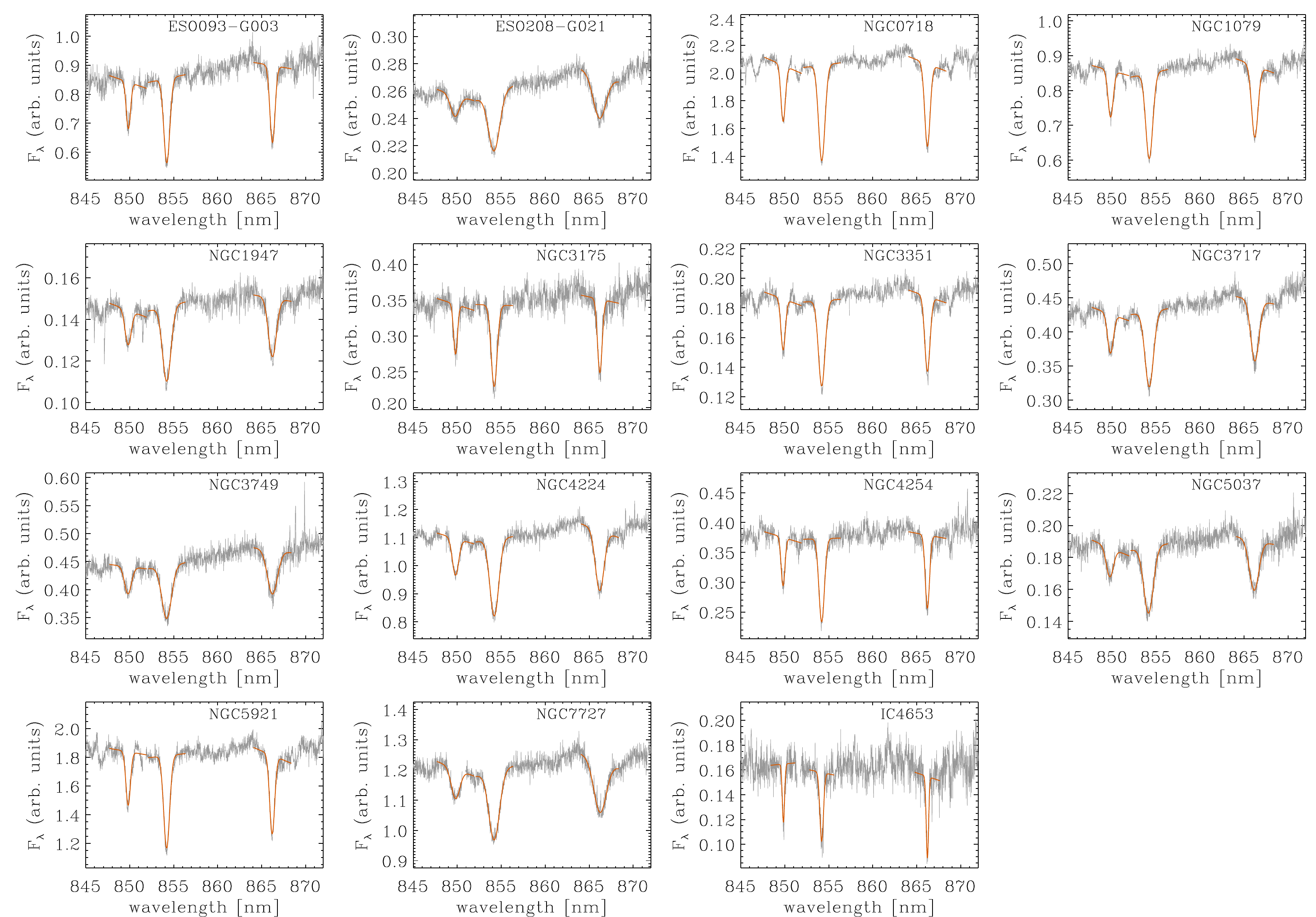}
\caption{Fits to the CaII triplet lines for one spectrum of each active galaxy (upper 11 panels) and inactive galaxy (lower 15 panels).
Only the data are shown for NGC\,1365 and NGC\,5506 because there were no clear absorption lines to fit. The data are shown in grey and the fits in red. A Gaussian line profile was fitted independently to each line in order to avoid bias due to the continuum shape and emission lines. The arbitrary units of $F_\lambda$ are the same for all panels.}
\label{fig:caii_linefits}
\end{figure*}

Specific fits to the CaII triplet lines for one spectrum of each of the active and inactive galaxies are shown in Fig.~\ref{fig:caii_linefits}. The respective spectral segments for NGC\,1365 and NGC\,5506 are shown without fits, because the absorption lines could not be convincingly identified as separate features.
In all cases, a single Gaussian fits the line profile very well, with no evidence for additional narrow components that might indicate a contribution from the ISM.

\begin{figure*}
\includegraphics[width=17.5cm]{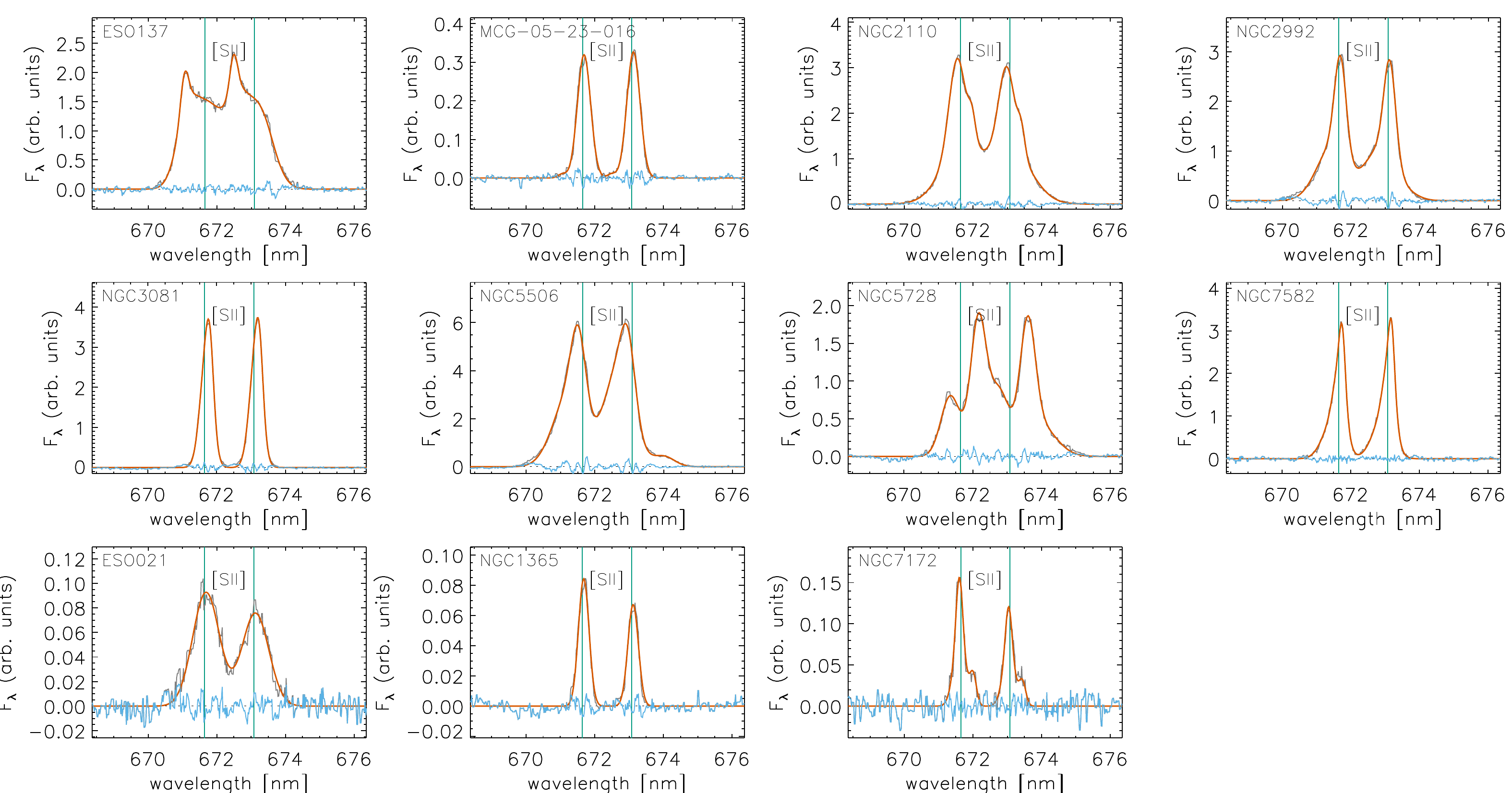}

\vspace{5mm}

\includegraphics[width=17.5cm]{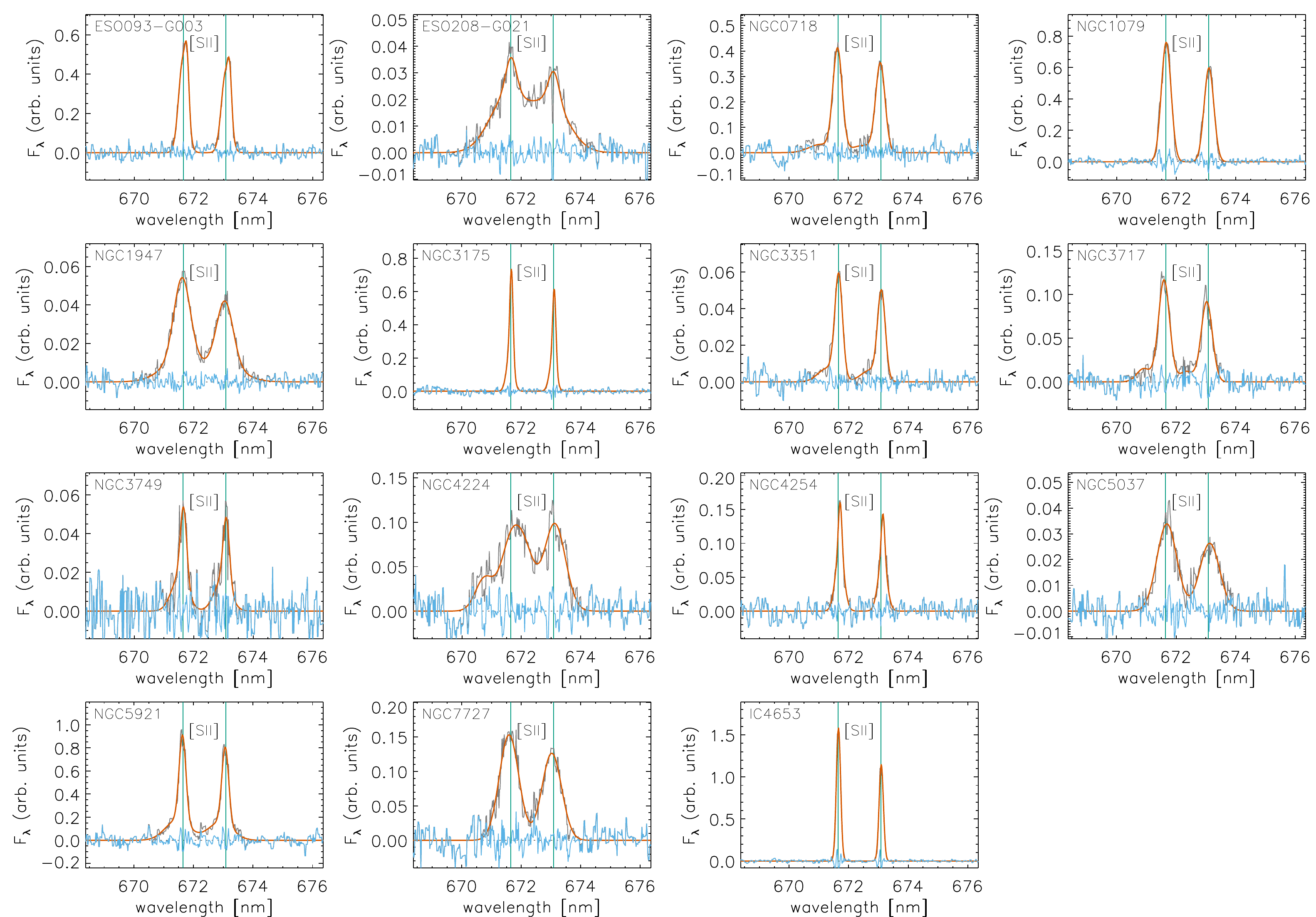}
\caption{Fits to the [SII] doublet in one spectrum for each active galaxy (upper 11 panels) and inactive galaxy (lower 15 panels). The data are shown in grey, the fits in red, and the residuals in blue. In these plots, the fitting process included a linear function to match the local continuum. The arbitrary units of $F_\lambda$ are the same for all panels.}
\label{fig:sii_linefits}
\end{figure*}

\begin{figure*}
\includegraphics[width=17.5cm]{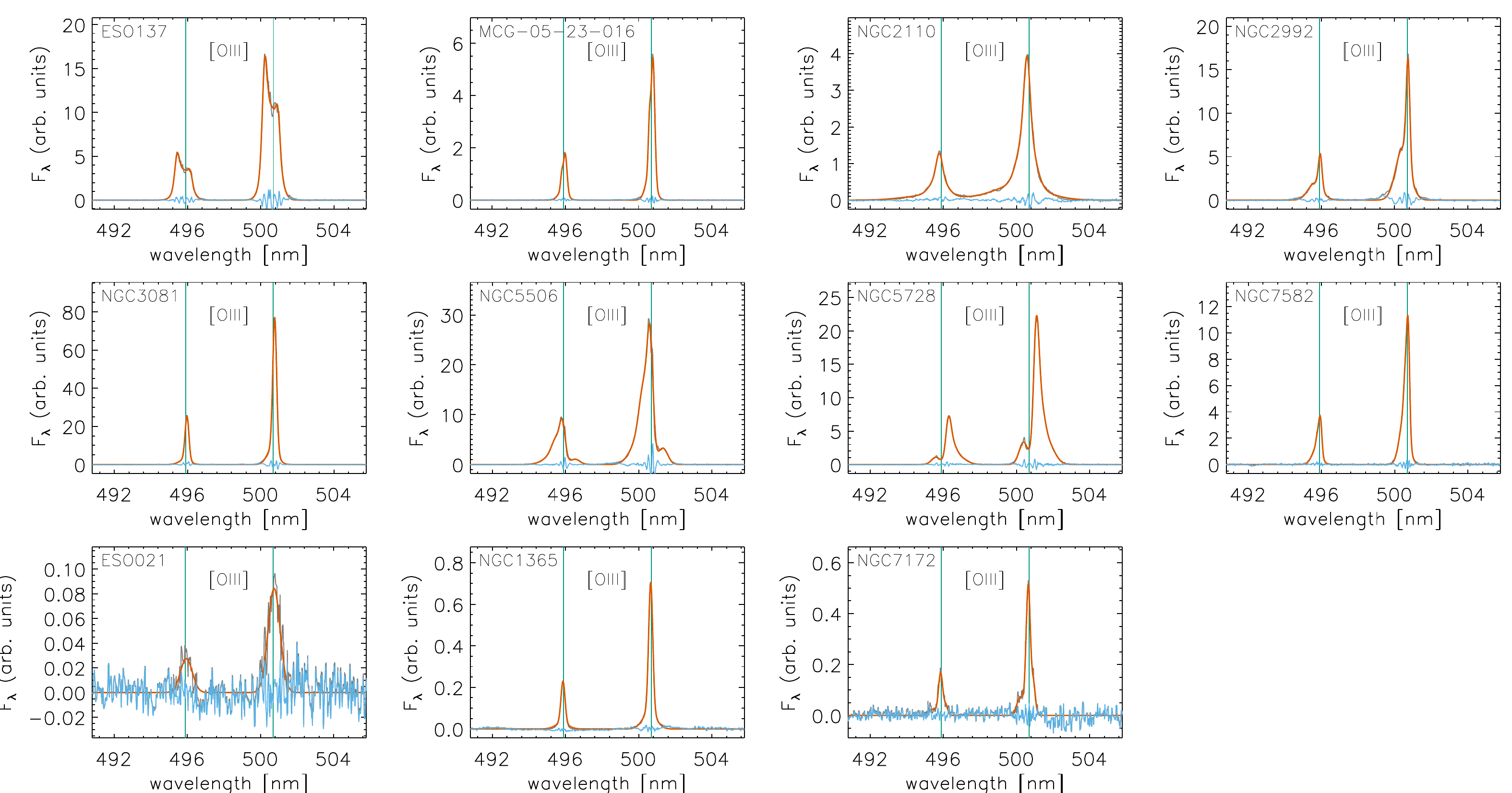}
\caption{Fits to the [OIII] doublet in one spectrum for each active galaxy. The data are shown in grey, the fits in red, and the residuals in blue. In these plots, the fitting process included a linear function to match the local continuum. The arbitrary units of $F_\lambda$ are the same for all panels.}
\label{fig:oiii_linefits_s}
\end{figure*}

Specific fits to the [SII] doublet for one spectrum of each of both the active and inactive galaxies are shown in Fig.~\ref{fig:sii_linefits}.
The fitting process is described in Sec.~\ref{sec:linefitting}.
Equivalent fits to the [OIII] doublet for the active galaxies are shown in Fig.~\ref{fig:oiii_linefits_s}.
These have been fitted in the same way as the [SII] doublet, but independently, because the [SII] and [OIII] profiles are not good matches to each other. 

\begin{figure*}
\includegraphics[width=18cm]{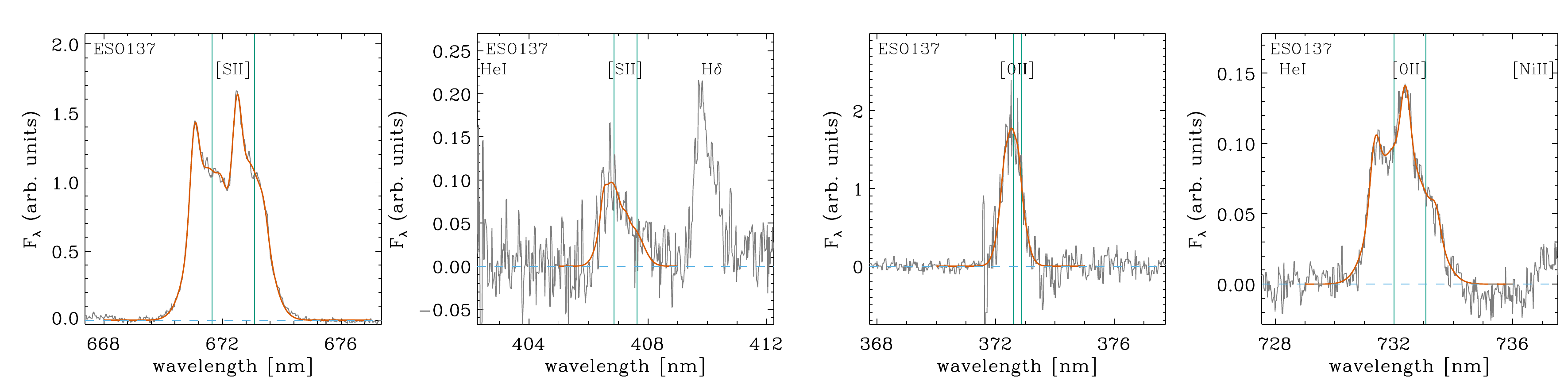}
\includegraphics[width=18cm]{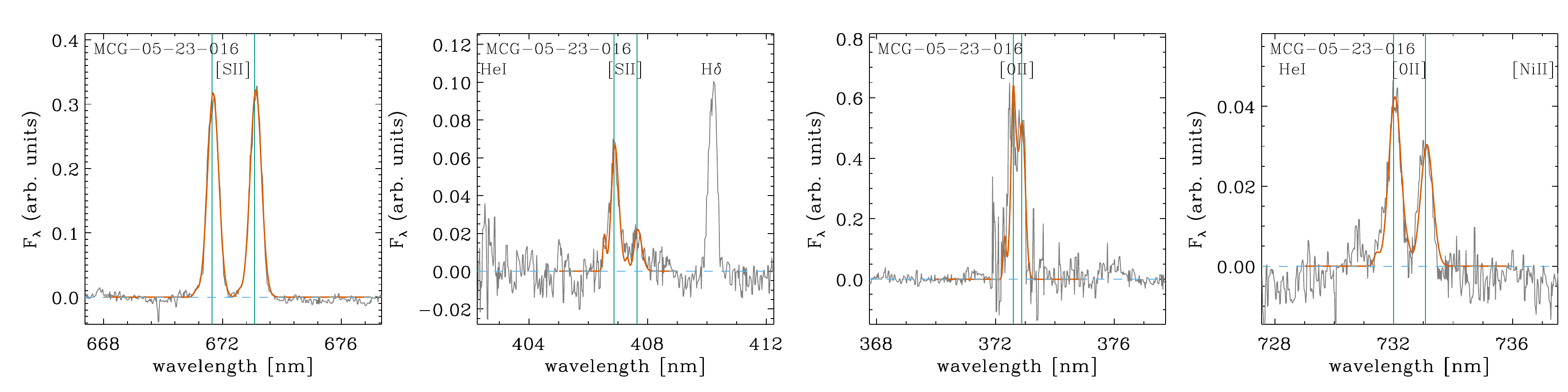}
\includegraphics[width=18cm]{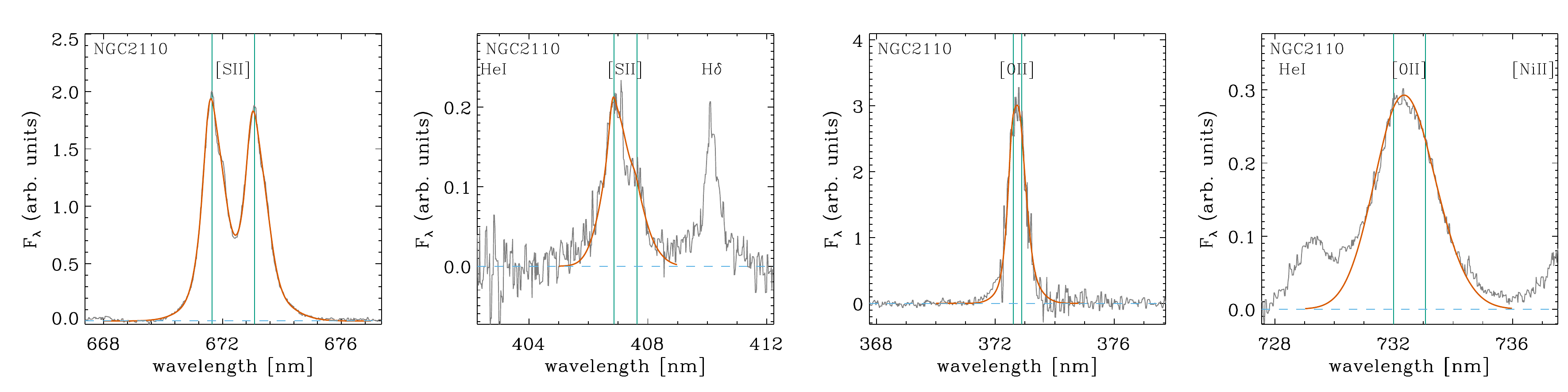}
\includegraphics[width=18cm]{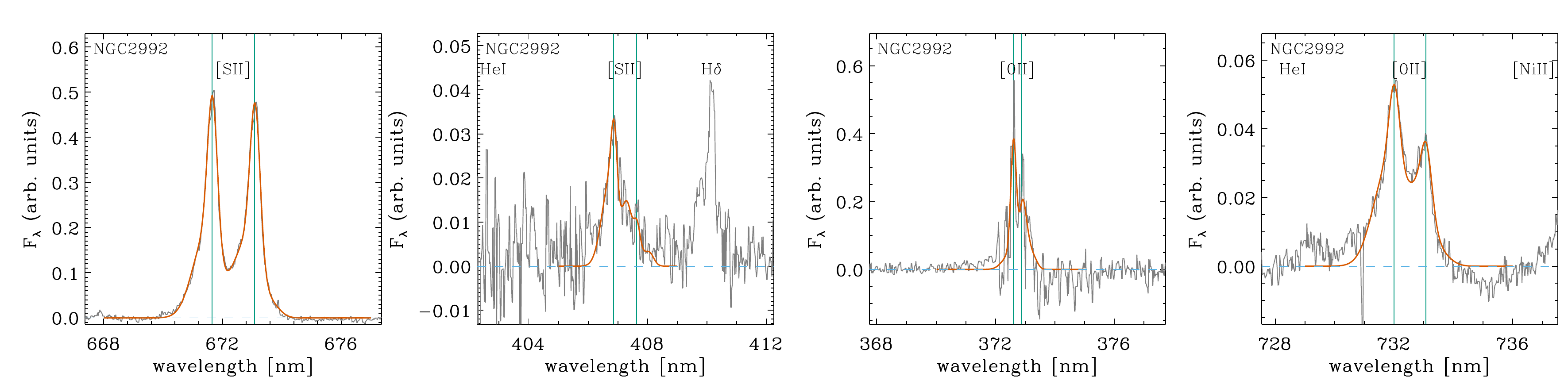}
\caption{Fits of the various [SII] and [OII] doublets in one spectrum for each active galaxy. In these plots, a linear function to match the local continuum was fitted separately to the emission lines using line-free regions either side of the line of interest.
The arbitrary units of $F_\lambda$ are the same for all panels.} 
\label{fig:ta_linefits1}
\end{figure*}

\begin{figure*}
\includegraphics[width=18cm]{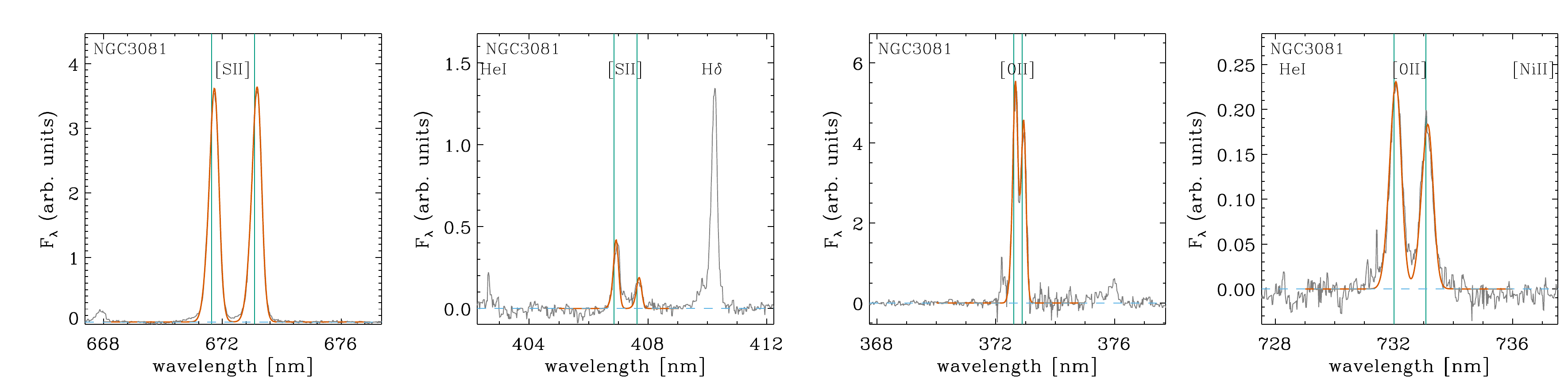}
\includegraphics[width=18cm]{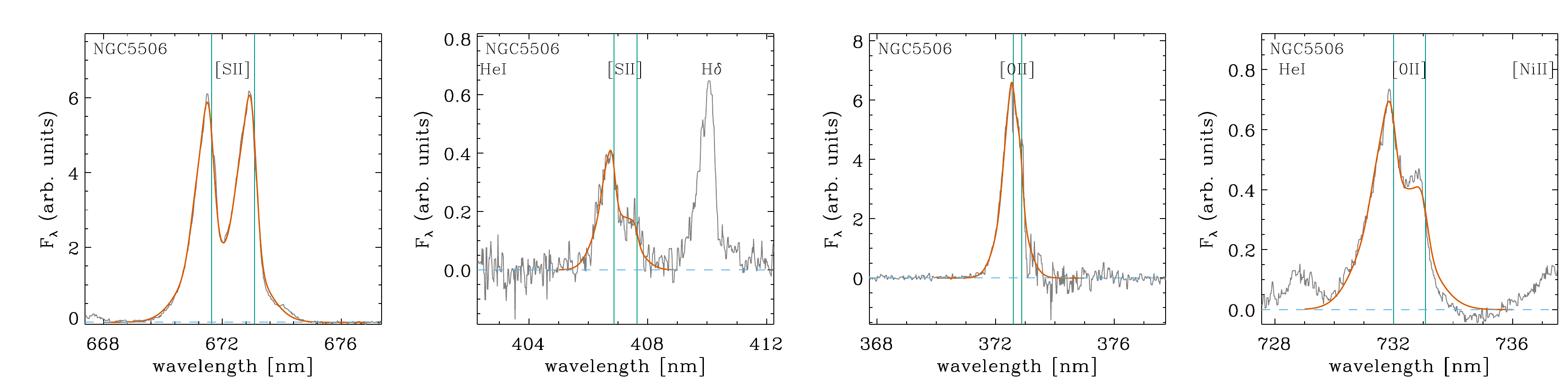}
\includegraphics[width=18cm]{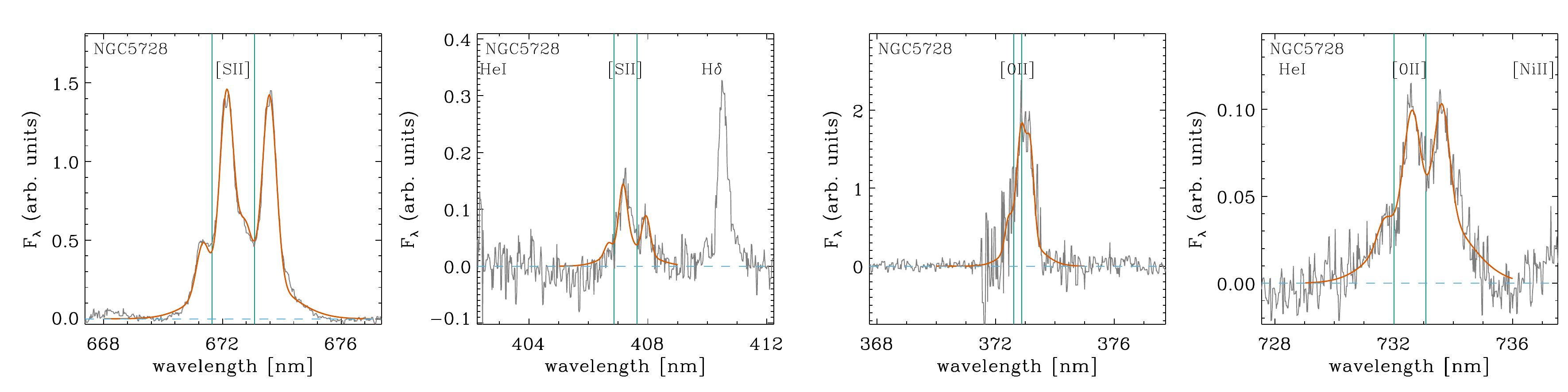}
\includegraphics[width=18cm]{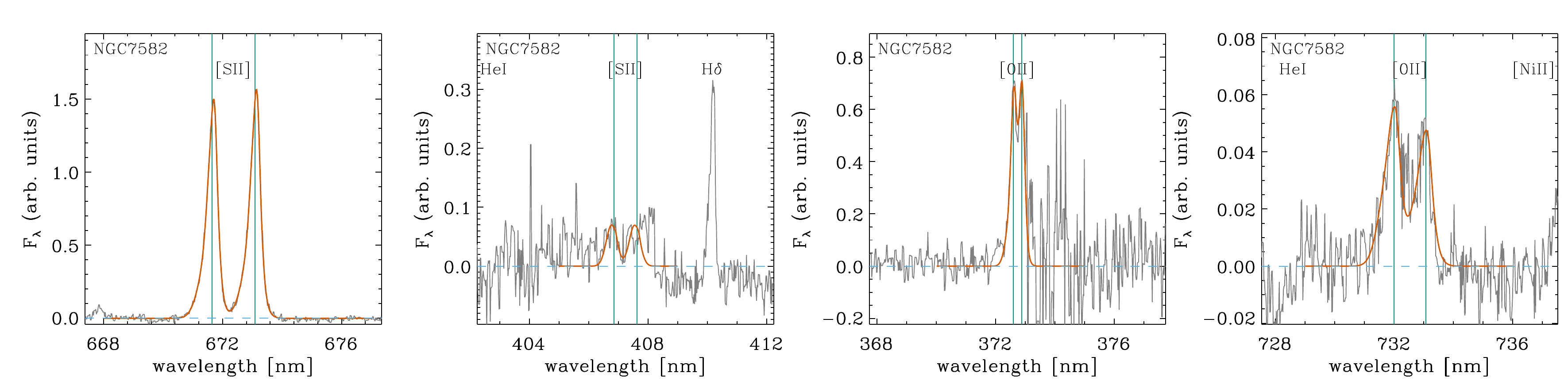}
\caption{Fits of the various [SII] and [OII] doublets in one spectrum for each active galaxy. In these plots, a linear function to match the local continuum was fitted separately to the emission lines using line-free regions either side of the line of interest.
The arbitrary units of $F_\lambda$ are the same for all panels.} 
\label{fig:ta_linefits2}
\end{figure*}

For the eight active galaxies in which we fit the auroral and transauroral lines, the fits to the strong [SII] doublet are repeated in the left-hand panels of Figs.~\ref{fig:ta_linefits1} and~\ref{fig:ta_linefits2}.
The remaining three panels in each row of these figures show fits to the other lines used to estimate the electron density in the auroral/transauroral method.
The line profiles have been constrained in a similar way to the [SII] doublet, with the additional criterion that only the scaling of each component is allowed to vary (the velocity offset and width are fixed to that derived from the [SII] doublet fit).
Especially for the weaker lines, fitting and subtracting the continuum to minimize the impact of stellar absorption features is an important preparatory step.
This has been done by fitting theoretical stellar templates to line-free regions over a wider baseline than shown in the panels, as described in Sec.~\ref{sec:contsub}, and checked by eye.
Locations where additional line emission is expected have been marked in the various panels.

\begin{figure*}
\includegraphics[width=17.5cm]{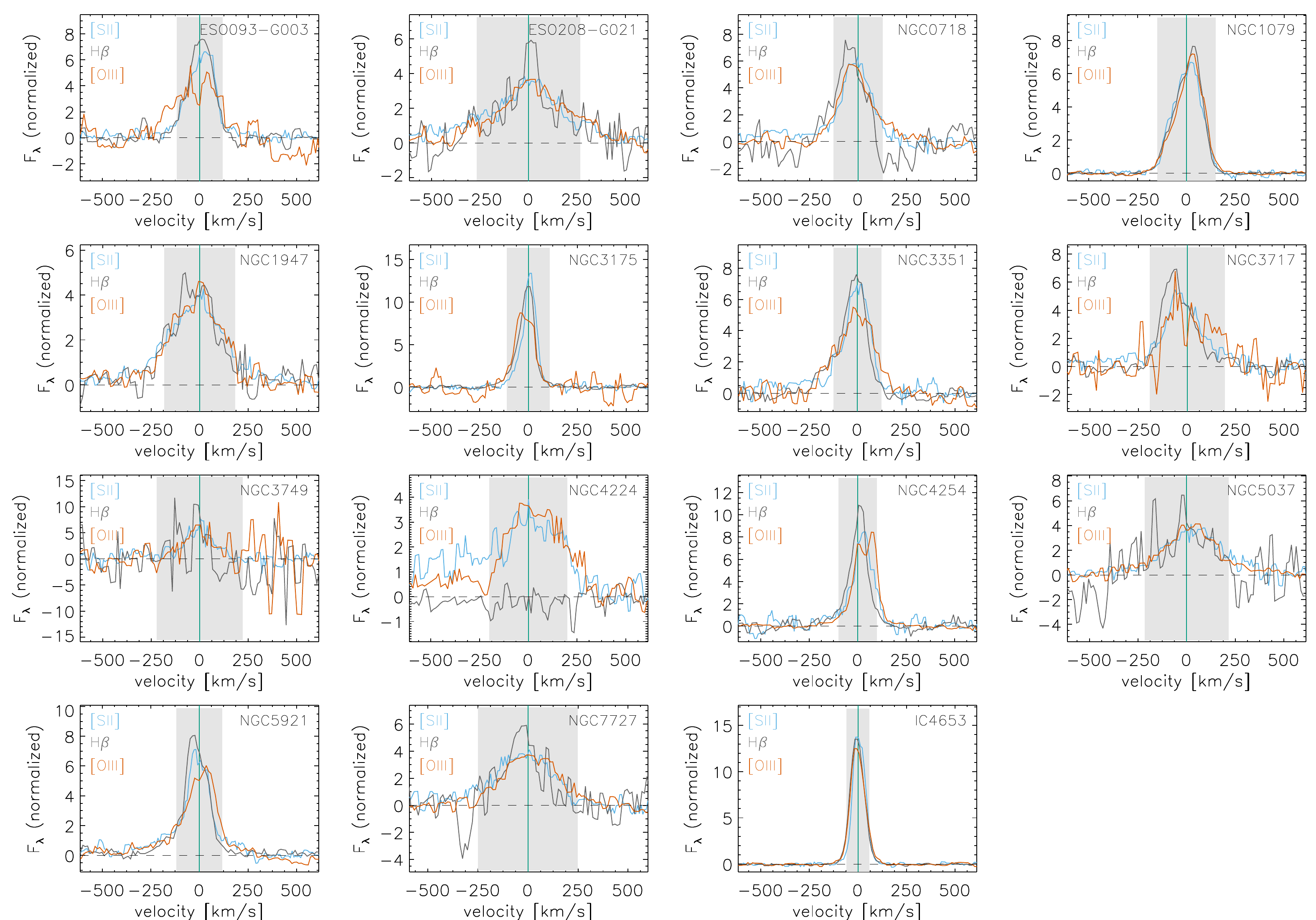}
\caption{Comparison of the central part of the line profiles of the inactive galaxies as a function of velocity, normalised so that they have the same flux within $\pm$250\,km\,s$^{-1}$.
The H$\beta$, [OIII], and [SII] profiles (data only) are shown in grey, red, and blue respectively; with the FWHM of the stellar absorption profile indicated by the shaded grey region for reference.
For visualisation purposes, the [SII] profile is a combination of the short side of the 6716\AA\ line and the long side of the 6731\,\AA\ line, scaled to match where they overlap.}
\label{fig:profile_inactive}
\end{figure*}

The profiles of the H$\beta$, [SII], and [OIII] lines for the inactive galaxies are shown in Fig.~\ref{fig:profile_inactive}. These show a remarkable variety of shapes, although the profiles tend to be centered close to the systemic velocity and asymmetries are less pronounced than for the active galaxies.
Although there is a variety of line widths among the objects, they appear to be consistent with the gravitational potential, as indicated by the FWHM of the stellar absorption features which is denoted by the shaded grey region.

\begin{figure*}
\includegraphics[width=17.0cm]{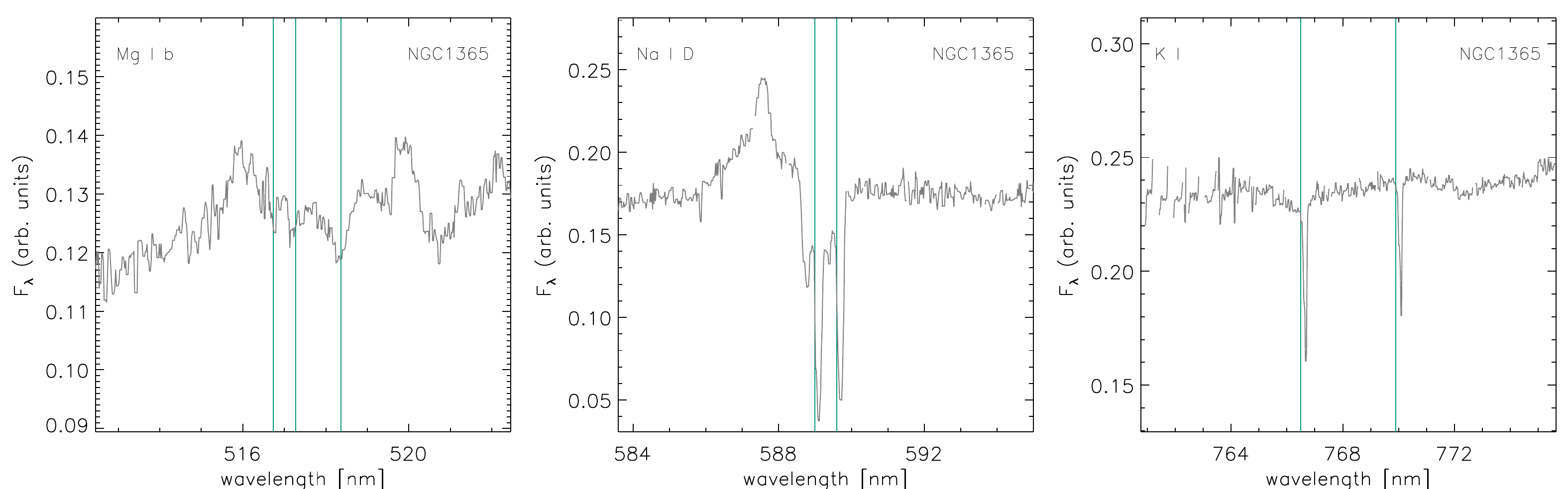}
\includegraphics[width=12cm]{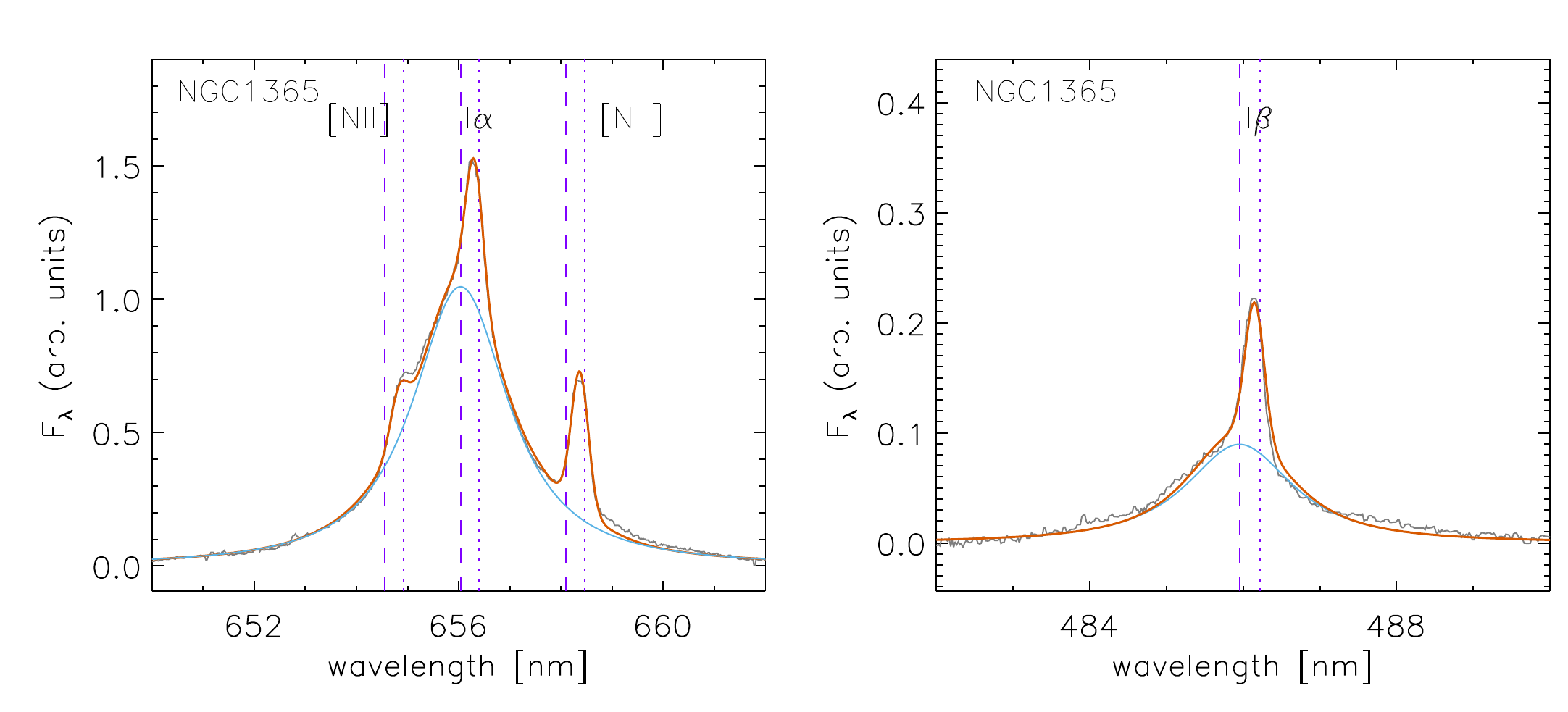}
\caption{Spectral segments and fits to features for NGC\,1365, the most complex of our AGN to fit. The arbitrary units of $F_\lambda$ are the same for all panels.
Top row: segments around the Mg\,I~b triplet, the Na\,I~D doublet, and the K\,I resonance doublet. The vertical green lines denote the adopted systemic velocity for each line. These are consistent with the Mg\,I lines which trace stellar absorption, but show that in this object the Na\,I~D and K\,I are tracing only the ISM. The Na\,I~D shows weaker blueshifted absorption and stronger redshifted absorption, while the K\,I lines show only the redshifted component. Redshifted absorption tracing outflow is possible for the receding bicone projected against an extended stellar continuum.
Bottom row: segments around the H$\alpha$ and H$\beta$ lines, showing the fits to the broad and narrow lines. The dashed and dotted purple lines indicate the velocities associated with the two components of Na\,I~D absorption. The broad line has been fit with a Moffat function, centered at -110\,km\,s$^{-1}$ with respect to systemic, which matches the blue-shifted absorption. The narrow lines are clearly offset and appear to be more associated with the redshifted absorption.}
\label{fig:n1365}
\end{figure*}

\begin{figure*}
\includegraphics[width=17.5cm]{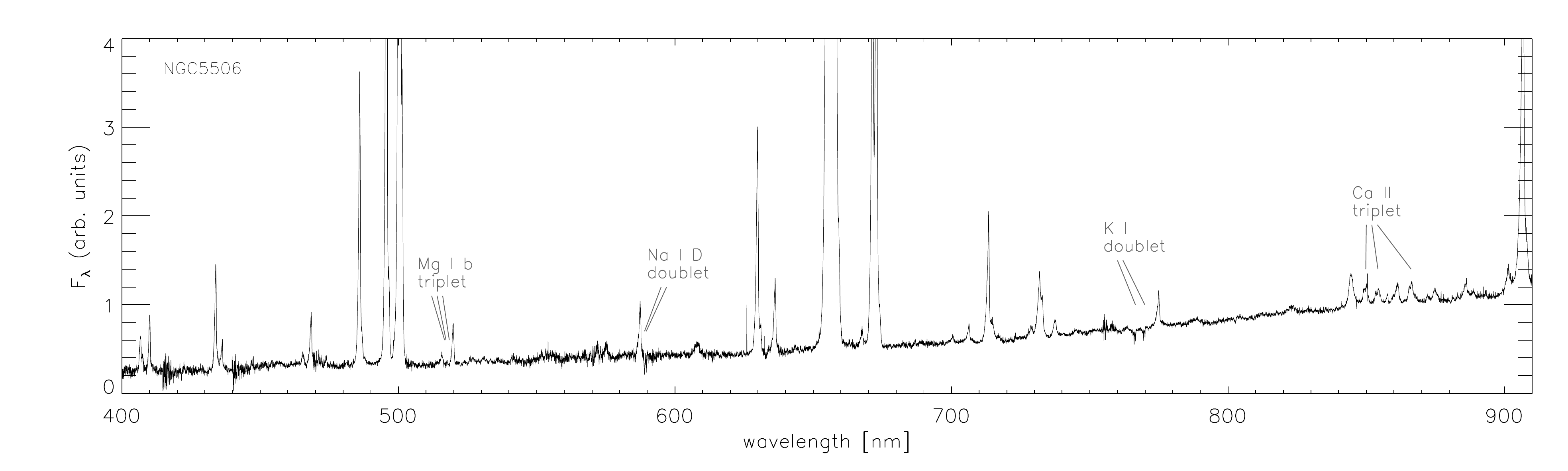}
\includegraphics[width=17.0cm]{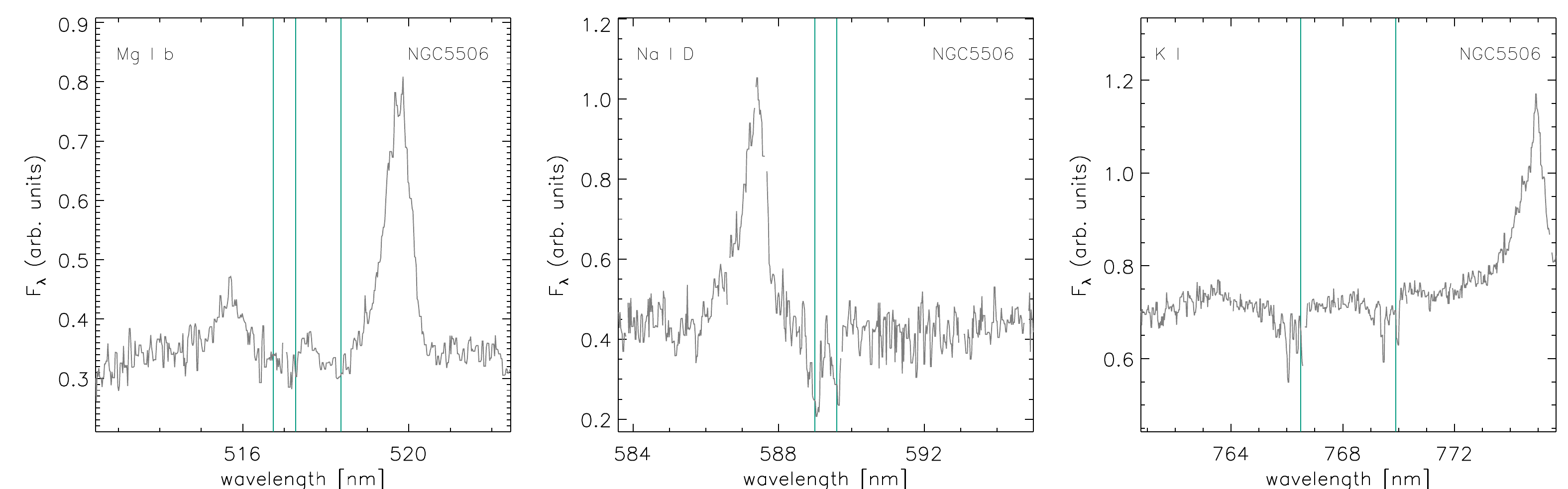}
\caption{Spectral segments for NGC\,5506.The arbitrary units of $F_\lambda$ are the same for all panels.
Top panel: there are very few absorption features measurable in the range 400--900\,nm. The most prominent are the Na\,I~D and K\,I doublets, which are dominated by the ISM. The location of the Mg\,I~b and Ca\,II triplets, which would trace the stellar kinematics, are also indicated.
Bottom row: Panels plot the Mg\,I triplet, showing it is barely measurable but plausibly at our adopted systemic velocity; as well as the Na\,I and K\,I doublets, showing that they are dominated by various components of ISM absorption at a range of velocities, mostly blue-shifted. The Ca\,II triplet region (shown in Fig.~\ref{fig:caii_linefits}) is filled with line emission. In all panels, the solid green lines indicate the wavelengths of the features for our adopted systemic velocity.}
\label{fig:ngc5506}
\end{figure*}

\bsp	
\label{lastpage}
\end{document}